\documentclass[a4paper,fleqn,usenatbib]{mnras}

\usepackage{etoolbox}
\makeatletter
\newcount\c@additionalboxlevel
\newcount\c@maxboxlevel
\patchcmd\@combinedblfloats{\box\@outputbox}{%
  \stepcounter{additionalboxlevel}%
  \box\@outputbox
}{}{\errmessage{\noexpand\@combinedblfloats could not be patched}}

\AtBeginShipout{%
  \ifnum\value{additionalboxlevel}>\value{maxboxlevel}%
    \typeout{Warning: maxboxlevel might be too small, increase to %
      \the\value{additionalboxlevel}%
    }%
  \fi 
  \@whilenum\value{additionalboxlevel}<\value{maxboxlevel}\do{%
    \typeout{* Additional boxing of page `\thepage'}%
    \setbox\AtBeginShipoutBox=\hbox{\copy\AtBeginShipoutBox}%
    \stepcounter{additionalboxlevel}%
  }%
  \setcounter{additionalboxlevel}{0}%
}
\makeatother

\usepackage[T1]{fontenc}
\usepackage{ae,aecompl}
\usepackage{footnote}
\usepackage{soul}
\usepackage[dvipsnames]{xcolor}

\usepackage{graphicx}	
\usepackage{amsmath}    
\usepackage{amssymb}	

\usepackage{booktabs,caption}
\usepackage[flushleft]{threeparttable}

\title[The dust content of the Crab Nebula]{The dust content of the Crab Nebula}

\author[Ilse De Looze et al.]{I. De Looze$^{1,2}$\thanks{E-mail: idelooze@star.ucl.ac.uk},
M.~J. Barlow$^{2}$,
R. Bandiera$^{3}$,
A. Bevan$^{2}$,
M.~F. Bietenholz$^{4,5}$,
\newauthor H. Chawner$^{6}$,
H.~L. Gomez$^{6}$,
M. Matsuura$^{6}$, 
F. Priestley$^{2}$,
R. Wesson$^{2}$ 
\\
$^{1}$Sterrenkundig Observatorium, Ghent University, Krijgslaan 281 - S9, 9000 Gent, Belgium \\
$^{2}$Dept. of Physics \& Astronomy, University College London, Gower Street, London WC1E 6BT, UK \\ 
$^{3}$Osservatorio Astrofisico di Arcetri, Largo E. Fermi 5, 50125 Firenze, Italy \\
$^{4}$Hartebeesthoek Radio Observatory, PO Box 443, Krugersdorp, 1740, South Africa \\
$^{5}$Department of Physics and Astronomy, York University, Toronto, M3J 1P3, Ontario, Canada \\
$^{6}$School of Physics and Astronomy, Cardiff University, Queens Buildings, The Parade, Cardiff CF24 3AA \\
}

\date{Accepted XXX. Received YYY; in original form ZZZ}

\pubyear{2018}

\begin{document}
\label{firstpage}
\pagerange{\pageref{firstpage}--\pageref{lastpage}}
\maketitle

\begin{abstract}
We have modelled the near-infrared to radio images of the Crab Nebula with a Bayesian SED model to simultaneously fit its synchrotron, interstellar and supernova dust emission. We infer an interstellar dust extinction map with an average $A_{\text{V}}=1.08\pm0.38$\,mag, consistent with a small contribution ($\lesssim$22$\%$) to the Crab's overall infrared emission. The Crab's supernova dust mass is estimated to be between 0.032 and 0.049\,M$_{\odot}$ (for amorphous carbon grains) with an average dust temperature $T_{\text{dust}}$=41$\pm$3\,K, corresponding to a dust condensation efficiency of 8-12$\%$. This revised dust mass is up to an order of magnitude lower than some previous estimates, which can be attributed to our different interstellar dust corrections, lower SPIRE flux densities, and higher dust temperatures than were used in previous studies. The dust within the Crab is predominantly found in dense filaments south of the pulsar, with an average $V$ band dust extinction of $A_{\text{V}}=0.20-0.39$\,mag, consistent with recent optical dust extinction studies. The modelled synchrotron power-law spectrum is consistent with a radio spectral index $\alpha_{\text{radio}}$=0.297$\pm$0.009 and an infrared spectral index $\alpha_{\text{IR}}$=0.429$\pm$0.021. We have identified a millimetre excess emission in the Crab's central regions, and argue that it most likely results from two distinct populations of synchrotron emitting particles. We conclude that the Crab's efficient dust condensation (8-12$\%$) provides further evidence for a scenario where supernovae can provide substantial contributions to the interstellar dust budgets in galaxies.
\end{abstract}




\section{Introduction}
Early dust formation in the Universe \citep{2015Natur.519..327W,2017ApJ...837L..21L,2018Natur.557..392H} has been suggested to result from an efficient condensation of dust species in the aftermaths of core-collapse supernovae (CCSNe), requiring each supernova remnant (SNR) to produce a net dust mass ranging between 0.1 and 1\,M$_{\odot}$ \citep{2003MNRAS.343..427M,2007ApJ...662..927D}. Core-collapse nucleation models are able to accommodate high dust condensation efficiencies (e.g., \citealt{2001MNRAS.325..726T,2010ApJ...713..356N,2015A&A...575A..95S,2018MNRAS.480.5580S,2019MNRAS.484.2587M}), but only during recent years have we been able to observationally confirm the hypothesis of CCSNe being dust factories by detecting the far-infrared (FIR) and sub-millimetre (submm) emission of up to 0.7\,M$_{\odot}$ of dust in several Galactic supernova remnants \citep{2010A&A...518L.138B,2012ApJ...760...96G,2014ApJ...786...55A,2017MNRAS.465.3309D,2017ApJ...836..129T,2018MNRAS.479.5101R,2019MNRAS.483...70C} and in SN\,1987A \citep{2011Sci...333.1258M,2014ApJ...782L...2I,2015ApJ...800...50M}, using the \textit{Herschel} Space Observatory \citep{2010A&A...518L...1P} and the Atacama Large Millimetre Array (ALMA). Recent dust masses inferred for a handful of mostly extra-galactic SNRs by probing the effects of dust absorption and scattering on the optical line emission profiles of supernova ejecta, were similarly high \citep{2016MNRAS.456.1269B,2017MNRAS.465.4044B,2019MNRAS.485.5192B}. 

\begin{figure}
	\includegraphics[width=8.5cm]{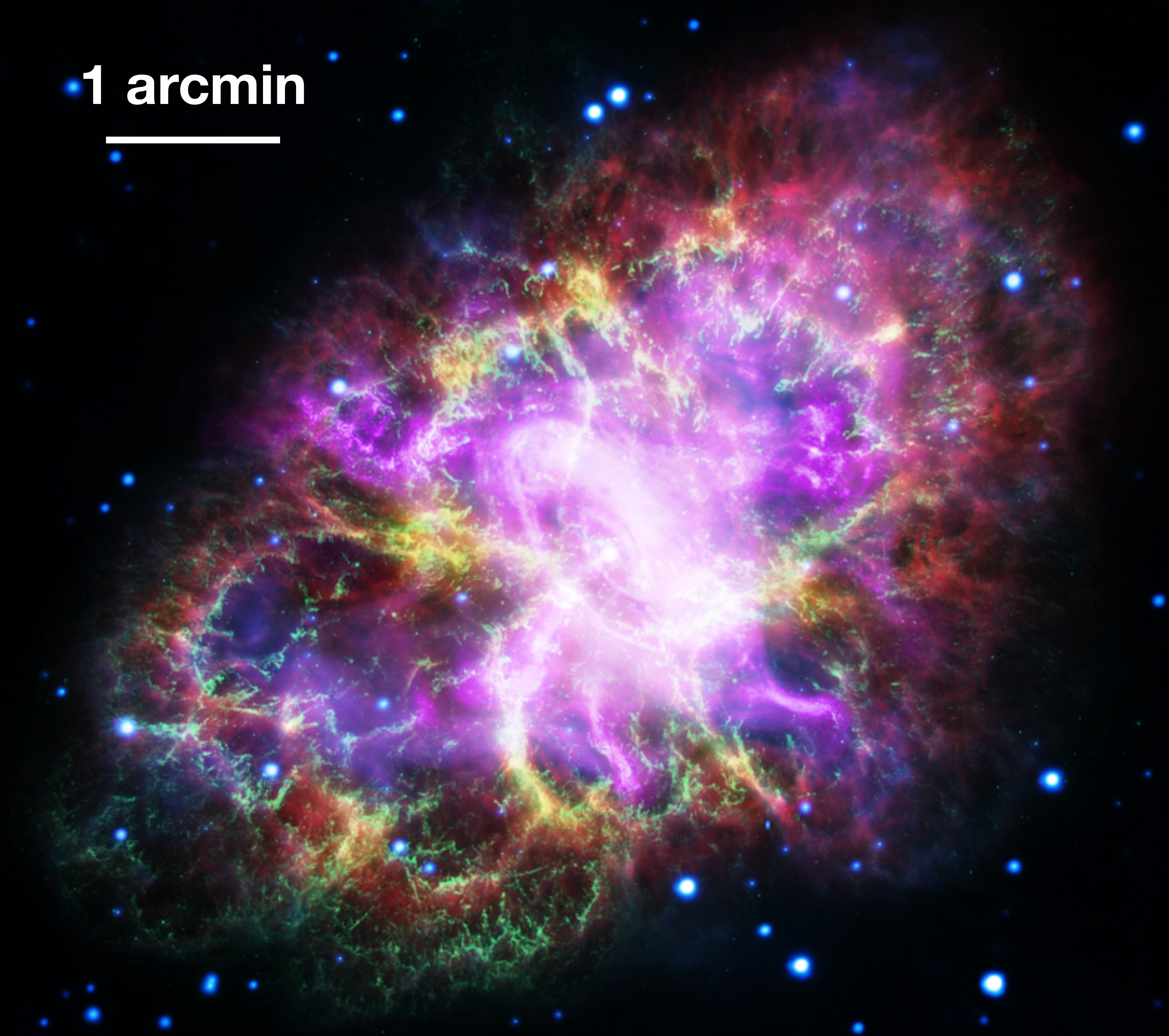}
    \caption{Composite image of the Crab Nebula combining data from five different wavelength domains: radio (red, Very Large Array), infrared (yellow, \textit{Spitzer} Space telescope), optical (green, \textit{Hubble} Space Telescope), ultraviolet (blue, XMM-Newton), and X-ray (purple, \textit{Chandra} X-ray Observatory). Image credits: NASA, ESA, NRAO/AUI/NSF and G. Dubner (University of Buenos Aires).}
    \label{Crab_composite}
\end{figure}

In this paper, we study the formation of dust in the Crab Nebula, a Galactic pulsar wind nebula (PWN) located at a distance of 2\,kpc\footnote{This distance measurement to the Crab is quite uncertain, and a recent GAIA study of the central pulsar has suggested that the distance to the Crab might be underestimated \citep{2019ApJ...871...92F}. However, since this GAIA measurement is also quite uncertain we adopt in this paper the canonical distance of 2\,kpc, which has most commonly been used.} \citep{1968AJ.....73..535T}. The Crab Nebula is believed to be the remnant of a supernova of type II-P from a 8-11\,M$_{\odot}$ progenitor star \citep{2008AJ....136.2152M,2013MNRAS.434..102S} which exploded in 1054AD. The Crab Nebula is one of the brightest radio sources and best studied objects in the sky, with most of the emission at various wavelengths powered by the enormous amounts of energy released by the Crab's central pulsar. The observable part of the Crab Nebula is shaped by a PWN ploughing into more extended supernova ejecta \citep{1977ASSL...66...53C,2008ARA&A..46..127H}. The Crab Nebula consists of the Crab pulsar, embedded in a synchrotron nebula, which emits from X-ray to radio wavelengths (see Fig. \ref{Crab_composite} for a composite image of the Crab Nebula) and consists of a magnetised relativistic plasma energised by the shocked pulsar wind. Embedded in this synchrotron nebula, there is a network of filaments composed of thermal ejecta which have been detected from optical to infrared wavebands in a range of ionisation states \citep{1979ApJ...228..179D,1990ApJ...352..172G,1992ApJ...399..611B,2006AJ....132.1610T,2012ApJ...753...72T}. Spectroscopic studies of these filaments \citep{1978ApJ...220..490M,2003MNRAS.346..885S,2007AJ....133...81M} have been interpreted with models dominated by helium, with a helium mass fraction of 85 to 90$\%$, and enhanced mass fractions for carbon, oxygen, neon, sulphur and argon, while nitrogen is depleted \citep{2012AJ....144...27S,2015ApJ...801..141O}. The C/O mass ratios above unity remain unexplained by current CCSN nucleosynthesis models, but suggest that the Crab Nebula is predominantly carbon-rich. 

These filaments also contain dust detected either in extinction \citep{1998ApJ...504..344S,2017A&A...599A.110G} or through thermal dust emission in the infrared \citep{1984ApJ...278L..29M,2004MNRAS.355.1315G,2006AJ....132.1610T,2012ApJ...753...72T,2012ApJ...760...96G}. Some of the dust also appears to be associated with dusty globules, spread across the outer part of the remnant, with sizes unresolved at current instrumental resolutions. In the pre-\textit{Herschel} era, the Crab dust masses inferred based on observations from \textit{IRAS} (0.005-0.03\,M$_{\odot}$, \citealt{1984ApJ...278L..29M}), \textit{ISO}+SCUBA (0.02-0.07\,M$_{\odot}$, \citealt{2004MNRAS.355.1315G}), and \textit{Spitzer} (1.2-5.6$\times$10$^{-3}$\,M$_{\odot}$ of silicate dust, \citealt{2006AJ....132.1610T,2012ApJ...753...72T}) remained below the dust mass production efficiency required to account for dusty galaxies observed in the early Universe. However, a total integrated analysis of \textit{Herschel} observations out to far-infrared (FIR) wavelengths, including a large set of ancillary near-infrared to radio observations to account for synchrotron contamination, resulted in $0.24^{+0.32}_{-0.08}\,M_{\odot}$ of T=28\,K silicate dust, or more likely $0.11\pm0.01\,M_{\odot}$ of T=34\,K carbon dust \citep{2012ApJ...760...96G}, or a mixture of 0.14\,M$_{\odot}$ and 0.08\,M$_{\odot}$ of silicate and carbon dust. \citet{2013ApJ...774....8T}, instead, estimated a dust mass of 0.019$^{+0.010}_{-0.003}$\,M$_{\odot}$ for a different type of carbonaceous grains with an average dust temperature of $T_{\text{dust}}$=56$\pm$2\,K. Based on a combination of photoionisation and dust radiative transfer modelling, \citet{2015ApJ...801..141O} inferred a mass of 0.18-0.27\,M$_{\odot}$ of clumped amorphous carbon dust, or mixed models with 0.11-0.13\,M$_{\odot}$ and 0.39-0.47\,M$_{\odot}$ of amorphous carbon and silicate dust, respectively, within the limits of their model uncertainties. 

In this paper, we present a spatially resolved analysis of the \textit{Herschel} observations, as well as a large ancillary dataset extending from near-infrared to radio wavelengths in order to disentangle the supernova dust emission from the synchrotron radiation that dominates the Crab's emission at nearly all wavelengths. In Section \ref{Data.sec}, we present the \textit{Herschel} observations and ancillary datasets, and the different pre-processing steps that were applied to the data. In Section \ref{GlobalSED.sec}, we present total integrated photometric measurements, describe the model used to fit the Crab's total integrated flux densities, and discuss the model contributions to various wavebands. Similarly, the resolved modelling approach is described in Section \ref{ResolvedSED.sec}. Our results for the synchrotron model, supernova dust mass and composition, and mm excess emission are discussed in detail in Sections \ref{Synchrotron.sec}, \ref{SNdustmasses.sec} and \ref{MillimetreExcess.sec}, respectively. Finally, our conclusions are summarised in Section \ref{Conclusions.sec}. In the Appendices, we outline the specifics of our Bayesian spectral energy distribution (SED) modelling method (see Appendix \ref{BayesianSED.sec}), discuss the Bayesian model residuals (see Appendix \ref{BayesianSEDresiduals.sec}) and present our model estimation of the interstellar medium (ISM) dust contribution (see Appendix \ref{GalacticDust.sec}). Appendix \ref{App_mmexcess} discusses alternative scenarios that could contribute to the mm excess emission in the Crab Nebula. In Appendix \ref{Alternative_synchrotron_model.sec}, we investigate whether the supernova dust model results depend on the assumed synchrotron spectrum, by exploring a more realistic evolutionary synchrotron model. Additional figures are assembled in Appendix \ref{Figures.sec}. 

\section{Observations and data handling}
\label{Data.sec}
\subsection{Herschel observations}
The Crab Nebula was observed with the \textit{Herschel} PACS and SPIRE instruments as part of the Guaranteed Time (GT) programme \texttt{MESS} (Mass-loss of Evolved StarS, PI: M. Groenewegen, \citealt{2011A&A...526A.162G}). The PACS and SPIRE photometric datasets from the MESS programme have been presented in detail by \citet{2012ApJ...760...96G} and \citet{2015ApJ...801..141O}. In addition, the Crab Nebula was observed as part of PACS photometric calibration observations (ObsIDs 1342183905 to 1342183912). Table \ref{DataOverview} provides an overview with details (i.e., ObsID, exposure time, central pointing) of the different sets of \textit{Herschel} photometric observations. 

The PACS photometric data from the MESS programme were obtained in scan-map mode with a scan speed of 20\,arcsec s$^{-1}$ and scan length of 22\,arcmin. A pair of two orthogonal cross-scans were observed in each waveband combination blue+red (70+160\,$\mu$m) and green+red (100+160\,$\mu$m)\footnote{Due to this observing mode, the PACS\,160\,$\mu$m map has twice the integration time of the 70 and 100\,$\mu$m maps.}. The PACS calibration observations covered a smaller region (16$\times$16 arcmin$^{2}$) compared to the 25$\times$25 arcmin$^{2}$ field of view of the MESS observations, and the scan-mapping was done at a slow speed of 10 arcsec s$^{-1}$. Due to the different observing modes used for both sets of PACS observations, and the different 1/f noise patterns in their maps, we reduced the PACS data from each observing program separately. The PACS data were reduced with the latest version of \texttt{HIPE} v14.0 \citep{2010ASPC..434..139O} using the standard \texttt{Scanamorphos} \citep{2013PASP..125.1126R} data reduction script that requires the Level 1 data from the \textit{Herschel} Science Archive (HSA) and reduces it to Level 2.5. The reduced maps from the two programs were combined into a single map for each waveband after rebinning of both maps to the same pixel grid.  The FWHM of the PACS beam in the combined image corresponds to 5.6$\arcsec$, 6.8$\arcsec$ and 11.4$\arcsec$ at 70, 100 and 160\,$\mu$m, respectively (see PACS Observers$'$ Manual\footnote{http://herschel.esac.esa.int/Docs/PACS/html/pacs$\_$om.html}). We have assumed a calibration uncertainty of 7$\%$ in each of the PACS wavebands \citep{2014ExA....37..129B}.

The SPIRE maps were obtained in ``Large Map" mode with a scan length of 30$\arcmin$ and scan speed of 30 arcsec s$^{-1}$ resulting in a map of size 32$\times$32 arcmin$^{2}$. Three pairs of orthogonal cross-scans were observed to reduce the 1/f noise in the combined maps. The SPIRE data were reduced with \texttt{HIPE} version v14.0.0 using the standard pipeline for the SPIRE Large Map Mode. As part of the data processing, the \textit{Planck} HFI maps at 857 and 545\,GHz (350 and 550\,$\mu$m) were used to determine the absolute scaling of the SPIRE maps with extended emission. The FWHM of the SPIRE beam in the final images corresponds to 18.2$\arcsec$, 24.9$\arcsec$ and 36.3$\arcsec$ at 250, 350 and 500\,$\mu$m, respectively (see SPIRE Observers$'$ Manual\footnote{http://herschel.esac.esa.int/Docs/SPIRE/html/spire$\_$om.html}). Calibration errors of 5.5$\%$ were assumed for SPIRE (SPIRE Observers$'$ Manual, \citealt{2013MNRAS.433.3062B}). More details on the various \textit{Herschel} PACS and SPIRE data reduction steps can be found in \citet{2017MNRAS.465.3309D}.


\begin{table*}
\centering
\caption{Overview of the observation identification numbers (ObsIDs), observing dates, central coordinate positions and total observing times for the \textit{Herschel} PACS and SPIRE photometric observations of the Crab Nebula.} 
\label{DataOverview}
\begin{tabular}{|l|cccccc|} 
\hline
Object & ObsID & Date & OD & RA (J2000) & DEC (J2000) & Obs. Time \\
  & & & [y-m-d] & [$^{h}$$^{m}$$^{s}$] & [$^{\circ}$ $\arcmin$ $\arcsec$] & [$s$] \\
\hline
\multicolumn{7}{|c|}{PACS photometry} \\ 
\hline
M\,1 & 1342183905 & 2009-09-15 & 124 &  05:34:31.27 & 22:01:05.46 & 2221 \\
M\,1 & 1342183906 & 2009-09-15 & 124 &  05:34:31.27 & 22:01:05.46 & 2221 \\
M\,1 & 1342183907 & 2009-09-15 & 124 &  05:34:31.32 & 22:00:47.06 & 2221 \\
M\,1 & 1342183908 & 2009-09-15 & 124 &  05:34:31.32 & 22:00:47.06 & 2221 \\
M\,1 & 1342183909 & 2009-09-15 & 124 &  05:34:31.31 & 22:01:03.79 & 2221 \\
M\,1 & 1342183910 & 2009-09-15 & 124 &  05:34:31.31 & 22:01:03.79 & 2221 \\
M\,1 & 1342183911 & 2009-09-15 & 124 &  05:34:31.40 & 22:00:47.90 & 2221 \\
M\,1 & 1342183912 & 2009-09-15 & 124 &  05:34:31.40 & 22:00:47.90 & 2221 \\
\hline
M\,1 & 1342204441 & 2010-09-13 & 487 &  05:34:31.86 & 22:01:03.98 & 1671 \\
M\,1 & 1342204442 & 2010-09-13 & 487 &  05:34:31.85 & 22:01:03.85 & 1671 \\
M\,1 & 1342204443 & 2010-09-13 & 487 &  05:34:31.86 & 22:01:03.98 & 1671 \\
M\,1 & 1342204444 & 2010-09-13 & 487 &  05:34:31.85 & 22:01:03.85 & 1671 \\
\hline
\multicolumn{7}{|c|}{SPIRE photometry} \\ 
\hline
M\,1 & 1342191181 & 2010-02-25 & 287 &  05:34:32.42 & 22:00:50.89 & 4555 \\
\hline
\end{tabular}
\end{table*}

\subsection{Ancillary data}

\subsubsection{Spitzer}
The Infrared Array Camera (IRAC, \citealt{2004ApJS..154...10F}), Multiband Imaging Photometer (MIPS, \citealt{2004ApJS..154...25R}) and Infrared Spectrograph (IRS, \citealt{2004SPIE.5487...62H}) on board the \textit{Spitzer} Space Telescope \citep{2004ApJS..154....1W} have observed the Crab Nebula as part of the Gehrz Guaranteed Time Observing Program (Program ID: 130). A detailed description of the observations, data reduction and analysis of the \textit{Spitzer} data have been presented by \citet{2006AJ....132.1610T}. We retrieved the IRAC and MIPS\,24\,$\mu$m final data products from the \textit{Spitzer} Heritage archive\footnote{http://sha.ipac.caltech.edu/applications/Spitzer/SHA/}. Extended source correction factors were applied to the IRAC images according to the recommendations of the IRAC Instrument Handbook. Flux calibration uncertainties for extended sources are assumed to be 10$\%$ in the IRAC bands (IRAC Instrument Handbook\footnote{http://irsa.ipac.caltech.edu/data/SPITZER/docs/irac/- iracinstrumenthandbook/}), and 4$\%$ and 10$\%$ in the MIPS 24 \citep{2007PASP..119..994E} and MIPS\,70\,$\mu$m \citep{2007PASP..119.1019G} bands.

The \textit{Spitzer} IRS low-resolution spectra were retrieved from the Combined Atlas of Sources with \textit{Spitzer} IRS Spectra (CASSIS, \citealt{2011ApJS..196....8L}) at the same three positions used by \citet{2012ApJ...753...72T} (with AORKEYs ``12634624", ``16200704" and ``16201216") to explore the effect of line contributions to the near- and mid-infrared continuum wavebands. Due to the extended nature of the Crab's emission, we opted for the ``tapered column" (default) extraction, which accounts for the increasing size of the point spread function (PSF) with wavelength while extracting fluxes. 

\subsubsection{WISE}
The Crab Nebula was observed as part of the all-sky Wide-field Infrared Survey Explorer (\textit{WISE}, \citealt{2010AJ....140.1868W}) in four photometric bands at 3.4, 4.6, 11.6 and 22.0\,$\mu$m at angular resolutions of FWHM = 6.1$\arcsec$,
6.4$\arcsec$, 6.5$\arcsec$, and 12$\arcsec$, respectively. We retrieved \textit{WISE} maps from the NASA/IPAC Infrared Science Archive. These were converted from DN units to Vega magnitudes using the photometric zero point magnitudes as specified in the image headers. The zero magnitude flux densities (see Explanatory Supplement to the NEOWISE Data Release Products) were applied to convert these images from magnitudes to flux densities in Jy. We assumed calibration uncertainties of 2.4$\%$, 2.8$\%$, 4.5$\%$ and 5.7$\%$ \citep{2013AJ....145....6J}, respectively. 

\subsubsection{Millimetre and radio observations}

In the millimetre wavelength range, we used the Goddard-IRAM Superconducting 2 Millimeter Observer (GISMO, \citealt{2006SPIE.6275E..1DS}) 2\,mm data observed on the IRAM\,30\,m telescope, and Multiplexed Squid TES Array at Ninety GHz (MUSTANG, \citealt{2008SPIE.7020E..05D}) 3.3\,mm observations obtained with the Green Bank Telescope (GBT). The data sets, observational details and data reduction strategy have been presented in detail in \citet{2011ApJ...734...54A}. At 2\,mm, \citet{2011ApJ...734...54A} reported a total integrated flux density of 244$\pm$24\,Jy. The total flux density at 3.3\,mm was less well constrained due to the small field of view (FOV=40$\arcsec\times40\arcsec$) and the loss of large scale emission. We relied on the total integrated millimetre-centrimetre-radio spectrum of the Crab Nebula and the best fitting synchrotron spectrum with spectral index of $\alpha$=0.297 (see Section \ref{Synchrotron.sec}) to calibrate the 2 and 3.3\,mm images to the same epoch as the \textit{Planck} observations. The total integrated flux densities at 2\,mm and 3\,mm were updated to 242.6$\pm$24.3\,Jy and 256.7$\pm$25.7\,Jy, respectively, after applying corrections for the Crab's decrease in flux (see later). We have assumed uncertainties of 10$\%$ on these flux densities to account for uncertainties in the background subtraction, absolute calibration, rescaling and expanding of the image to a recent epoch (see Section \ref{CorrectionFlux.sec}), and the possible lack of flux detected on large scales. 

In addition, we used the MAMBO\,1.3\,mm map obtained with the IRAM\,30\,mm telescope by \citet{2002A&A...386.1044B}. Aperture photometry, rescaling and expanding the 1.3\,mm map to the present epoch resulted in a total integrated flux density of 254.2\,Jy at 1.3\,mm. We have assumed a conservative uncertainty factor of 20$\%$ following the recommendations of \citet{2002A&A...386.1044B}.

A radio map at 1.4\,GHz of the Crab Nebula was assembled from observations between 1987 and 1988 with the Very Large Array (VLA) in all four configurations \citep{1990ApJ...357L..13B}. More details on the observations and data reduction can be retrieved from \citet{1990ApJ...357L..13B}. We updated the VLA 1.4\,GHz image from B1950 to J2000 coordinates. Due to the expansion of the supernova remnant, we have furthermore expanded the SNR's emission spatially by 3.11\,$\%$ accounting for the expansion rate of 0.135\,$\%$ per year \citep{2015MNRAS.446..205B} between the observing date (1987) and the reference epoch of \textit{WISE}, \textit{Herschel} and \textit{Planck} observations (2010)\footnote{The expansion does not affect the total integrated photometry measurements.}. Total integrated flux densities of 599.3\,Jy and 833.77\,Jy were inferred at 4.8 and 1.4\,GHz, respectively. Uncertainty factors of 20$\%$ were applied to account for uncertainties in the primary beam correction, absolute calibration, and rescaling and resizing of the image to the 2010 reference epoch. Table \ref{AncillaryDataOverview} provides an overview of the observational details for these millimetre and radio observations.

\subsubsection{Planck}

The \textit{Planck} ``aperture photometry" measurements for the Crab Nebula presented by \citet{2016A&A...586A.134P} were retrieved. Alternative \textit{Planck} flux measurements for the Crab Nebula are reported in the Second \textit{Planck} Catalogue of Compact Sources (PCCS2, \citealt{2016A&A...594A..26P}), but were considered inadequate due to the extended nature of the supernova remnant with respect to the \textit{Planck} beam sizes.

\begin{table*}
\centering
\caption{Overview of the ancillary datasets of the Crab Nebula with references: (1) \citet{2002A&A...386.1044B}; (2) \citet{2011ApJ...734...54A}; (3) \citet{1990ApJ...357L..13B}; (4) \citet{1991ApJ...368..231B}. }
\label{AncillaryDataOverview}
\begin{tabular}{|cccccc|} 
\hline
Instrument & Wavelength & Telescope & Obs Date &  FWHM & Reference \\
\hline
\multicolumn{6}{|c|}{Millimetre+radio data} \\ 
\hline
MAMBO & 1.3\,mm & IRAM\,30\,m & Dec 1998+Feb 2000 & 10.5$\arcsec$ & (1) \\
GISMO & 2\,mm & IRAM\,30\,m & Nov 2007 & 16.7$\arcsec$ & (2) \\
MUSTANG & 3.3\,mm & GBT & Feb 2008 & 9$\arcsec$ & (2) \\
VLA & 4.8\,GHz & VLA & 1987 & 5$\arcsec^{ *}$ & (3,4) \\
VLA & 1.4\,GHz & VLA & 1987 & 5$\arcsec^{ *}$ & (3,4) \\
\hline
\multicolumn{6}{c}{$^{*}$VLA images had been smoothed from their original resolution of $1.8\arcsec\times2.0\arcsec$ to a resolution of $5\arcsec$.}
\end{tabular}
\end{table*}

\subsection{Correction for flux decay}
\label{CorrectionFlux.sec}
Due to changes in the energy distribution of the underlying relativistic particles, the radio synchrotron emission decreases with time. To compare the Crab Nebula's observations obtained at different epochs, we scaled the observed images and/or flux densities to account for the secular decay of the Crab Nebula's emission by -0.202$\%$ per year to a reference epoch of 2010. This flux decay rate was inferred by \citet{2015MNRAS.446..205B} as the weighted average of earlier estimates from \citet{1985ApJ...293L..73A}, \citet{2007ARep...51..570V} and \citet{2011ApJS..192...19W}. Our applied rate assumes that the flux decay is wavelength independent, which seems to be largely supported by recent studies \citep{2007ARep...51..570V,2011ApJS..192...19W}. To correct the resolved maps, we assume that the flux decay does not have any spatial dependence. We neglect any secular variations on local scales which appear to be present predominantly within a 1$\arcmin$ region around the pulsar and can account for changes up to 10$\%$ of the peak flux \citep{2015MNRAS.446..205B}.

\subsection{Line contamination}
We corrected the MIPS\,24\,$\mu$m images and total integrated flux densities for line contamination based on the recommended values reported by \citet{2012ApJ...760...96G}, who accounted for line contributions of 43$\pm$6$\%$\footnote{This value reported by \citet{2012ApJ...753...72T} accounts for the contribution of line emission after synchrotron subtraction. Based on our best-fit synchrotron model, this value translates into a 24$\pm$6$\%$ contribution of line emission to the total integrated MIPS\,24\,$\mu$m flux density.} in the MIPS\,24\,$\mu$m waveband, and 4.9$\pm$0.05$\%$ and 8.7$\pm$0.3$\%$ in the PACS\,70 and PACS\,100\,$\mu$m wavebands. We assumed the same correction factor (43$\pm$6) for the WISE\,22\,$\mu$m waveband. We estimated an average contribution of 16$\pm$8$\%$ from [Ar~{\sc{iii}}] 8.99\,$\mu$m and [Ar~{\sc{ii}}] 6.99\,$\mu$m line emission to the IRAC\,8\,$\mu$m waveband based on the \textit{Spitzer} IRS spectra extracted from three dense filaments in the southern half of the Crab. We attempted to similarly quantify the line contributions to the WISE\,12\,$\mu$m waveband, and found that the [Ne~{\sc{ii}}] 12.81\,$\mu$m, [Ne~{\sc{V}}] 14.32\,$\mu$m, [Ne~{\sc{iii}}] 15.56\,$\mu$m and [S~{\sc{iv}}] 10.51\,$\mu$m lines account for a non-negligible portion of the WISE\,12\,$\mu$m emission, with contributions that significantly vary throughout the remnant (possibly due to the sensitivity of these lines to the local ionisation conditions). We therefore omitted the WISE\,12\,$\mu$m images from our analysis to prevent the introduction of a bias in the fitting results driven by poorly constrained line contributions in the WISE\,12\,$\mu$m waveband. The PACS\,160\,$\mu$m and three SPIRE wavebands were found to have a negligible ($<$1$\%$) contribution from line emission to the broadband continuum flux densities. Due to the limited spatial coverage of \textit{Spitzer} IRS spectra, \textit{Herschel} PACS and SPIRE spectroscopic observations, and the insufficient spatial resolution of ISO observations, we were unable to infer spatial variations in line-to-continuum ratios across the remnant. 

\subsection{Image processing}

To compare observations obtained by various instruments with different intrinsic resolutions, we subtracted a background from all \textit{Herschel} and ancillary images using a set of background apertures with radius equal to 4$\times$FWHM randomly placed in the background regions of the respective images. Convolution of the images to the SPIRE\,500\,$\mu$m image resolution was performed using the kernels from \citet{2011PASP..123.1218A}, after which images were regridded to the size (14$\arcsec$$\times$14$\arcsec$) of pixels in the SPIRE\,500\,$\mu$m image. 
The \textit{Spitzer} and \textit{WISE} images were corrected for Galactic foreground extinction. We adopted a reddening of E(B-V) = 0.39$\pm$0.03 along the line of sight to the Crab Nebula at the assumed distance of the remnant (i.e., 2\,kpc) from the 3D dust reddening map constructed by \citet{2015ApJ...810...25G} based on Pan-STARRS1 (PS1) and 2MASS photometry for 800 and 200 million stars, respectively. For a Galactic reddening law with $R_{\text{V}}$ = 3.1 \citep{1999PASP..111...63F}, we derive a $V$ band dust extinction of $A_{\text{V}}$=1.21, which compares reasonably well to the reddening ($A_{\text{V}}$=1.6$\pm$0.2) inferred from optical spectrophotometric measurements of two Crab filaments \citep{1973ApJ...180L..83M}. We apply dust extinction correction factors of 1.06, 1.04, 1.03 and 1.02 to the IRAC\,3.6, 4.5, 5.8 and 8.0\,$\mu$m bands. The latter factors are somewhat lower compared to the correction factors of 1.11 and 1.08 applied to the IRAC\,3.6 and 4.5\,$\mu$m channels by \citet{2012ApJ...753...72T}, and was based on the hydrogen column density of the foreground gas derived from soft X-ray absorption \citep{2001A&A...365L.212W}. 

\subsection{Colour correction}
We did not apply colour corrections to the observed flux densities, as the model SEDs are convolved with the appropriate filter response curves\footnote{We were not able to retrieve filter response curves for the millimetre and radio observations of the Crab Nebula, and have determined monochromatic flux densities in those wavebands.} before comparing the model to the observations during every step of the Bayesian fitting algorithm. We did apply a correction factor for the dependence of the effective SPIRE beam area on the shape of the spectrum due to the absolute SPIRE calibration in units of flux density per beam. From the different model components contributing to the SPIRE wavebands, we have calculated the spectral index, $\alpha_{\text{S}}$, in $I_{\text{S}}$($\nu$)~=~$I_{\text{S}}$($\nu_{0}$)($\nu$/$\nu_{0}$)$^{\alpha_{\text{S}}}$, where $I_{\text{S}}$($\nu_{0}$) is the surface brightness at a reference frequency $\nu_{0}$. We then applied the SPIRE beam area correction factors for different values of $\alpha_{\text{S}}$ based on the tabulated values in Table 5.4 of the SPIRE Observers$'$ Manual. 

\begin{table*}
	\centering
				\caption{Overview of the total integrated photometry for the Crab Nebula measured within an elliptical aperture with semi-major and -minor axes $245\arcsec\times163\arcsec$ centred on the position of the pulsar (RA, DEC)~=~(RA: 05$^{h}$34$^{m}$31.94$^{s}$ DEC: 22$^{\circ}$0$\arcmin$52.18$\arcsec$) with a position angle of 40$^{\circ}$ (to match the same flux region as assumed in \citealt{2012ApJ...760...96G}). Columns 1, 2 and 4 list the waveband, total and emission line flux densities, respectively. Total flux densities were extinction corrected, and were already scaled to our 2010 reference epoch. Columns 5, 7, 8 and 10 report the emission due to synchrotron radiation, interstellar (IS) dust emission, SN dust emission and mm excess emission, respectively. The SN dust flux densities were inferred from the total integrated model for amorphous carbon grains ``a-C" (see Section \ref{GlobalSED.sec}). The values in parentheses represent the contributions of the different emission components relative to the total model flux, which have also been displayed in Figure \ref{Crab_schematic}. Because the Bayesian model is fitting the observed flux densities within the limits of uncertainty, the fractional contributions might not exactly add up to a 100$\%$. Columns 3, 6 and 9 present the total, synchrotron and SN dust flux densities, inferred by \citet{2012ApJ...760...96G} (G12), for comparison. A dash in the Table indicates that the contributions are insignificant (i.e., lower than 0.01$\%$), while a forward slash is used for wavebands that were not used to constrain our model.} %
	\begin{tabular}{|l|ccccccccc} 
		\hline
		        (1) & (2) & (3) & (4) & (5) & (6) & (7) & (8) & (9) & (10) \\
                Waveband & Total & \textcolor{purple}{Total} & Line & Synchr. & \textcolor{purple}{Synchr.}  & IS dust & SN dust & \textcolor{purple}{SN dust} & Mm excess \\
                 & Obs. $F_{\nu}$ & \textcolor{purple}{Obs. $F_{\nu}$} & Obs. $F_{\nu}$ & Model $F_{\nu}$ & \textcolor{purple}{Model $F_{\nu}$} & Model $F_{\nu}$ & Model $F_{\nu}$ & \textcolor{purple}{Model $F_{\nu}$} & Model $F_{\nu}$ \\
                 &  Jy & \textcolor{purple}{Jy (G12)} &  Jy [$\%$] & Jy [$\%$] & \textcolor{purple}{Jy (G12)} &  Jy [$\%$] & Jy [$\%$] & \textcolor{purple}{Jy (G12} & Jy [$\%$]  \\
		\hline 
               IRAC\,3.6\,$\mu$m & 12.0$\pm$1.3 & \textcolor{purple}{12.6$\pm$0.2} & - & 11.9$^{+0.8}_{-0.6}$ & \textcolor{purple}{13.2} & - & - & - & - \\ 
                &  &  &  & [99.2$\%$] & & &  &  &  \\                
               IRAC\,4.5\,$\mu$m & 14.5$\pm$1.6 & \textcolor{purple}{14.4$\pm$0.3} & - & 13.1$^{+0.8}_{-0.6}$ & \textcolor{purple}{14.5} & - & - & - & - \\  
                &  &  &  & [90.3$\%$] & & &  &  &  \\                 
               IRAC\,5.8\,$\mu$m & 15.3$\pm$2.2 & \textcolor{purple}{16.8$\pm$0.1} & - & 14.5$^{+1.0}_{-0.6}$ & \textcolor{purple}{16.1} & - & - & - & - \\ 
                & & &  & [94.8$\%$] & & &  &  &  \\   
               IRAC\,8\,$\mu$m & 19.7$\pm$2.0 & \textcolor{purple}{18.3$\pm$0.1} & - & 16.6$^{+1.2}_{-0.7}$ & \textcolor{purple}{18.5} & - & 0.05$^{+0.04}_{-0.05}$ & - & - \\ 
                & & &  & [84.3$\%$] & & & [0.3$\%$] &  &  \\                               
               \textit{WISE}\,3.4\,$\mu$m  & 11.2$\pm$0.3 & \textcolor{purple}{12.9$\pm$0.6}  & - & 11.6$^{+0.8}_{-0.6}$ & \textcolor{purple}{13.1} & - & - & - & -  \\ 
                &  &  &  & [103.6$\%$] & & &  &  & \\                
               \textit{WISE}\,4.6\,$\mu$m  & 14.1$\pm$0.4 & \textcolor{purple}{14.7$\pm$0.8} & - & 13.2$^{+0.8}_{-0.6}$ & \textcolor{purple}{14.6} & - & - & - & - \\   
                & & & & [93.6$\%$] & & &  &  &  \\                  
               \textit{WISE}\,12\,$\mu$m & 32.2$\pm$1.4 & \textcolor{purple}{/} & / & / & \textcolor{purple}{/} & / & / & / & / \\   
                & & & & & & &  &  &  \\                 
               \textit{WISE}\,22\,$\mu$m  & 62.0$\pm$3.7 & \textcolor{purple}{60.3$\pm$3.5} & 14.6$\pm$3.7 & 25.6$^{+2.8}_{-1.0}$ & \textcolor{purple}{28.1} & 0.5$^{+0.5}_{-0.3}$ & 17.8$^{+10.3}_{-11.4}$ & - & -   \\
                & & & [23.6$\%$] & [41.3$\%$] & & [0.8$\%$] & [28.7$\%$] &  &    \\               
               MIPS\,24\,$\mu$m & 58.7$\pm$2.9 & \textcolor{purple}{59.8$\pm$6.0} & 13.9$\pm$3.5 & 26.3$^{+3.0}_{-1.0}$ & \textcolor{purple}{29.2} & 0.5$^{+0.6}_{-0.3}$ & 20.9$^{+13.9}_{-11.8}$ & \textcolor{purple}{17.2} & -  \\
                &  &  & [23.6$\%$] & [44.8$\%$] & & [0.9$\%$] & [35.6$\%$] &  &   \\               
               PACS\,70\,$\mu$m & 220.9$\pm$19.5 & \textcolor{purple}{212.8$\pm$21.3} & 10.8$\pm$0.1 & 41.8$^{+7.4}_{-1.5}$ & \textcolor{purple}{45.6} & 2.6$^{+4.2}_{-1.6}$ & 168.2$^{+58.4}_{-63.2}$ & \textcolor{purple}{156.8} & -  \\ 
                &  &  & [4.9$\%$] & [18.9$\%$] & & [1.2$\%$] & [76.1$\%$] &  &   \\   
                MIPS\,70\,$\mu$m & 183.9$\pm$18.6 & \textcolor{purple}{208.0$\pm$33.3} & / & / & \textcolor{purple}{45.6} & / & / & / & / \\   
                & & & & & & &  &  &  \\         
               PACS\,100\,$\mu$m & 208.2$\pm$18.5 & \textcolor{purple}{215.2$\pm$21.5} & 18.1$\pm$0.6 & 48.4$^{+9.5}_{-1.7}$ & \textcolor{purple}{52.9} & 9.6$^{+13.9}_{-5.9}$ & 142.2$^{+40.0}_{-55.9}$ & \textcolor{purple}{143.6} & 0.03$^{+1.10}_{-0.03}$ \\  
                &  &  & [8.7$\%$] & [23.2$\%$] & & [4.6$\%$] & [68.3$\%$] &  & [0.01$\%$] \\                 
               PACS\,160\,$\mu$m & 166.7$\pm$14.0 & \textcolor{purple}{141.8$\pm$14.2} & - & 58.9$^{+13.2}_{-2.2}$ & \textcolor{purple}{64.3} & 23.9$^{+24.9}_{-13.3}$  & 69.9$^{+35.6}_{-35.7}$ & \textcolor{purple}{77.5} & 0.3$^{+3.4}_{-0.3}$ \\ 
                &  &  &  & [35.3$\%$] & & [14.3$\%$] & [41.9$\%$] & & [0.2$\%$] \\                
               SPIRE\,250\,$\mu$m  & 110.7$\pm$6.7 & \textcolor{purple}{103.4$\pm$7.2} &- & 70.2$^{+17.6}_{-2.7}$ & \textcolor{purple}{77.5} & 23.9$^{+19.1}_{-12.3}$ & 25.1$^{+22.2}_{-14.3}$ & \textcolor{purple}{25.9} & 1.6$^{+7.7}_{-1.6}$ \\   
                & & &  & [63.4$\%$] & & [21.6$\%$] & [22.7$\%$] & & [1.4$\%$] \\                  
               SPIRE\,350\,$\mu$m  & 106.3$\pm$6.0 & \textcolor{purple}{102.4$\pm$7.2} &- & 81.0$^{+22.0}_{-3.3}$ & \textcolor{purple}{89.2} & 14.7$^{+10.0}_{-7.2}$ & 10.4$^{+12.8}_{-6.2}$ & \textcolor{purple}{13.2} & 5.0$^{+12.5}_{-5.0}$  \\ 
                & & & & [76.2$\%$] & & [13.8$\%$] & [9.8$\%$] & & [4.7$\%$]  \\                
               SPIRE\,500\,$\mu$m  & 107.0$\pm$6.0 & \textcolor{purple}{129.0$\pm$9.0} & - & 93.6$^{+27.7}_{-4.1}$ & \textcolor{purple}{103.5} & 6.7$^{+4.0}_{-3.1}$ & 3.7$^{+6.6}_{-2.2}$ & \textcolor{purple}{10.1} & 14.0$^{+17.8}_{-14.0}$ \\ 
                & & & & [87.5$\%$] & & [6.3$\%$] & [3.5$\%$] & & [13.1$\%$] \\                
               GISMO\,2\,mm  & 242.6$\pm$24.3 & - & - & 170.6$^{+58.3}_{-9.0}$ & - & 0.1$^{+0.1}_{-0.1}$ & 0.06$^{+0.26}_{-0.04}$ & - & 78.9$^{+34.8}_{-78.7}$   \\ 
                &  &  &  & [70.3$\%$] & & [0.04$\%$] &  [0.02$\%$] &  & [32.5$\%$]  \\                
               VLA\,4.8\,GHz  & 599.3$\pm$125.7 & - & - & 599.1$^{+22.0}_{-25.5}$ & - & - & - & - & - \\ 
               & & & & [100.0$\%$] & &  &  &  &  \\ 
               VLA\,1.4\,GHz  & 833.8$\pm$174.9 & - & - & 861.9$^{+54.6}_{-62.1}$ & - & - & - & - & -  \\
                &  &  &  & [103.4$\%$] & & &  &  &  \\       
		\hline
	\end{tabular}
		\label{Table_Fluxfraction}
\end{table*}

\section{Total integrated spectrum of the Crab Nebula}
\label{GlobalSED.sec}
\subsection{Total integrated flux densities}
\label{GlobalFluxes.sec}
We have measured the total emission of the Crab Nebula in various wavebands from the near-infrared to radio wavelength domain within an elliptical aperture ($245\arcsec\times163\arcsec$) centred on the Crab pulsar (RA: 05$^{h}$34$^{m}$31.94$^{s}$ DEC: 22$^{\circ}$0$\arcmin$52.18$\arcsec$) with a position angle of 40$^{\circ}$ (see Table \ref{Table_Fluxfraction}, second column). Due to our choice of aperture, we can compare our \textit{Herschel} photometric measurements to the flux densities reported in \citet{2012ApJ...760...96G} (see Table \ref{Table_Fluxfraction}). For most wavebands, both sets of flux densities are consistent within the errors, with the exception of the PACS\,160\,$\mu$m and SPIRE\,500\,$\mu$m bands. Our PACS\,160\,$\mu$m flux density is 17.6$\%$ higher, while the SPIRE\,500\,$\mu$m flux density inferred here is 17$\%$ lower compared to the \citet{2012ApJ...760...96G} results. At PACS\,160\,$\mu$m, the (minor) difference is thought to arise from the deeper imaging of the nebula (with the inclusion of the deep PACS calibration images), as the applied data reduction techniques were similar in both cases. The slightly lower SPIRE\,500\,$\mu$m flux density could result from updated SPIRE beam size measurements (resulting in a decrease in flux density of 8$\%$) or the use of \textit{Planck} data to recover the absolute calibration for extended structures, or a combination of both effects. \citet{2019arXiv190303389N} infer a SPIRE\,500\,$\mu$m flux density (103.3$\pm$8.4\,Jy) based on a ``flux threshold" method to identify the Crab's emission (rather than aperture photometry) and after subtracting any background and interstellar dust emission along the Crab's line of sight, which is consistent with our background-subtracted and ISM dust-corrected SPIRE\,500\,$\mu$m flux density (100.3$\pm$6.8\,Jy).

\subsection{Model description}
\label{GlobalModel.sec}
The near-infrared to radio emission of the Crab Nebula is dominated by four emission components: synchrotron, supernova dust, interstellar dust, and an (as yet) unidentified mechanism at millimetre wavebands. 

\subsubsection{Interstellar dust contribution}
We have estimated the contribution from interstellar dust based on 2033 neighboring pixels dominated by ISM dust emission, and found a maximum contribution of 22$\%$ in the SPIRE\,250\,$\mu$m waveband, and lower contributions in all other bands (see Appendix \ref{GalacticDust.sec}). We have subtracted an average ISM dust contribution from the total integrated flux densities and from the individual pixels in the maps prior to the SED modelling. Uncertainties inherent to these ISM dust corrections were added in quadrature to the measurement uncertainties. The residual flux densities were then modelled with a three component model aiming to reproduce the Crab's synchrotron radiation, warm and cold SN dust emission, and excess emission in the millimetre wavelength range. We attempted to run models without a mm excess component, but those synchrotron+SN dust models tend to overestimate the SPIRE flux densities by a factor of 2 or more.

\subsubsection{Synchrotron model}
Synchrotron radiation is generally characterised by a non-thermal power-law spectrum ($F_{\nu}$ $\propto$ $\nu^{\alpha}$ for a given spectral index $\alpha$). With the optical spectral index ($\alpha$=0.6 to 1.0, \citealt{1993A&A...270..370V}) being steeper than the radio one ($\alpha$=0.3, \citealt{1977A&A....61...99B}), a spectral break between 1 and 1000\,$\mu$m was inferred \citep{1984ApJ...278L..29M,1987ASIC..195..209W}, which is representative of the upper limit on the energy of the electron population. We have therefore modelled the synchrotron spectrum as a broken power law spectrum: 
\begin{equation}
\begin{aligned}
F_{\nu}  ={} & F_{\nu_{0}} \times \left(\frac{\nu}{\nu_{0}}\right)^{-\alpha_{\text{radio}}}~~~~~~~~~~~~\left(\text{if}~\lambda\ge\lambda_{\text{break}}\right) \\
             & F_{\nu_{0}} \times \left(\frac{\nu}{\nu_{0}}\right)^{-\alpha_{\text{IR}}} \times a ~~~~~~~~~~\left(\text{if}~\lambda<\lambda_{\text{break}}\right)
\end{aligned}
\end{equation}
with 
\begin{equation}
a=\left(\frac{\nu_{\text{break}}}{\nu_{\text{0}}}\right)^{-\alpha_{\text{radio}}} \times \left(\frac{\nu_{\text{break}}}{\nu_{\text{0}}}\right)^{+\alpha_{\text{IR}}}
\end{equation}
being a scaling factor that guarantees the continuity of the two power law slopes, and where the reference frequency is chosen as $\nu_{\text{0}}$=1.4\,GHz.
The radio and IR spectral indices, $\alpha_{\text{radio}}$ and $\alpha_{\text{IR}}$, are allowed to vary uniformly between 0.1 and 0.4, and between 0.3 and 1.0, respectively. The model prior furthermore excludes any models for which $\alpha_{\text{IR}}$$<$$\alpha_{\text{radio}}$, to avoid unphysical models with a flattening of the synchrotron spectrum at IR wavelengths. As earlier values for the synchrotron break wavelength were poorly constrained, we assume a flat prior for $\lambda_{\text{break}}$ within a large parameter space ranging from 20\,$\mu$m to 2\,cm.  In addition, the normalisation of the synchrotron spectrum, $F_{\text{1.4\,GHz}}$ has a flat prior ranging between 300 and 3000\,Jy. 

As a monochromatic break, which is currently assumed in our broken power-law synchrotron model, will be too sharp to mimic an evolutionary synchrotron spectrum, we have tested how our modelling results (in particular the supernova dust masses) are affected by our simple synchrotron model assumption (see Appendix \ref{Alternative_synchrotron_model.sec} for more info). More specifically, we have replaced our broken power-law spectrum with a synchrotron spectrum with a smooth evolutionary break. The evolutionary synchrotron model results in slightly worse fits to the total integrated Crab spectrum, in particular at submm wavelengths, with a reduced $\chi^{2}_{\text{dust}}$=8.2 compared to 1.2 for our broken power-law synchrotron model. Further refinement of this evolutionary synchrotron model will be deferred to future work, due to the complexity of the spatial and secular variations in the energy distribution of relativistic particles in the Crab. However, we have verified that our choice of synchrotron model does not affect the supernova dust model parameters, and therefore decided to apply a broken power-law synchrotron spectrum (which requires no prior assumptions about the energy distribution of electrons and its evolution with time in the Crab) for the modelling presented in the remainder of this work.

\begin{figure*}
	\includegraphics[width=16.5cm]{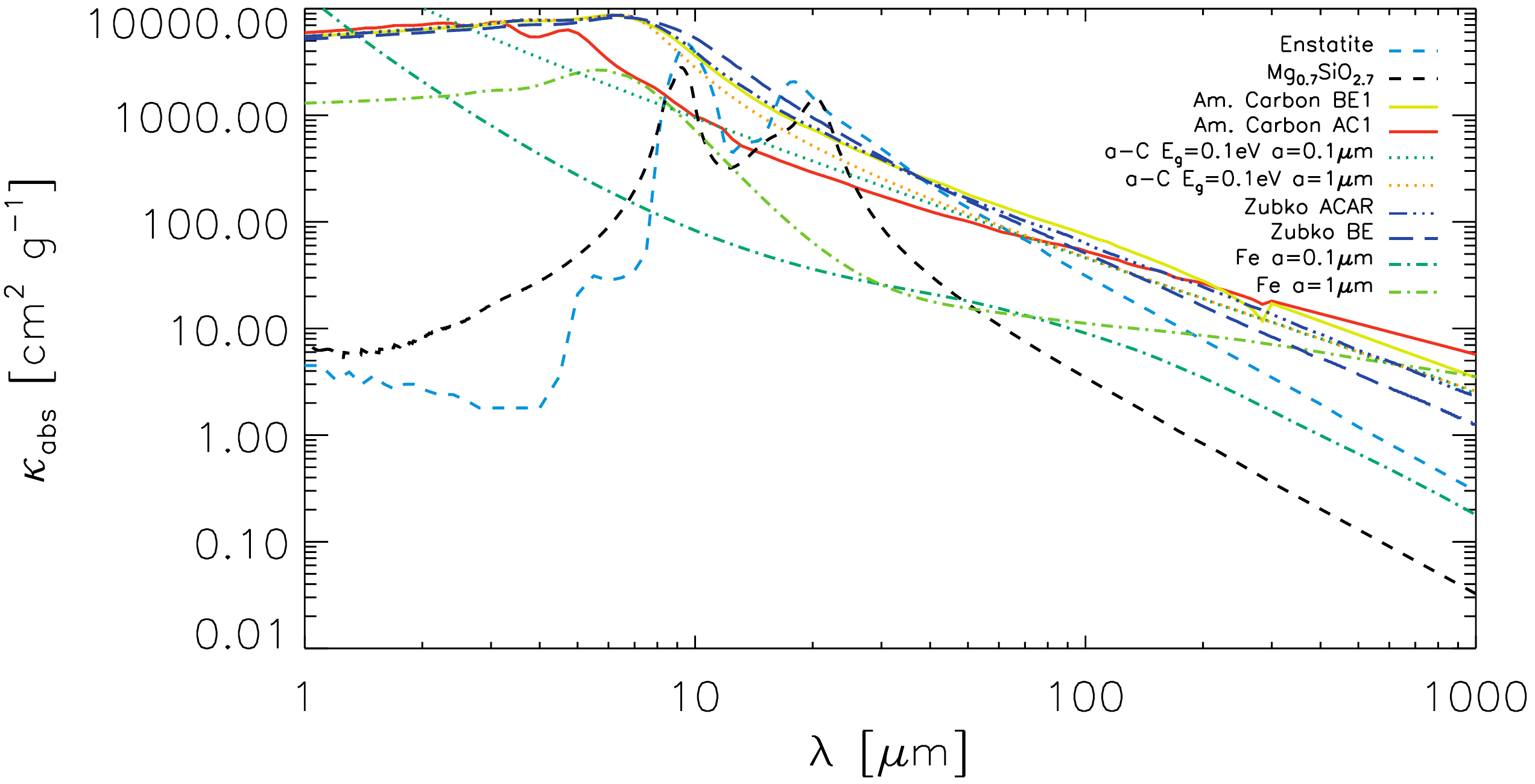}
    \caption{The variation as a function of wavelength in dust mass absorption coefficients, $\kappa_{\text{abs}}$, calculated based on Mie theory for spherical grains with sizes $a$=0.1 and 1\,$\mu$m, for different dust species. The legend in the top right corner clarifies the composition and grain size of the different dust species (if not specified, the grain size is assumed to be a=1\,$\mu$m).}
    \label{Crab_dustspecies_kabs}
\end{figure*} 

\subsubsection{Supernova dust model}
We have modelled the SN dust emission with a two-component warm+cold optically thin modified blackbody (MBB) model with a fixed dust composition, where the dust emission is given by:
\begin{equation}
F_{\nu}~=~\frac{M_{\text{dust}}}{D^{2}}~\kappa_{\nu}~B_{\nu}(T_{\text{dust}}).
\end{equation}
Rather than assuming a single dust mass absorption coefficient $\kappa_{\nu_{0}}$, and assuming a power law distribution $\kappa_{\nu}=\kappa_{\nu_{0}}\times(\nu/\nu_{0})^{\beta}$ with dust emissivity index $\beta$, we have $``$modified$"$ the black body function based on the $\kappa_{\nu}$ spectrum inferred for several specific grain species. Due to the high abundance of carbon in the Crab Nebula \citep{2015ApJ...801..141O}, we have attempted to model the total integrated SED using various carbonaceous grain species: amorphous carbon ``AC1" and ``BE1" grains from \citet{1991ApJ...377..526R}, ``ACAR" and ``BE" grains from \citet{1996MNRAS.282.1321Z} and amorphous carbon a-C grains with band gap $E_{\text{g}}$=0.1\,eV from \citet{2012A&A...540A...1J,2012A&A...540A...2J,2012A&A...542A..98J}. We have furthermore explored a variety of other grain species: silicate-type grains (MgSiO$_{3}$, \citealt{1995A&A...300..503D}; Mg$_{0.7}$SiO$_{2.7}$, \citealt{2003A&A...408..193J}) and pure iron grains (with sizes of a=0.1\,$\mu$m and 1\,$\mu$m; priv. comm. with T. Nozawa and E. Dwek). For silicate-type and carbonaceous grains, we have assumed a grain size of 1\,$\mu$m. To verify the effect of our grain size assumption, we have also modelled the total integrated SED using amorphous carbon ``a-C" grains with a size $a$=0.1\,$\mu$m. Figure \ref{Crab_dustspecies_kabs} provides an overview of how the dust mass absorption coefficients vary as a function of infrared and submm wavelengths for the various grain species explored in this work, covering a wide range of dust emissivities and absolute dust opacities. 
The warm and cold dust temperatures are inferred from a flat prior with $T_{\text{warm}}\in$[40K,100K] and $T_{\text{cold}}\in$[12K,60K] with $T_{\text{warm}}>T_{\text{cold}}$. The logarithmic values of the cold and warm dust masses are sampled from $\log\,M_{\text{warm}}\in[-7,1]\log\,M_{\odot}$ and $\log\,M_{\text{cold}}\in[-4,1]\log\,M_{\odot}$.
 
\subsubsection{Millimetre excess model}
To model the millimetre excess emission, we assume the following spectrum:
\begin{equation}
\label{Eq_mmexcess}
F_{\nu}~=~F_{\text{mm,peak}}*\exp(-\frac{\log(\nu/\nu_{\text{peak}})^{2}}{2\sigma^{2}}),
\end{equation}
which is fully characterised by its width ($\sigma$, with a flat prior between 0.1 and 1.0), peak frequency ($\nu_{\text{peak}}$, with a flat prior between 30 and 400\,GHz) and peak amplitude ($F_{\text{mm,peak}}$, with a flat prior between 1 and 300\,Jy). Due to the uncertain nature of the excess emission (see Section \ref{MillimetreExcess.sec}), we have opted for a rather generic functional form to fit the millimetre excess. 

In summary, we have a total of 11 free parameters in our model: four free parameters to model the synchrotron spectrum ($\alpha_{\text{radio}}$, $\alpha_{\text{IR}}$, $\lambda_{\text{break}}$, $F_{\text{1.4\,GHz}}$), four to model the supernova dust emission ($T_{\text{warm}}$, $\log$\,$M_{\text{warm}}$, $T_{\text{cold}}$, $\log$\,$M_{\text{cold}}$), and three to model the excess millimetre emission ($\nu_{\text{peak}}$, $\sigma$, $F_{\text{mm,peak}}$). Due to the large number of parameters, we have employed a Bayesian inference method coupled to a Markov Chain Monte Carlo (MCMC) algorithm to search the entire parameter space in an efficient way, and to reveal any parameter degeneracies (see Appendix \ref{BayesianSED.sec} for a detailed description of the method). 

\begin{figure*}
	\includegraphics[width=16.5cm]{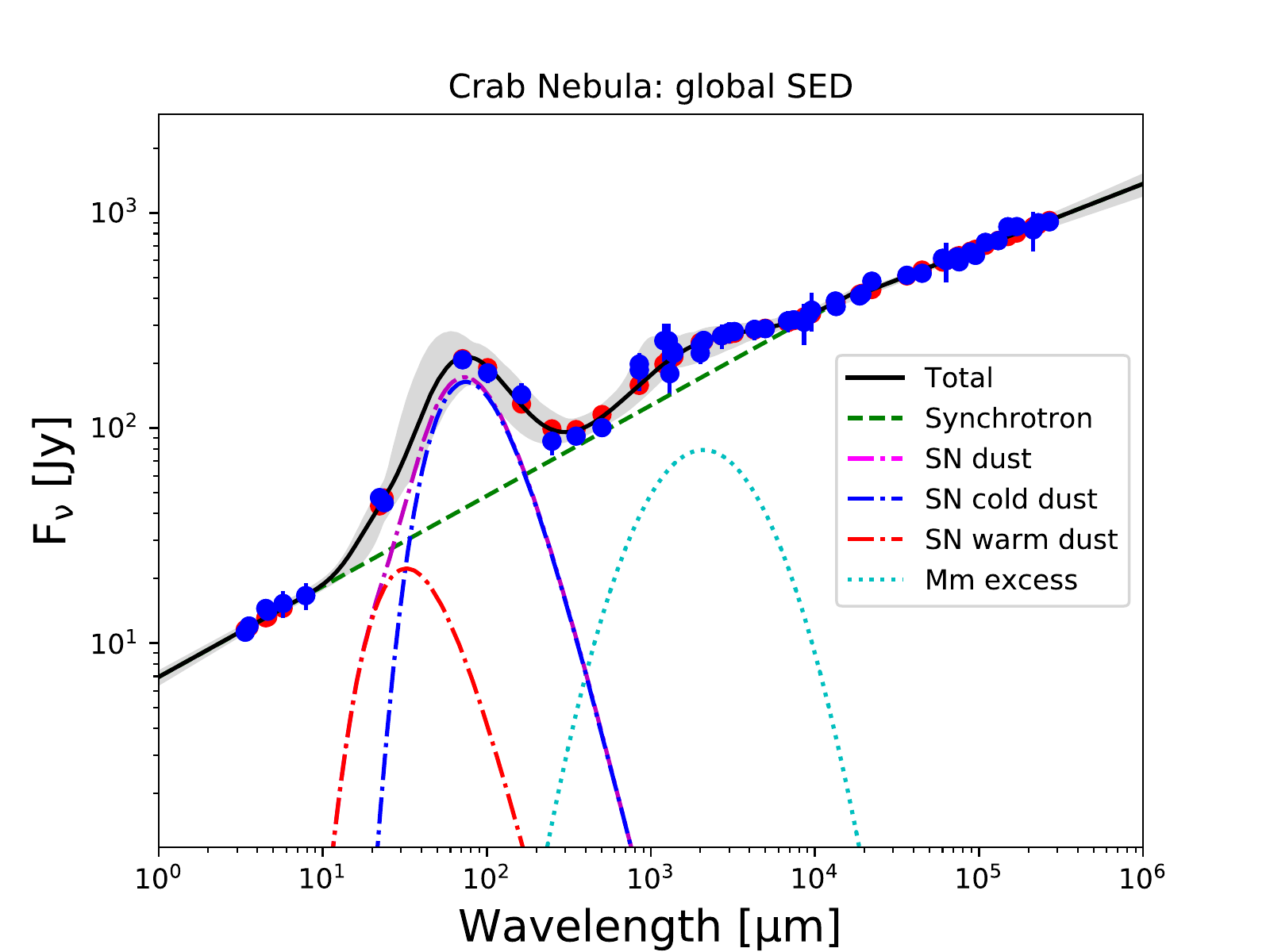}	
    \caption{Total integrated SED for the Crab Nebula extending from near-infrared to radio wavebands. The best-fit SED model is indicated with a black solid line, with the grey shaded region corresponding to the 16th and 84th percentiles of the N-dimensional likelihood (i.e., corresponding to the 1$\sigma$ upper and lower bounds to the model). The blue dots correspond to the observed datapoints (with the uncertainties shown as vertical lines), while the red circles indicate the model flux densities in those wavebands. We have also included the emission of individual model components: synchrotron radiation (green dashed curve), mm excess emission (blue dashed curve), cold and warm SN dust SEDs (blue and red dot-dashed curves, respectively), and the combined SN dust emission (purple dot-dashed curve).} 
    \label{Crab_global_spectrum}
\end{figure*}

Figure \ref{Crab_global_spectrum} shows the best-fit (i.e., maximum likelihood) model spectrum (black solid line) for 1\,$\mu$m-sized amorphous carbon ``a-C" grains used to model the Crab's supernova dust emission, and with the emission of individual components (synchrotron, cold+warm SN dust and excess emission) indicated. The 1D and 2D posterior distributions for each of the parameters are presented in Figure \ref{Crab_global_spectrum_errors}. We discuss the total integrated model results for synchrotron radiation, supernova dust emission and millimetre excess emission in more detail in Sections \ref{Synchrotron.sec}, \ref{SNdustmasses.sec} and \ref{MillimetreExcess.sec}, respectively. 

\begin{table*}
\centering
\caption{Overview of the median likelihood dust parameters inferred from a Bayesian SED modelling procedure on total integrated (top part) and resolved (bottom part) scales for a variety of distinct models assuming a single grain species. The model parameters describing the Crab's synchrotron radiation and mm excess emission are detailed in Table \ref{SynchrotronOverview}. Column 1 presents the different dust species and corresponding references (in superscript) from which their dust optical constants were adopted: (a) \citet{1991ApJ...377..526R} (b) \citet{2012A&A...540A...1J,2012A&A...540A...2J,2012A&A...542A..98J} for a-C grains with a band gap $E_{\text{g}}$=0.1\,eV (c) \citet{1995A&A...300..503D} (d) \citet{2003A&A...408..193J} (e) Takashi Nozawa and Eli Dwek (priv. comm.). We adopted mass densities of 1.6 g cm$^{-3}$ for amorphous carbon grains, 2.5 g cm$^{-3}$ for silicate-type grains \citep{2013A&A...558A..62J} and 7.89 g cm$^{-3}$ for iron grains \citep{2006ApJ...648..435N}. The reduced $\chi^{2}$ values inferred from all near-infrared to radio observational constraints, and the dust-only $\chi^{2}_{\text{red}}$ (calculated for WISE\,22 to SPIRE\,500\,$\mu$m wavebands, which have non-negligible supernova dust contributions) are reported in Columns 2 and 3, respectively. Columns 4 to 7 report the temperatures and masses for warm and cold supernova dust components. The total (warm+cold) dust masses are summarised in Column 8 (assuming a distance to the Crab of D=2\,kpc). These dust masses could go up by a factor of 2.8 based on the revised distance estimate for the Crab inferred from GAIA data \citep{2019ApJ...871...92F}. The maximum expected dust masses for different grain species are provided in Column 9 based on a condensation efficiency of 100$\%$ and the estimated yields from nucleosynthesis models from \citet{1995ApJS..101..181W} for a progenitor mass of 11\,M$_{\odot}$. The resolved model parameters correspond to the median dust temperatures calculated over all modelled pixels, while the dust masses correspond to the sum of the model dust masses over all pixels. The results obtained for our preferred grain model (a-C a=1\,$\mu$m grains) are indicated in boldface.}
\label{DustMassesOverview}
\begin{tabular}{lcccccccc} 
\hline
Parameters: & & & \multicolumn{2}{|c|}{Warm dust} & \multicolumn{2}{|c|}{Cold dust} & Total dust & 100$\%$ cond. eff. \\  
                     (1) & (2) & (3) & (4) & (5) & (6) & (7) & (8) & (9)  \\
                     & & & $T_{\text{warm}}$ & $\log$\,$M_{\text{warm}}$ & $T_{\text{cold}}$ & $\log$\,$M_{\text{cold}}$ & $M_{\text{total}}$ & $M_{\text{max}}$   \\
                     & & & [K] & [$\log$$M_{\odot}$] & [K] & [$\log$$M_{\odot}$] & [$M_{\odot}$] & [$M_{\odot}$]   \\     

\hline
\hline
\multicolumn{9}{|c|}{Total integrated SED fitting} \\ 
\hline
\hline
Dust species: & $\chi^{2}_{\text{red}}$ & $\chi^{2}_{\text{red}}$(dust) & & & & & & \\ 
\hline
Zubko "ACAR"$^{~\text{a}}$ & 1.9 & 1.3 & 69$_{-12}^{+21}$ & -3.44$_{-1.70}^{+1.43}$ & 48$_{-14}^{+8}$ & -1.89$_{-0.44}^{+0.30}$ & 0.013$_{-0.009}^{+0.022}$ & 0.054 \\ %
Zubko "BE"$^{~\text{a}}$ & 1.9 & 1.0 & 72$_{-17}^{+20}$ & -3.44$_{-1.02}^{+1.48}$ & 43$_{-9}^{+9}$ & -1.61$_{-0.28}^{+0.32}$ & 0.025$_{-0.012}^{+0.037}$ & 0.054 \\ %
Am. carbon ``AC1"$^{~\text{a}}$ & 1.9 & 1.4 & 65$_{-4}^{+22}$ & -2.30$_{-1.90}^{+0.41}$ & 50$_{-22}^{+9}$ & -1.88$_{-1.24}^{+0.18}$ & 0.018$_{-0.017}^{+0.015}$ & 0.054 \\ %
Am. carbon ``BE1"$^{~\text{a}}$ & 1.9 & 1.6 & 68$_{-11}^{+21}$ & -3.26$_{-1.77}^{+1.22}$ & 48$_{-17}^{+9}$ & -1.94$_{-0.68}^{+0.31}$ & 0.012$_{-0.010}^{+0.021}$ & 0.054 \\ %
a-C a=0.1\,$\mu$m$^{~\text{b}}$ & 1.9 & 1.3 & 69$_{-10}^{+21}$ & -3.13$_{-1.78}^{+1.22}$ & 49$_{-17}^{+9}$ & -1.83$_{-0.61}^{+0.29}$ & 0.016$_{-0.012}^{+0.026}$ & 0.054 \\
\textbf{a-C a=1\,$\mu$m}$^{~\text{b}}$ & \textbf{1.9} & \textbf{1.2} & \textbf{69$_{-12}^{+21}$} & \textbf{-3.34$_{-1.93}^{+1.45}$} & \textbf{50$_{-14}^{+7}$} & \textbf{-1.80$_{-0.51}^{+0.26}$} & \textbf{0.016$_{-0.011}^{+0.026}$} & \textbf{0.054} \\
a-C a=1\,$\mu$m, ev. synchr.$^{~\text{b}}$ & 2.3 & 8.2 & 66$_{-10}^{+22}$ & -3.59$_{-2.18}^{+1.70}$ & 53$_{-18}^{+4}$ & -1.86$_{-0.88}^{+0.14}$ & 0.014$_{-0.012}^{+0.018}$ & 0.054 \\
MgSiO$_{3}^{~~\text{c}}$ & 2.1 & 3.4 & 79$_{-3}^{+3}$ & -1.93$_{-0.08}^{+0.06}$ & 35$_{-17}^{+19}$ & -3.06$_{-0.65}^{+0.79}$ & 0.013$_{-0.003}^{+0.006}$ & 0.050 \\ %
Mg$_{0.7}$SiO$_{2.7}^{~~~~\text{d}}$ & 1.9 & 0.9 & 61$_{-18}^{+25}$ & -2.90$_{-1.12}^{+2.55}$ & 38$_{-6}^{+4}$ & -0.07$_{-0.19}^{+0.26}$ & 0.85$_{-0.30}^{+1.14}$ & 0.063  \\ %
Fe a=0.1\,$\mu$m$^{~\text{e}}$ & 2.1 & 3.4 & 79$_{-3}^{+3}$ & -0.67$_{-0.11}^{+0.07}$ & 37$_{-18}^{+19}$ & -2.34$_{-1.10}^{+1.34}$ & 0.22$_{-0.05}^{+0.13}$ & 0.105 \\ 
Fe a=1\,$\mu$m$^{~\text{e}}$ & 2.1 & 3.2 & 79$_{-3}^{+3}$ & 0.48$_{-0.08}^{+0.06}$ & 35$_{-16}^{+19}$ & -0.87$_{-0.77}^{+0.86}$ & 3.14$_{-0.63}^{+1.32}$ & 0.105 \\ 
\hline
\hline
\multicolumn{9}{|c|}{Resolved SED fitting} \\ 
\hline
\hline
a-C a=1\,$\mu$m$^{\text{b}}$ &  - & - & 70 (fixed) & -2.77$_{-0.07}^{+0.06}$ & 41$_{-2}^{+3}$ & -1.42$_{-0.09}^{+0.09}$ & 0.039$_{-0.007}^{+0.010}$ & 0.054 \\ %
\hline
\end{tabular}
\end{table*}

\subsection{Dominant emission mechanisms}
\label{GlobalContributions.sec}
\begin{figure*}
	\includegraphics[width=17.5cm]{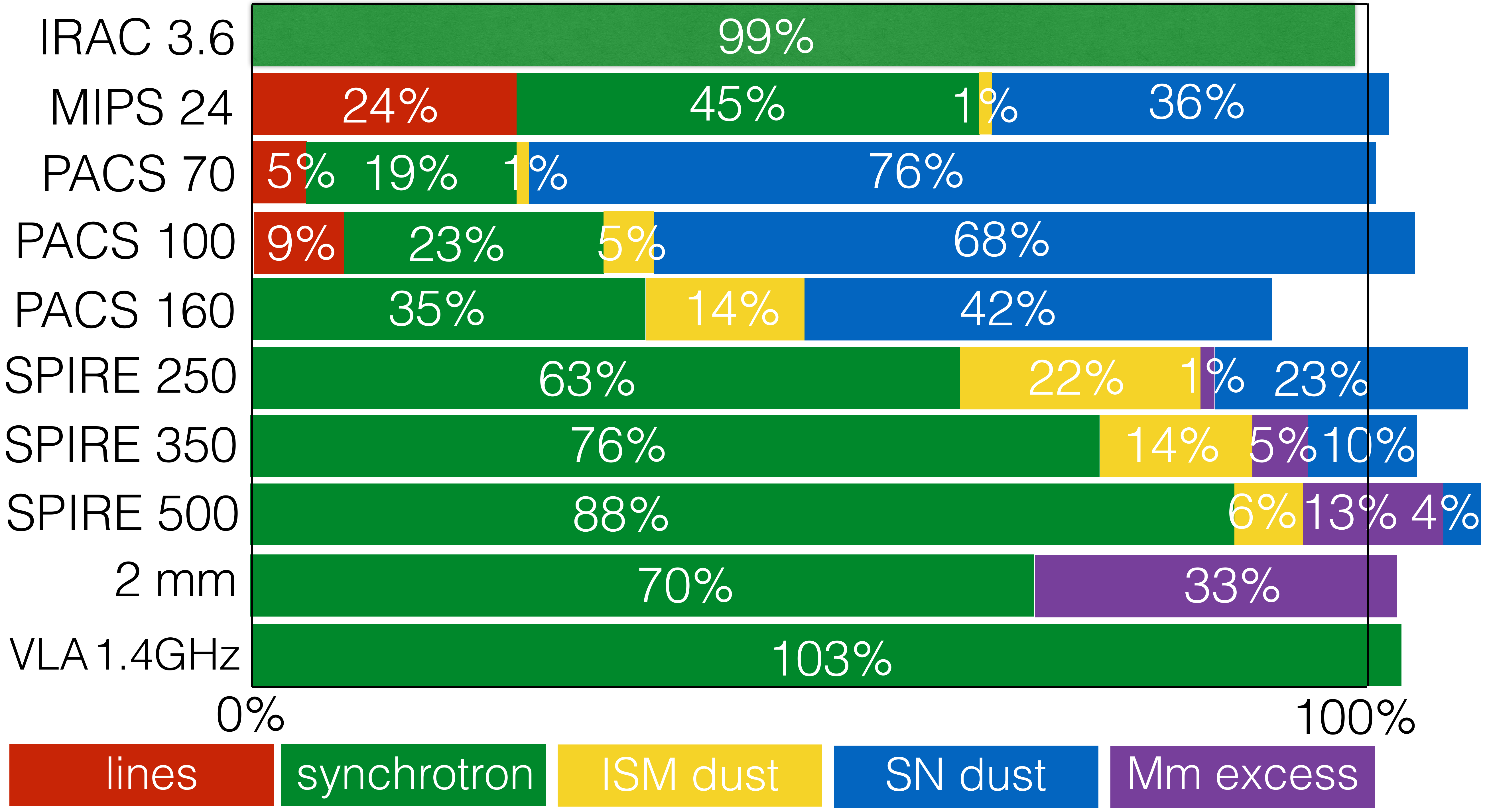}
    \caption{Overview of the contributions of different components (line emission, synchrotron radiation, ISM dust emission, mm excess, and SN dust emission) to the observed flux densities in each waveband, as inferred from the total integrated best-fit SED model assuming amorphous carbon ``a-C" grains for the SN dust component. The graph shows the results for a selected number of representative wavebands;  the individual contributions for omitted wavebands can be retrieved from Table \ref{Table_Fluxfraction}. The ISM dust contributions were inferred as described in Appendix \ref{GalacticDust.sec}. Because the Bayesian model is fitting the observed flux densities within the limits of uncertainty, the fractional contributions do not always exactly add up to a 100$\%$.}
    \label{Crab_schematic}
\end{figure*}

The total integrated SED modelling results enable us to infer the model contributions of different emission mechanisms (line emission, synchrotron radiation, ISM dust emission, mm excess, SN dust emission) to each waveband. Figure \ref{Crab_schematic} provides an overview of the estimated contributions for our total integrated model with amorphous carbon ``a-C" grains for the SN dust. Most wavebands, from the near-infrared to submillimetre and radio wavelengths are dominated by synchrotron emission, except at mid- and far-infrared wavelengths. The IRAC\,3.6\,$\mu$m and VLA 1.4\,GHz emission (and also VLA\,4.8\,GHz emission, not shown here) is modelled to originate from synchrotron radiation. At mm wavelengths, there is a non-negligible contribution from excess emission with a contribution of 33\,$\%$ at 2\,mm. Modelling the excess emission introduces a tail of excess emission that extends to submm wavelengths with 1, 5 and 13$\%$ of excess emission contributions to the SPIRE\,250, 350 and 500\,$\mu$m wavebands, respectively. In addition, synchrotron radiation accounts for 63, 76 and 88$\%$ of all emission in those respective wavebands, while SN dust emission is found to be responsible for 23$\%$, 10$\%$ and 4$\%$ of the submm emission. The total observed SPIRE flux densities are systematically overestimated by 9, 5, and 10$\%$ in our model at 250, 350 and 500\,$\mu$m, respectively. The overestimation might (in part) result from an overestimation of the mm excess contribution to the SPIRE wavebands due to the assumption of our spectrum (see Eq. \ref{Eq_mmexcess}) to model the excess emission. The SPIRE\,500\,$\mu$m emission is mostly arising from synchrotron radiation, with a negligible contribution from ISM dust and SN dust emission. The PACS 70, 100 and 160\,$\mu$m wavebands are dominated by SN dust emission (contributing 76, 68 and 42\,$\%$, respectively), with a secondary contribution from synchrotron radiation (19, 23 and 35$\%$, respectively) and minor contributions from line emission at PACS\,70\,$\mu$m (5\,$\%$) and PACS\,100\,$\mu$m (9\,$\%$). At mid-infrared wavelengths, the SN dust emission drops to 36$\%$; synchrotron radiation (45$\%$) is the dominant emission mechanism at MIPS\,24\,$\mu$m, together with a non-negligible contribution from line emission (24$\%$). The contributions of the different components to the MIPS and PACS wavebands do not perfectly add up to 100$\%$, but remain within model and observational uncertainties. 

\section{Resolved modelling of the Crab Nebula}
\label{ResolvedSED.sec}

\begin{figure*}
	\includegraphics[width=16.5cm]{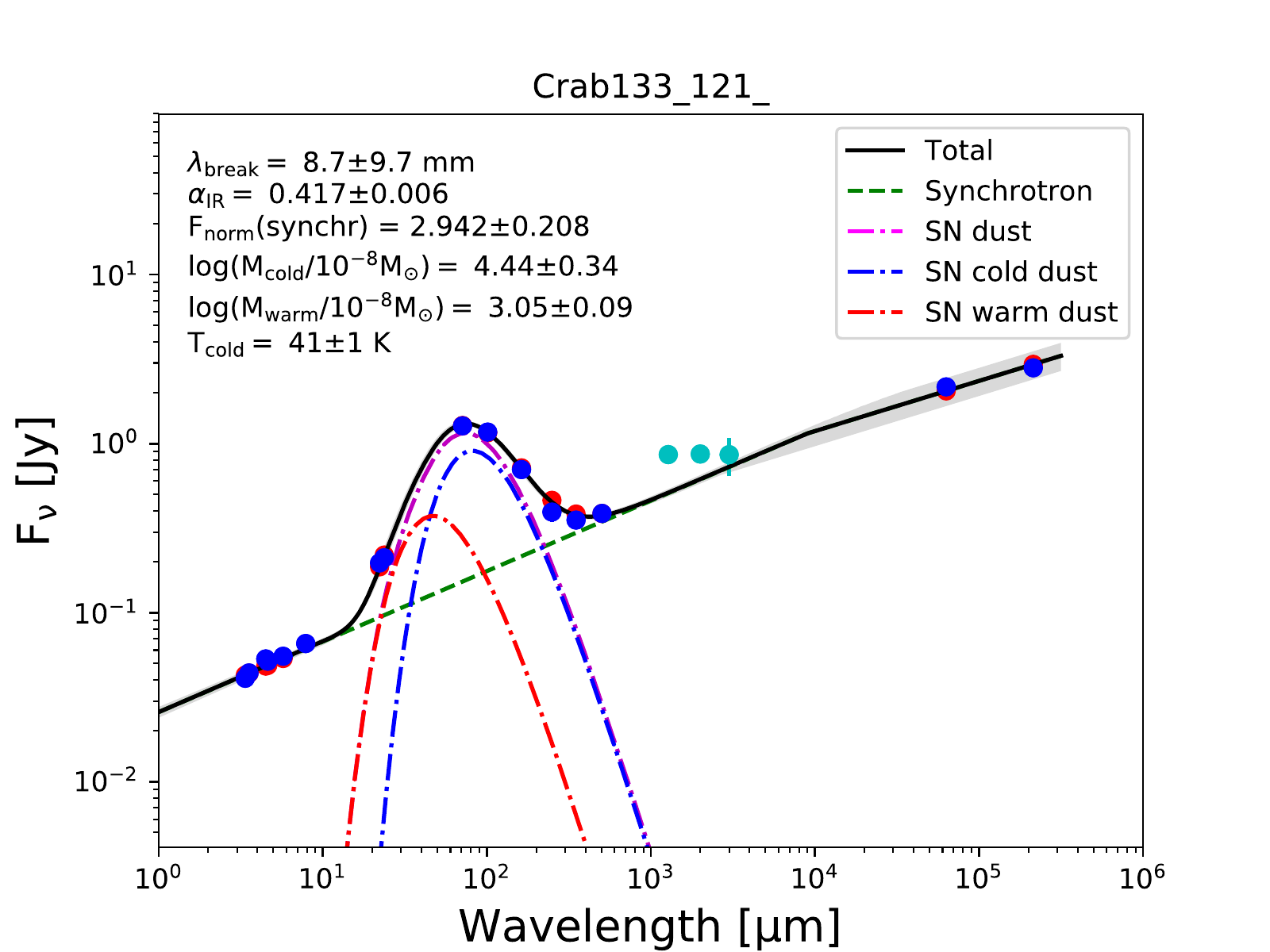}	
    \caption{Representative example of a resolved SED fit for a central pixel in the Crab Nebula. The best-fit SED model is indicated with a black solid line, with the grey shaded region corresponding to the 16th and 84th percentiles of the N-dimensional likelihood (i.e., corresponding to the 1$\sigma$ upper and lower bounds to the model). We have also included the emission of individual model components: synchrotron radiation (green dashed curve), cold and warm SN dust SEDs (purple dot-dashed curves), and the combined SN dust emission (cyan dot-dashed curve). The best-fit model parameters (maximum-likelihood and standard deviation) of each of the parameters are summarised in the top left corner of the plot. The blue dots correspond to the observed datapoints (with the uncertainties shown as vertical lines), while the red circles indicate the model flux densities in those wavebands. The cyan dots correspond to the mm flux densities in those pixels, which are shown for comparison, as these were not used to constrain the models.}
    \label{Crab_resolved_spectrum}
\end{figure*} 

To study spatial variations in the synchrotron spectral indices, and the temperature and mass of SN dust, we have modelled the mid-infrared to radio observations of the Crab Nebula in our resolved map with 336 individual 14$\arcsec\times$14$\arcsec$ (or 0.136$^{2}$ pc$^{2}$) pixels. Hereto, we have convolved sixteen images (IRAC\,3.6, 4.5, 5.8, 8.0\,$\mu$m; WISE\,3.4, 4.6 and 22\,$\mu$m, MIPS\,24\,$\mu$m, PACS\,70, 100 and 160\,$\mu$m, SPIRE\,250, 350, 500\,$\mu$m, 4.8 and 1.4\,GHz) to the SPIRE\,500\,$\mu$m resolution, and corrected all images for line emission and an average contribution from ISM dust emission.

We applied a Bayesian inference method similar to the one used for the total integrated SED fit (see Section \ref{GlobalModel.sec}). Due to the fewer flux density constraints and lower signal-to-noise on a pixel-by-pixel basis, we restricted the model to include a two-component SN dust MBB model and a broken-power-law synchrotron component (see Table \ref{BayesianModelParameters}). Due to insufficient sampling of the mm spectrum on resolved scales, the parameters of the synchrotron and mm excess model components could not be constrained simultaneously. 

We did not use the resolved images at mm wavelengths (1.3\,mm, 2\,mm, 3\,mm) to constrain the Bayesian models. Instead, we produced model images at each of these wavelengths, and used those to study the resolved distribution of model excess emission in Section \ref{ResolvedMillimetreExcess.sec}. We furthermore fixed the radio spectral index, which is otherwise hard to constrain with only two data points in the mm to radio wavelength domain. We relied on the total integrated SED fitting results to fix the value for $\alpha_{\text{radio}}$=0.297, which corresponds to the median likelihood value inferred from the total integrated fit (see Table \ref{SynchrotronOverview}), and is independent of the assumed SN dust grain species. This fixed value for $\alpha_{\text{radio}}$ is justified based on the spatially uniform radio spectral index map inferred by \citet{1997ApJ...490..291B} on resolved scales.

Due to the computational cost of SED modelling on a pixel-by-pixel basis, we only performed the modelling for a single SN dust species. Based on the best-fitting models to the total integrated flux densities (see Table \ref{DustMassesOverview}), and the large fractional carbon abundance in the gas phase in the Crab, we have opted for amorphous carbon ``a-C" grains with radii a=1\,$\mu$m, which rely on the optical constants inferred from recent laboratory studies, from \citet{2012A&A...540A...1J,2012A&A...540A...2J,2012A&A...542A..98J}\footnote{Although the best-fit was obtained for Mg$_{0.7}$SiO$_{2.7}$ grains, the total dust mass required to fit the total integrated emission of the Crab (by far) exceeds the maximum dust mass expected to condense out of the available metals. The lack of any dust continuum features in the \textit{Spitzer} spectra furthermore rules out a dominant silicate dust population in the Crab.}. The total integrated fitting results suggest that the SN dust emission is dominated by a single temperature component with a well characterised dust temperature, while the dust temperature of the other component is poorly constrained. Since the total integrated fits for carbonaceous grains tend to be dominated by cold ($\sim$43-50\,K) dust, we fix the warm dust temperature $T_{\text{warm}}$=70\,K in the resolved models. The prior range for the break wavelength was extended from 10\,$\mu$m to 60\,mm for the resolved modelling procedure. Figures \ref{Crab_resolved_spectrum} and \ref{Crab_resolved_spectrum_errors} show a representative Bayesian SED model, and corresponding 1D and 2D posterior distributions inferred for a central pixel in the Crab Nebula, respectively.

\begin{figure*}
	\includegraphics[width=16.5cm]{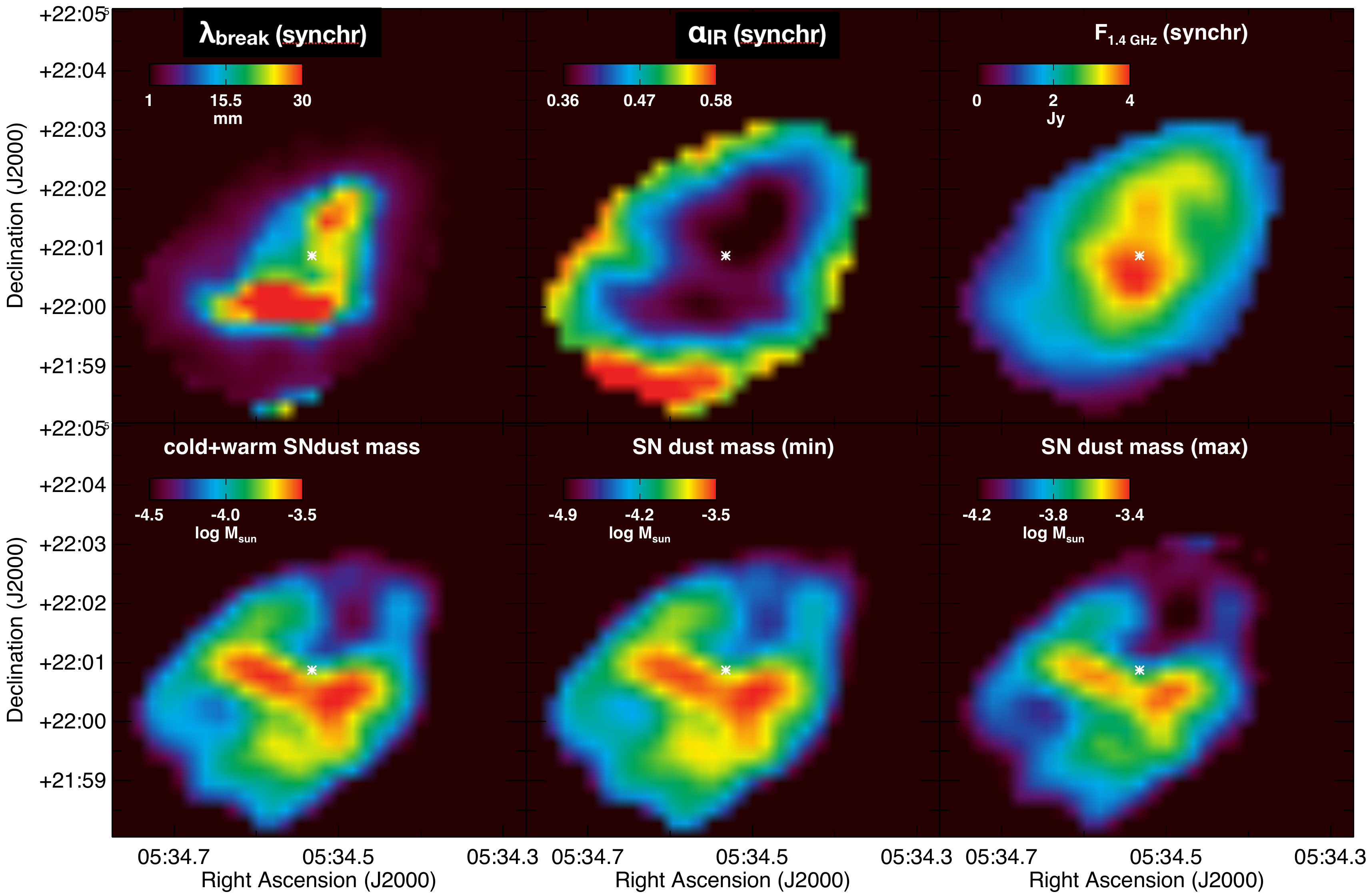}	
    \caption{Resolved maps of the modelled wavelength break, $\lambda_{\text{break}}$ (top row, left), IR spectral index, $\alpha_{\text{IR}}$ (top row, middle), 1.4\,GHz synchrotron flux density (top row, right), total SN dust mass per pixel (bottom row, left), and lower and upper limits on the SN dust mass per pixel (bottom row, middle and right, respectively), as inferred from a Bayesian SED fitting algorithm on resolved scales of 14$\times$14\,$\arcsec^{2}$ (or 0.136$^{2}$ pc$^{2}$, for a distance of 2\,kpc).}
    \label{Images_Crab_SED_resolved}
\end{figure*}
Resolved maps of the synchrotron spectral index, 1.4\,GHz synchrotron flux density, SN dust temperatures and masses are presented in Figure \ref{Images_Crab_SED_resolved}. We will now discuss the total integrated and resolved modelling results related to the Crab's synchrotron radiation (Section \ref{Synchrotron.sec}), supernova dust emission (Section \ref{SNdustmasses.sec}) and millimetre excess emission (Section \ref{MillimetreExcess.sec}). \\

\begin{table*}
\centering
\caption{Overview of the median likelihood parameters describing the Crab's synchrotron radiation and mm excess emission as inferred from a Bayesian SED modelling procedure on total integrated (top part) and resolved (bottom part) scales for amorphous carbon ``a-C" grains. The model parameters for other Bayesian models (assuming different supernova grain species) were omitted due to the independence of the synchrotron and mm excess model parameters on the SN grain model. Columns 2, 3, 4 and 5 present the synchrotron model parameters inferred for the radio spectral index, $\alpha_{\text{radio}}$, infrared spectral index, $\alpha_{\text{IR}}$, wavelength of the spectral break, $\lambda_{\text{break}}$, and the synchrotron flux density, $F_{\text{1.4\,GHz}}$, respectively. The peak frequency, $\nu_{\text{peak}}$, the peak flux density, $F_{\text{mm, peak}}$, and width, $\sigma$, of the spectrum describing the mm excess emission are presented in Columns 6, 7 and 8, respectively. The resolved model parameters correspond to the luminosity-weighted values, while the flux density $F_{\text{1.4\,GHz}}$ corresponds to the sum of the synchrotron model flux densities over all pixels.}
\label{SynchrotronOverview}
\begin{tabular}{lccccccc} 
\hline
Parameters: & \multicolumn{4}{|c|}{Synchrotron} & \multicolumn{3}{|c|}{Excess emission}  \\ 
(1) & (2) & (3) & (4) & (5) & (6) & (7) & (8) \\
Dust species: & $\alpha_{\text{radio}}$ & $\alpha_{\text{IR}}$ & $\lambda_{\text{break}}$ & $F_{\text{1.4\,GHz}}$ & $\nu_{\text{peak}}$ & $F_{\text{mm,peak}}$ & Width $\sigma$ \\
            & & & [mm] & [Jy] & [GHz] & [Jy] & \\                   
\hline
\hline
\multicolumn{8}{|c|}{Total integrated SED fitting} \\ 
\hline
\hline
a-C a=1\,$\mu$m  & 0.297$^{+0.009}_{-0.009}$ & 0.429$^{+0.021}_{-0.009}$ & 9.0$^{+8.4}_{-5.1}$ & 862$^{+17}_{-16}$ & 163$^{+42}_{-22}$ & 69$^{+14}_{-23}$ & 0.27$^{+0.06}_{-0.08}$ \\
a-C a=1\,$\mu$m, ev. synchr. & 0.359$^{+0.016}_{-0.049}$ & - & 1.1$^{+8.3}_{-1.0}$ & 921$^{+20}_{-25}$ & 68$^{+43}_{-21}$ & 69$^{+12}_{-21}$ & 0.62$^{+0.15}_{-0.30}$ \\
\hline
\hline
\multicolumn{8}{|c|}{Resolved SED fitting} \\ 
\hline
\hline
a-C a=1\,$\mu$m & 0.297 [fixed] & 0.433$^{+0.007}_{-0.007}$ & 11.5$^{+2.9}_{-1.7}$ & 725$^{+45}_{-47}$ & - & - & - \\
\hline
\end{tabular}
\end{table*}

\section{The Crab's synchrotron radiation}
\label{Synchrotron.sec}
The most probable value of the integrated synchrotron spectral index in the radio wavelength domain is peaked around $\alpha_{\text{radio}}$=0.297 (see Fig.\,\ref{Crab_global_spectrum_errors}), consistent with the index of $\alpha$=0.3 derived about 40 years ago by \citet{1977A&A....61...99B} and the radio spectral index $\alpha$=0.299 inferred by \citet{2012ApJ...760...96G}. The posterior distribution for the IR spectral index peaks around $\alpha_{\text{IR}}$=0.42, which resembles the value $\alpha$=0.417 inferred by \citet{2012ApJ...760...96G}. However, the wide tail in the 1D posterior distribution extending towards $\alpha_{\text{IR}}$=0.46 suggests that a steeper slope can not be ruled out by the data. In fact, the parameter value for $\alpha_{\text{IR}}$ strongly depends on the position of the synchrotron break: the IR spectral index is shallower ($\alpha_{\text{IR}}$=0.42) for a break at longer wavelengths ($\lambda_{\text{break}}\sim$17\,mm); while a steeper slope ($\alpha_{\text{IR}}$=0.45) is required for a short wavelength break ($\lambda_{\text{break}}\sim$4\,mm). Other than for $\alpha_{\text{IR}}$ and $\lambda_{\text{break}}$, a hint of a bimodality is also seen in the 1D posterior distribution of the peak millimetre emission, $F_{\text{mm,peak}}$ (see Section \ref{MillimetreExcess.sec}). There is a degeneracy between the contribution of mm excess emission to the total integrated SED, the wavelength of the break in the synchrotron spectrum and the IR spectral index, which can not be disentangled based on our current set of observations. However, the highest probability (by far) of the Bayesian model is given to the model solution with a synchrotron spectral break around 17\,mm, an IR spectral index $\alpha_{\text{IR}}\sim$0.42 and a strong excess peaked around 150\,GHz (or 2\,mm), regardless of the assumed SN dust species. The model synchrotron emission at 1.4\,GHz is peaked around $\sim$860\,Jy, which corresponds well with the observed flux density (within the error bars) at the same frequency (see Table \ref{Table_Fluxfraction}). 

In our resolved SED models, the infrared spectral index $\alpha_{\text{IR}}$ varies from 0.36 in the inner regions up to 0.58 in the outer regions (see Fig. \ref{Images_Crab_SED_resolved}, top right panel); these spatial variations are significant with respect to the average model uncertainties ($<$0.01) on these $\alpha_{\text{IR}}$ values. The average luminosity-weighted IR index ($\alpha_{\text{IR}}$=0.433) is consistent with the total integrated value ($\alpha_{\text{IR}}$=0.429) for the Crab (see Table \ref{SynchrotronOverview}). The IR spectral index is flatter in the inner regions, in particular around the torus and jet structures, and steepens in the outer regions, which is consistent with the spectral index map derived based on optical and near-infrared images \citep{1993A&A...270..370V,2012ApJ...753...72T,2018arXiv181101767L}. This steepening of the spectrum towards the outer regions has been interpreted in terms of a scenario in which electrons emitting at optical and near-infrared wavelengths have fast cooling times, shorter than the age of the Crab itself (e.g. \citealt{1984ApJ...283..710K,1996MNRAS.278..525A,2002A&A...386.1044B,2010A&A...523A...2M,2013MNRAS.433.3325S,2014MNRAS.438..278P,2018arXiv181101767L}). We show in this paper that this trend continues into the far-infrared and (sub-)millimeter wavelength regime, and that radiative losses also affect these longer wavelengths. The average IR spectral index at the position of the Crab's torus and the jet is close to 0.36, and thus somewhat steeper than the average radio spectral index of 0.297. According to our models, the infrared and (sub-)millimetre regime correspond to a fast cooling regime, while the radio synchrotron spectrum is dominated by a slow cooling regime. These differences between the central radio and infrared spectral index, and radial variations inferred for the infrared spectral index suggest the presence of two mechanisms responsible for the acceleration of these particles \citep{2002A&A...386.1044B,2018arXiv181101767L}.

We have also attempted to constrain the position of the synchrotron break on resolved scales (see Fig. \ref{Images_Crab_SED_resolved}, top left panel). Although the range of possible ``break" wavelengths is rather wide (see Fig. \ref{Crab_resolved_spectrum_errors}), we observe a clear dichotomy between the inner and outer regions of the Crab Nebula. The inner regions correspond to a break occurring at longer wavelengths (20-30\,mm), while the break occurs at shorter wavelengths (2-4\,mm) in the outer Crab regions. These two regimes are also reflected in the bi-modality of the break wavelength in the total integrated SED (see Fig. \ref{Crab_global_spectrum_errors}), where a high likelihood was attributed to a synchrotron break at cm wavelengths (as observed in the emission-dominating inner regions). This scenario is consistent with the cm and longer wavelength radio wavelengths being significantly affected by synchrotron losses in the central Crab regions, while the outer regions are also prone to synchrotron losses at shorter wavelengths. Similar inferences were made by \citet{2002A&A...386.1044B}, who derived a longer wavelength break in dense filaments, and interpreted it in terms of the presence of stronger magnetic fields.

It is important to note that the absolute wavelengths inferred for the position of the break will sensitively depend on the assumed synchrotron model (i.e., a broken-power law or evolutionary break model). Earlier studies demonstrated that a spectral break occurs around 20\,$\mu$m in the integrated spectrum of the Crab Nebula, consistent with estimates of the average Crab's magnetic field B of 200\,$\mu$G \citep{2011MNRAS.410..381B}. The best-fit model for an evolutionary synchrotron spectrum (see Appendix \ref{Alternative_synchrotron_model.sec}) similarly suggests a break in the spectrum around 24\,$\mu$m, in line with previous estimates. Due to the sensitivity of the break position to the assumed synchrotron model, we focus the discussion here on the relative differences observed in our resolved synchrotron spectral break map.

The synchrotron brightness at 1.4\,GHz (see Fig. \ref{Images_Crab_SED_resolved}, top right panel) peaks near the pulsar and the toroidal structures around the pulsar (see Fig. \ref{Crab_composite} for comparison), with an extension towards the north and north-west of the PWN, similar to the morphology of radio emission observed for the Crab Nebula (e.g., \citealt{2015MNRAS.446..205B}). The sum over all modelled pixels at 1.4\,GHz (725$^{+45}_{-47}$\,Jy, see Table \ref{SynchrotronOverview}) is consistent with the summed emission in the observed 1.4\,GHz image (675$\pm$135\,Jy). Note that the sum of the resolved flux densities is lower than the total integrated flux density at 1.4\,GHz (834$\pm$175\,Jy, see Table \ref{Table_Fluxfraction}) due to the total integrated measurement encompassing a larger aperture than the resolved pixel-by-pixel analysis (which required pixels to attain a certain signal-to-noise).

It is important to note that re-running the models on resolved scales with the inclusion of the 1.3, 2 and 3.3\,mm data, we were unable to simultaneously fit the FIR/submm \textit{Herschel} and mm observations in the central regions of the SNR, as these models would significantly overestimate the SPIRE fluxes with a synchrotron spectrum constrained by the IRAC and \textit{WISE} near-infrared fluxes and mm data points, and thus seem to require the addition of a mm excess component in the model. It can however not be ruled out that models including spatial variations in the spectral break, multiple spectral breaks and/or multiple synchrotron components would be able to remedy this situation. In Section \ref{MillimetreExcess.sec} and Appendix \ref{App_mmexcess}, we will discuss these and other alternative scenarios, respectively, to account for the mm excess observed in the Crab Nebula.

Overall, the excellent agreement between the spectral indices and the spatial variations within the remnant derived by \citet{2012ApJ...753...72T} and in this paper, makes us confident that our model is appropriate to reproduce the Crab's synchrotron radiation in the IR wavelength domain, and demonstrates the need for a spatially resolved synchrotron model to infer the residual emission that can be attributed to SN dust (see Section \ref{SNdustmasses.sec}). 

\section{Supernova dust emission and extinction}
\label{SNdustmasses.sec}
The SN dust emission in the total integrated SED of the Crab was modelled using a combination of warm and cold SN dust temperature components. Table \ref{DustMassesOverview} presents the warm and cold dust masses and temperatures, and combined dust masses for a range of different grain species inferred from the same modelling technique. The best dust SED fits (with the lowest reduced $\chi^{2}_{\text{dust}}$ values) were obtained for amorphous carbon grains (``AC1",``BE1", ``ACAR", ``BE" and ``a-C") and silicate-type grains with low Mg-to-O ratios (Mg$_{0.7}$SiO$_{2.7}$). The model fits deteriorate for Mg protosilicates (MgSiO$_{3}$) and iron grains (with sizes a=0.1 and a=1\,$\mu$m) due to an over-prediction of the observed SPIRE fluxes by the model. 

The posterior distributions (see Fig.\ref{Crab_global_spectrum_errors}) demonstrate that the SN dust SED is dominated by a well-constrained single temperature (warm or cold) component, while the posterior distribution for the dust temperature of the other component (with minimal contribution to the overall SN dust emission) tends to be flat. Although a better fit is still obtained using a two-component MBB model, reasonably good fits can thus be obtained with a single-$T_{\text{dust}}$ component. The cold dust temperature in the model ranges from $38^{+4}_{-6}$\,K (for Mg$_{0.7}$SiO$_{2.7}$ grains) to 48$^{+9}_{-17}$\,K, 50$^{+9}_{-22}$\,K, 48$^{+8}_{-14}$\,K, 43$^{+9}_{-9}$\,K, and 50$^{+7}_{-14}$\,K (for amorphous carbon ``BE1", ``AC1", ``ACAR", ``BE" and ``a-C" grains, respectively). These dust temperatures are high compared to the average cold dust temperatures inferred by \citet{2012ApJ...760...96G} for silicates ($T_{\text{d,cold}}$=28$^{+6}_{-3}$\,K) and amorphous carbon ``BE" grains ($T_{\text{d,cold}}$=34$^{+2}_{-2}$\,K), but somewhat more similar to their one-component fits for silicates ($T_{\text{dust}}$=34\,K) and carbonaceous ``BE" grains ($T_{\text{dust}}$=40\,K). Our inferred dust temperature are also in agreement with the dust temperature ($T_{\text{dust}}$=42.1$\pm$1.1\,K) inferred by \citet{2019arXiv190303389N}, and with the dust temperatures $T_{\text{dust}}$=45\,K (for silicates) and $T_{\text{dust}}$=45\,K (for graphite grains) inferred by \citet{2004MNRAS.355.1315G}. Models dominated by a warm component have dust temperatures $T_{\text{warm}}$=79$^{+3}_{-3}$\,K (for iron grains), which are higher than the warm dust temperatures for silicate ($T_{\text{warm}}$=56$^{+8}_{-3}$\,K) and carbon ($T_{\text{warm}}$=63$^{+5}_{-3}$\,K) grains from \citet{2012ApJ...760...96G}. 

The total (cold+warm) supernova dust masses required to reproduce the Crab's total integrated infrared dust SED vary significantly from one grain species to another. Average carbon dust masses of 0.018$^{+0.015}_{-0.017}$\,M$_{\odot}$ ("AC1" grains), 0.012$^{+0.021}_{-0.010}$\,M$_{\odot}$ ("BE1" grains), 0.013$^{+0.022}_{-0.009}$\,M$_{\odot}$ (``ACAR" grains), 0.025$^{+0.037}_{-0.012}$\,M$_{\odot}$ (``BE" grains) and 0.016$^{+0.026}_{-0.011}$\,M$_{\odot}$ (``a-C" grains) are required to fit the infrared SED, which are all consistent within the error bars regardless of the assumed carbon dust emissivities (see Fig. \ref{Crab_dustspecies_kabs}). These inferred dust masses are within the limits of the amount of carbon produced in the progenitor star (0.054\,M$_{\odot}$). Models for supernova dust with an iron composition would require 0.22$^{+0.13}_{-0.05}$\,M$_{\odot}$ (a=0.1\,$\mu$m) or 3.14$^{+1.32}_{-0.63}$\,M$_{\odot}$ (a=1\,$\mu$m) of dust to reproduce the emission, which greatly exceeds the maximum amount of iron expected to be present (0.105\,M$_{\odot}$\footnote{The estimated iron grain mass from nucleosynthesis models \citep{1995ApJS..101..181W} accounts for the amount of iron produced by the star, but also the formation route of iron through radiative decay of nickel.}). Similarly, we would require 0.85$^{+1.14}_{-0.30}$\,M$_{\odot}$ of Mg$_{0.7}$SiO$_{2.7}$ grains to fit the supernova dust emission, which is at least an order of magnitude higher than the available metal content for grain condensation (0.063\,M$_{\odot}$). On the other hand, a reasonable mass (0.013$^{+0.006}_{-0.003}$\,M$_{\odot}$) of MgSiO$_{3}$ grains suffices to fit the SED. The nearly two orders of magnitude difference in dust masses inferred for these silicate-type dust species results from the lower FIR/submm dust emissivities (by a factor of 10) of Mg$_{0.7}$SiO$_{2.7}$ grains compared to MgSiO$_{3}$ (see Fig. \ref{Crab_dustspecies_kabs}). In addition, the Bayesian model reproduces most of the Crab's dust emission using warm (79\,K) MgSiO$_{3}$ grains, while 38\,K Mg$_{0.7}$SiO$_{2.7}$ grains are preferred.

Based on nucleosynthesis arguments, we can rule out that iron (a=0.1 or 1\,$\mu$m) or Mg$_{0.7}$SiO$_{2.7}$ grains dominate the supernova dust emission in the Crab Nebula. The nearly featureless dust continuum in \textit{Spitzer} IRS spectra \citep{2012ApJ...753...72T} is also consistent with the absence of a significant mass of Mg$_{0.7}$SiO$_{2.7}$ (or any other silicate-type) grains. We can however not rule out that some iron ($<$0.105\,M$_{\odot}$) and Mg$_{0.7}$SiO$_{2.7}$ ($<$0.063\,M$_{\odot}$) grains have formed in the Crab Nebula. While warm Mg$_{0.7}$SiO$_{2.7}$ dust has been detected in several Galactic SNRs (Cas\,A, G54.1; \citealt{2008ApJ...673..271R,2017ApJ...836..129T}), we inferred that it is impossible for these silicate-type grains to be responsible for most of the (cold) dust mass in SNRs (at least for the Crab and Cassiopeia\,A, see \citealt{2017MNRAS.465.3309D} for the latter) due to their low grain emissivities at submm wavelengths (see Fig. \ref{Crab_dustspecies_kabs}). 

We assumed a grain size of 1\,$\mu$m for most dust species explored in this model, based on a series of recent (independent) studies consistent with large ($>$0.1\,$\mu$m) dust grains growing in SNRs at late times (e.g., \citealt{2014Natur.511..326G,2015ApJ...801..141O,2015MNRAS.446.2089W,2016MNRAS.456.1269B}, Priestley et al.\,in prep.). For most grain species, our grain size assumption will not affect the inferred dust masses due to the weak dependence of the dust mass absorption coefficient, $\kappa_{\text{abs}}$, at infrared and submm wavelengths long wards of the typical grain radii (i.e., $>$1\,$\mu$m). We tested this assumption for a-C grains with radii a=0.1 and 1\,$\mu$m; no significant differences in the inferred SN dust model parameters were inferred (see Table \ref{DustMassesOverview}). However, iron grains form an exception with a=1\,$\mu$m grains being up to ten times more emissive at submm wavelengths compared to a=0.1\,$\mu$m grains (\citealt{2006ApJ...648..435N}; see Fig. \ref{Crab_dustspecies_kabs}), which has an impact on the inferred iron dust masses for different grain sizes (see Table \ref{DustMassesOverview}).

Based on earlier \textit{Herschel} studies, our dust mass estimates are up to an order of magnitude lower than dust mass predictions from \citet{2012ApJ...760...96G} (M$_{\text{dust}}$=0.24$^{+0.32}_{-0.08}$\,M$_{\odot}$ for silicates and M$_{\text{dust}}$=0.11$\pm$0.01\,M$_{\odot}$ for amorphous carbon ``BE" grains), and from \citet{2015ApJ...801..141O} (M$_{\text{dust}}$=0.18-0.27\,M$_{\odot}$ for clumps of carbon grains), but overlap with the estimated dust mass range from \citet{2013ApJ...774....8T} (0.016-0.061\,M$_{\odot}$, for amorphous carbon ``AC" grains) and \citet{2019arXiv190303389N} (M$_{\text{dust}}$=0.056$\pm$0.037\,M$_{\odot}$, assuming a single modified blackbody body function with dust emissivity index $\beta$=1.5 and dust mass absorption coefficient $\kappa_{\text{100}\,\mu\text{m}}$=40 cm$^{2}$ g$^{-1}$). The supernova dust masses inferred by Priestley et al.\,(in prep.) from modelling our supernova dust SED (see Table \ref{Table_Fluxfraction}) with amorphous carbon ``ACAR" (0.026-0.039\,M$_{\odot}$) and ``BE" (0.032-0.076\,M$_{\odot}$) grains\footnote{The uncertainties on their supernova dust mass measurements are driven by the assumed distance to the heating sources.}, allowing the grain size distribution to vary, heated by the PWN radiation field and by collisions with electrons and ions in the ambient gas, fall within the range of values reported in this paper for the same type of grains. Only for MgSiO$_{3}$ (0.076-0.218\,M$_{\odot}$ silicate grains do their supernova dust masses tend to be significantly higher compared to ours due to their lower inferred dust temperatures ($T_{\text{dust}}$=25\,K) for 1\,$\mu$m-sized grains. \citet{2013ApJ...774....8T} attributed their lower dust masses mainly to the use of more emissive carbon grains. However, their large error bars on observed SPIRE\,350 and 500\,$\mu$m datapoints made it hard to accurately constrain the Crab's dust mass. Instead, we argue that our lower dust masses can mostly be attributed to our different correction for ISM dust emission (see Appendix \ref{GalacticDust.sec}), which has resulted in lower SPIRE\,350 and 500\,$\mu$m ``supernova dust" fluxes (see Table \ref{Table_Fluxfraction}). These lower submm fluxes have resulted in the derivation of higher dust temperatures and thus lower dust masses, for a diverse set of grain species characterised by different dust optical constants, compared to estimates reported in previous papers. 

Summing the dust masses from each pixel, we infer a dust mass of 0.039$^{+0.010}_{-0.007}$\,M$_{\odot}$ for amorphous carbon ``a-C" grains. The cold dust component accounts for most mass (0.038$^{+0.009}_{-0.007}$\,M$_{\odot}$) and has an average dust temperature of 41$^{+3}_{-2}$\,K. The warm dust component (with a warm dust temperature fixed at 70\,K) accounts for 0.0017$^{+0.0003}_{-0.0003}$\,M$_{\odot}$ of mass. Our resolved supernova dust mass of 0.032-0.049\,M$_{\odot}$ (for amorphous carbon ``a-C" grains with size a=1\,$\mu$m) spans a smaller range of possible values compared to the dust mass inferred from the total integrated flux densities for the same grain species (0.005-0.042\,M$_{\odot}$), which we argue can be attributed to an accurate modelling of spatial variations in the synchrotron spectrum and the supernova dust temperature and mass distribution on resolved scales, which was averaged over for the total integrated SED fits. Compared to the heavy element mass predicted by nucleosynthesis models (0.42\,M$_{\odot}$, for a 11\,M$_{\odot}$ progenitor star, \citealt{1995ApJS..101..181W}), the modelled dust masses for ``a-C" grains correspond to a dust condensation efficiency of 8-12$\%$. The inferred dust condensation efficiency is consistent with estimates of the dust condensation efficiency for Cassiopeia\,A (0.1-0.17, \citealt{2010ApJ...713..356N,2018ApJ...866..128R,2019MNRAS.485..440P}).

In our resolved model, most supernova dust is located south of the pulsar (see Fig. \ref{Images_Crab_SED_resolved}, bottom left panel), coincident with the dense filaments of thermally excited ejecta observed in the optical and IR (see Fig. \ref{Crab_composite}; \citealt{1979ApJ...228..179D,1990ApJ...352..172G,1992ApJ...399..611B}), and with the detection of dense knots of molecular gas \citep{1990ApJ...352..172G,2010ApJ...716L...9L,2011ApJS..194...30L,2012MNRAS.421..789L}. The overall dust mass distribution peaks at the locations of dense thermal filaments, and shows a deficit towards the north-west of the remnant. The dust mass in the north-west part of the remnant may be lower due to the lack of sufficiently dense filaments where elements could have more easily accreted onto grain seeds and grown dust. The absence of a dense circum- or interstellar medium has resulted in higher shock velocities towards the NW \citep{1995AJ....109.2635L} and has prevented the gas from cooling sufficiently after passage of the shock, with cooling time scales longer than the age of the Crab Nebula \citep{1997ApJ...491..796S,2008ARA&A..46..127H}, and may provide an explanation for the lower supernova dust masses towards the NW. The most prominent Crab's filaments are thought to have formed through Rayleigh-Taylor instabilities about 100 to 200 years after the explosion \citep{1998ApJ...499..282J,2014MNRAS.443..547P}, implying that the dust was only later assembled in these filaments as the onset of dust formation is thought to take place 300-500 days after explosion with the build up of dust mass lasting several tens of years \citep{2014Natur.511..326G,2015MNRAS.446.2089W,2016MNRAS.456.1269B}. It is of interest that dust is not only present in dense filaments but appears to be distributed throughout the remnant, which presumably indicates that dust is present in smaller filaments or globules, as observed by e.g., \citet{2017A&A...599A.110G}, which remain unresolved by our infrared observations. 

\begin{figure}
	\includegraphics[width=8.25cm]{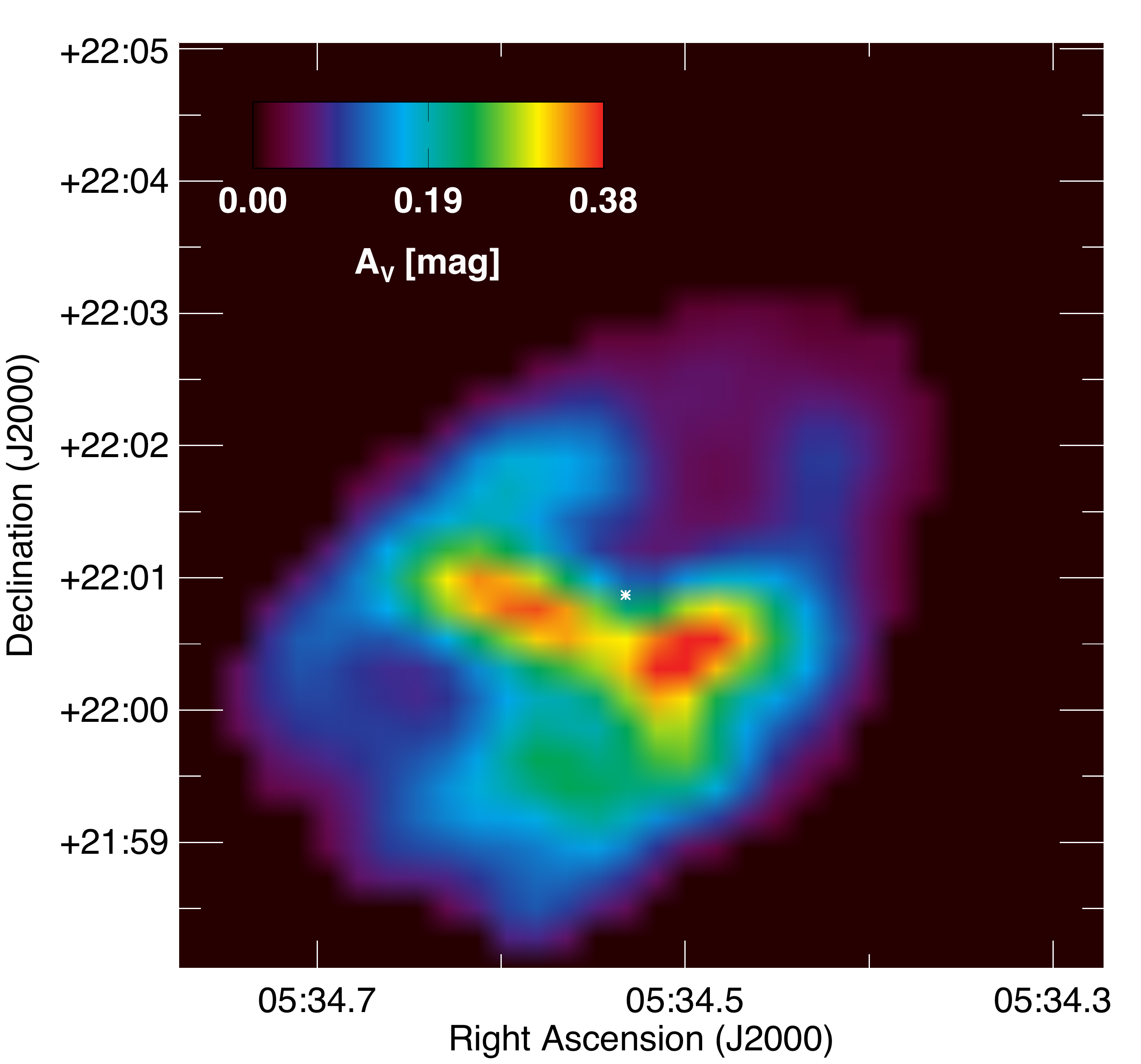} \\
    \caption{Map of the $V$ band extinction ($A_{\text{V}}$, right) along the line of sight to the Crab Nebula, as inferred from the cold+warm supernova dust mass (see Figure \ref{Images_Crab_SED_resolved}, bottom right panel). The white cross indicates the position of the Crab pulsar.}
    \label{Crab_AV_map}
\end{figure}
We converted our resolved supernova dust mass map to a measure of the $V$ band extinction, $A_{\text{V}}$ (see Figure \ref{Crab_AV_map}), for which we relied on the $A_{\text{V}}$-to-dust column density ratio inferred for ``a-C" grains. The highest values ($A_{\text{V}}$=0.20-0.39\,mag) correspond to the dense filaments, while $A_{\text{V}}$ ranges between 0.05 and 0.10\,mag in most parts of the nebulae. The lowest $A_{\text{V}}$=0.05\,mag values are inferred for the outskirts of the Crab. Our $V$ band extinction estimates ($A_{\text{V}}$=0.20-0.39\,mag) in the dense filaments are in excellent agreement with the $V$ band extinction ($A_{\text{V}}$=0.20-0.34\,mag) measured directly from optical images by \citet{2017A&A...599A.110G} for 7 dark and distinct globules, and makes us confident that our inferred dust masses are accurate. We should note that the median $A_{\text{V}}$ (0.10$\pm$0.08\,mag) along the line-of-sight to the Crab is significantly lower than the $A_{\text{V}}$ (1.08$\pm$0.39\,mag) inferred from the interstellar dust emission in the vicinity of the nebula, which suggests that there is about an order of magnitude more interstellar extinction along the line of sight to the Crab compared to supernova dust. Nonetheless, the emission in the FIR wavebands is dominated by thermal emission from supernova dust, and the contributions from interstellar and supernova dust emission are comparable at submm wavelengths, because the average supernova dust temperature ($T_{\text{dust}}$=41\,K) is significantly higher than the median ISM value ($T_{\text{dust}}$=15\,K, see Appendix \ref{GalacticDust.sec}).

We conclude that the dust mass (0.032-0.049\,M$_{\odot}$) formed in the Crab Nebula is consistent with a scenario of efficient dust condensation (8-12$\%$) in the supernova ejecta; this result confirms that dust grains can efficiently condense in the post-explosion ejecta of supernova. Based on an updated distance to the Crab pulsar inferred from GAIA data, the Galactic remnant may be located significantly further from the Sun (3.37\,kpc, \citealt{2019ApJ...871...92F}) than previously thought (2\,kpc, \citealt{1968AJ.....73..535T}). At this distance, the Crab's dust mass would increase to 0.091-0.138\,M$_{\odot}$ with a dust condensation efficiency of 22-33\,$\%$. Although the Crab's dust mass is lower than the average dust mass (0.1-1\,M$_{\odot}$) per single supernova event required to account for the dust budgets observed in the Early Universe \citep{2003MNRAS.343..427M}, we might not expect a supernova from a progenitor star with mass 8-11\,M$_{\odot}$ to produce 1\,M$_{\odot}$ of dust (simply because only 0.42\,M$_{\odot}$ of metals are expected to be present in the post-explosion ejecta). Most dust will instead be produced by higher mass progenitors (e.g., Cas\,A, SN\,1987A). The dust mass inferred for the Crab Nebula is lower than the ones inferred for other Galactic PWNe \citep{2019MNRAS.483...70C}. Due to the careful selection of dust structures correlated with the PWNe and the temperature components warmer than the average interstellar dust by \citet{2019MNRAS.483...70C}, their dust mass estimated have effectively been corrected for the background interstellar dust emission, in a similar way as was done in \citet{2017MNRAS.465.3309D} and in this work, and is thus unlikely to account for the difference in inferred dust masses. Instead, the dissimilar dust masses are thought to be caused by a combination of effects. First of all, due to a difference in assumed dust properties with $\kappa_{\text{850}\,\mu\text{m}}$=0.7\,m$^{2}$ kg$^{-1}$ (assumed in \citealt{2019MNRAS.483...70C}), which is about a factor of 3 lower compared to the standard ``a-C" dust model used to infer the Crab's resolved dust mass. Secondly, the infrared images in \citet{2019MNRAS.483...70C} were not corrected for possible line contributions; but given that line emission tends to have the highest contributions to mid- and far-infrared wavebands, it is not clear whether the cold dust mass estimates would be effected much. Finally, it is likely that the three PWNe from \citet{2019MNRAS.483...70C} result from progenitors with a higher mass (than that of the Crab progenitor) for which we expect to find an elevated mass of condensed grains in accordance to their more massive reservoirs of metals. A consistent comparison of PWN dust masses would require an accurate knowledge of the supernova dust composition (and its variation with SN explosion energy and progenitor mass) in PWNe. The Crab's dust production does confirm that PWNe are efficient dust factories (e.g., \citealt{2019MNRAS.484.5468O}). 

\section{Millimetre-excess emission}
\label{MillimetreExcess.sec}
In our total integrated SED model, the observed mm excess emission in the Crab Nebula is best described by a spectrum with peaks around $\nu_{\text{peak}}$=163$^{+42}_{-22}$\,GHz (or around 2\,mm), with a width of $\sigma$=0.27$^{+0.06}_{-0.08}$ and amplitude of $F_{\text{mm,peak}}$=69$^{+14}_{-23}$\,\,Jy. The model parameters to describe the excess emission at mm wavebands hold regardless of the assumed dust species. We note that a model with a broken power-law synchrotron spectrum was not capable of reproducing the observed fluxes in the mm wavelength range without accounting for a mm excess in the model. A synchrotron model with an evolutionary break might reduce the need for a separate component to account for mm excess emission, but degrades the quality of the fits at submm wavelengths. We note that the excess emission, as fitted by our total integrated SED model, accounts for 33$\%$ of the emission at 2\,mm, and can therefore not be attributed to short-term variability in radio brightness due to instabilities within the nebular flow (which is thought to result in variations up to 10$\%$ at most, \citealt{2015MNRAS.446..205B}). An excess at millimetre wavebands in the Crab has been identified before by several other authors \citep{2002A&A...386.1044B,2010ApJ...711..417M}. In Section \ref{ResolvedMillimetreExcess.sec}, we address the spatially resolved distribution of the mm excess emission, while we expand on the possible origin of the excess emission in Section \ref{OriginMillimetreExcess.sec}. 

\subsection{Spatial distribution of excess emission}
\label{ResolvedMillimetreExcess.sec}

\begin{figure*}
	\includegraphics[width=17.5cm]{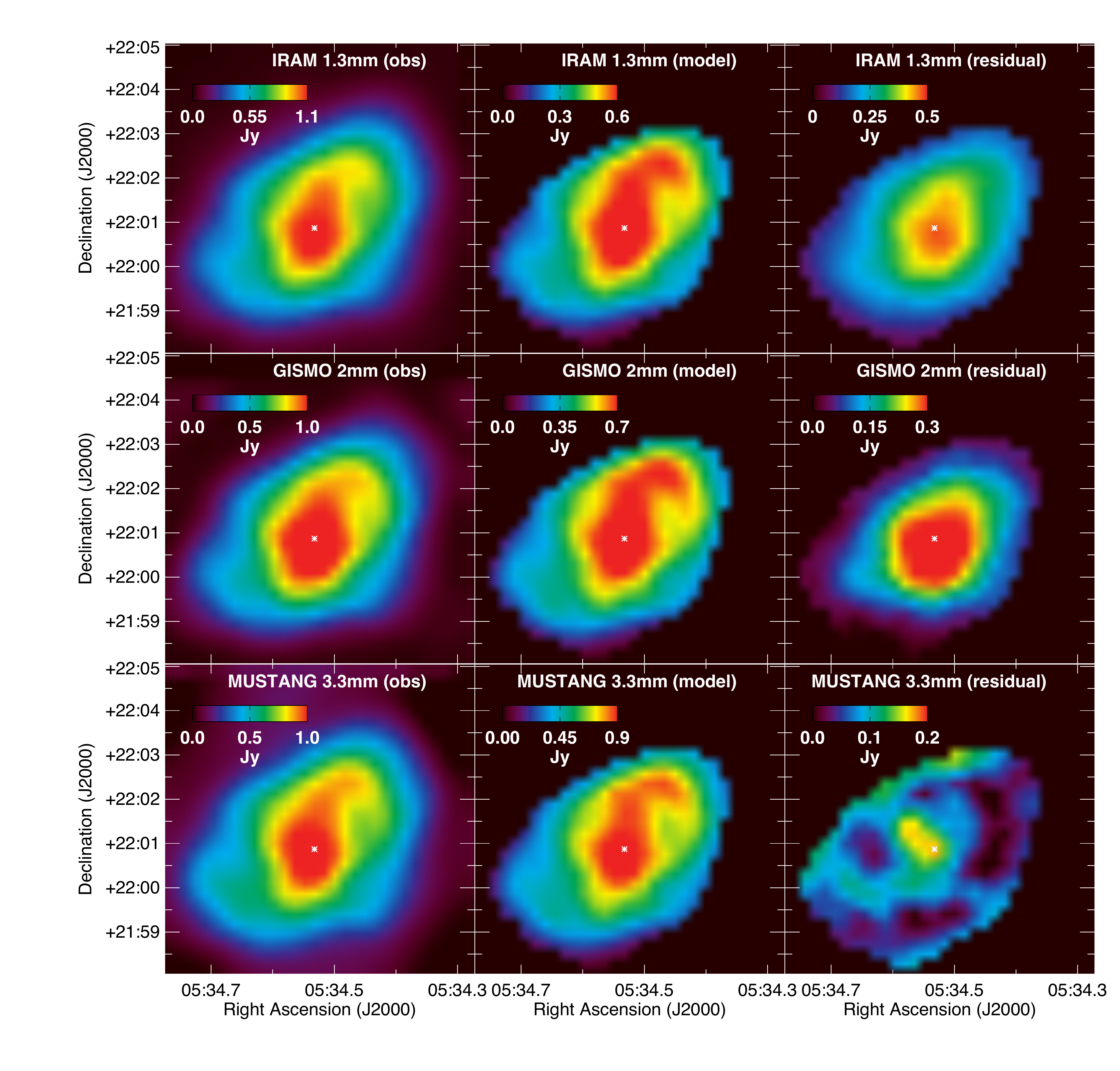}	
    \caption{Resolved maps of the observed (left column), modelled (middle column), and residual (i.e., observations-model, right column) emission in three mm wavebands: MAMBO 1.3\,mm (top row), GISMO 2\,mm (middle row) and MUSTANG 3.3\,mm (bottom row). The model emission in those mm wavebands was inferred from a Bayesian SED model on resolved scales of 14$\times$14 arcsec$^{2}$ (or 0.136$^{2}$ pc$^{2}$, for a distance of 2\,kpc).}
    \label{Images_Crab_mm_resolved}
\end{figure*}

We employ an interpolation for every pixel of the resolved Bayesian SED models (see Section \ref{ResolvedSED.sec}) to predict the millimetre emission inferred by our combined synchrotron and supernova dust model. Figure \ref{Images_Crab_mm_resolved} compares the observed 1.3, 2 and 3.3\,mm images (left column) with the model interpolations (middle column). The final column presents the residual after subtracting the model emission from the observed mm wave maps. The residual emission peaks in the centre, and appears to roughly follow the structure of the torus and the jet as seen in X-ray \citep{2006ApJ...652.1277S,2015ApJ...801...66M} and in radio images \citep{1992MNRAS.255..210V,2015MNRAS.446..205B,2017ApJ...840...82D} of the Crab Nebula (see Fig. \ref{Crab_composite}), which had also been remarked on by \citet{2002A&A...386.1044B}.

\subsection{Origin of the mm-excess emission}
\label{OriginMillimetreExcess.sec}

To explain the mm excess emission in the Crab Nebula, we explore four different scenarios: (1.) spatial variations in the synchrotron spectrum and/or a secondary synchrotron component; (2.) free-free emission from a hot plasma; (3.) spinning dust grains; or (4.) magnetic (Fe-bearing) grains. We believe a synchrotron origin for the mm-excess emission is most likely, and discuss this scenario in more detail below. The other possible scenarios are discussed and compared with the expected location of the mm excess emission as inferred from residuals after subtracting the best-fit resolved synchrotron model from the observed mm images (see Section \ref{ResolvedMillimetreExcess.sec}), in Appendix \ref{App_mmexcess}. 
\smallskip \\
\textbf{(1.) Spatial variations or multiple synchrotron components:}
While the mm excess emission could not be reproduced with a single broken power-law synchrotron spectrum, it is possible that spatial variations in the wavelength break of the spectrum and/or a secondary synchrotron component would be able to account for the millimetre excess emission. Spatial variations in the break wavelength are expected to result in a synchrotron spectrum with multiple breaks on nebular scales. And, that is also what can be expected based on physical arguments, as the break in the synchrotron spectrum corresponds to the transition from a regime affected by synchrotron losses (shortwards of the break) to a regime without such losses. The separation between those two regimes is known to vary spatially, affecting the position of the break (see top left panel of Fig. \ref{Images_Crab_SED_resolved} in this paper, and \citealt{2002A&A...386.1044B}). These spatial variations of the position of the break could, however, also further smear out any bump (or excess) in the spectrum. A bump in the synchrotron spectrum depends mostly on the particle population (with a minor dependence on the source geometry and shock dynamics) and results from the pile up of energetic electrons just below the spectral break \citep{2009ApJ...703..662R}. In our case, this would suggest a spectral break shortwards of 2\,mm in the central regions of the Crab (where the excess is observed, see Section \ref{ResolvedMillimetreExcess.sec}). A spectral break at mm wavelengths would imply a magnetic field several times stronger than estimated for the Crab, and demonstrates the dependence of our break wavelength on the simple broken power-law spectrum assumed to model the synchrotron emission. Attempts to model the Crab's synchrotron radiation with an evolutionary break did not improve the overall fits (see Appendix \ref{Alternative_synchrotron_model.sec}). Due to the complexity of spatial and temporal variations in the Crab's synchrotron spectrum, we defer any further investigation of the optimal synchrotron parameters to future work.  

More likely than an evolution in the energy distribution of synchrotron-emitting particles, the synchrotron radiation might originate from two different synchrotron components \citep{1996MNRAS.278..525A,2002A&A...386.1044B,2010A&A...523A...2M,2013MNRAS.433.3325S,2014MNRAS.438..278P,2018arXiv181101767L}, which are thought to spatially correlate with the different particles emitting at radio and optical wavelengths. The transition between those two particle distributions is likely to occur between infrared and radio wavelengths, which might coincide with the excess emission observed in the Crab. It is however not clear whether two synchrotron components can account for the bump in the spectrum observed at mm wavelengths. Future work with increased photometric coverage in the mm, cm and radio wavelength ranges will be required to verify whether two distinct synchrotron components could be responsible for the mm excess emission. 

We emphasize that the origin of the mm excess emission will not affect the dust mass inferred for the Crab Nebula in this paper. The supernova dust emission is restricted to wavebands shortwards of the SPIRE\,500\,$\mu$m band, while the mm excess emission hardly contributes at sub-mm wavelengths (see Table \ref{Table_Fluxfraction}). 

\section{Conclusions}
\label{Conclusions.sec}
We have modelled the total integrated and resolved near-infrared to radio wavelength spectrum of the Crab Nebula with a Bayesian SED model to simultaneously account for the synchrotron radiation, interstellar and supernova dust emission and millimetre excess emission as observed from this pulsar wind nebula by \textit{Spitzer}, \textit{WISE}, \textit{Herschel} and several mm and radio ground based facilities. Here, we summarise our new results from this analysis.
\begin{itemize}
\item The contribution from interstellar dust emission along the line-of-sight to the Crab Nebula was estimated based on modelling of the ISM dust emission for regions in the vicinity of the Crab. A maximum contribution of 22$\%$ was found for the SPIRE\,250\,$\mu$m waveband, with lower percentage contaminations in the other \textit{Herschel} bands. The average ISM dust column corresponds to a $V$-band extinction of $A_{\text{V}}=1.08\pm0.38$\,mag along the line of sight (see Appendix \ref{GalacticDust.sec}), in agreement with the reddening inferred based on Pan-STARRS1 and 2MASS photometry \citep{2015ApJ...810...25G}.
\item The resolved supernova dust mass of 0.032-0.049\,M$_{\odot}$ (for amorphous carbon ``a-C" grains) implies that less dust has formed in the Crab Nebula than previously derived (0.11-0.24\,M$_{\odot}$, \citealt{2012ApJ...760...96G}). Our dust mass estimates are consistent within the uncertainties with the carbon dust masses (0.016-0.061\,M$_{\odot}$) from \citet{2013ApJ...774....8T}, and the dust mass (0.056$\pm$0.037\,M$_{\odot}$) reported by \citet{2019arXiv190303389N}. Unlike \citet{2013ApJ...774....8T}, who ascribed their lower values to a difference in assumed dust properties, we attribute our lower dust mass estimates (compared to \citealt{2012ApJ...760...96G}) to the corrections applied to account for interstellar dust emission and a reduced SPIRE\,500\,$\mu$m flux, which together have resulted in lower SN dust contributions at SPIRE\,350 and 500\,$\mu$m and consequently an average dust temperature ($T_{\text{dust}}\sim41^{+3}_{-2}$\,K) for the Crab higher than was previously inferred ($T_{\text{dust}}\sim$28-34\,K, \citealt{2012ApJ...760...96G}). 
\item The $V$ band extinction ($A_{\text{V}}$=0.20-0.39\,mag, see Fig. \ref{Crab_AV_map}) inferred from the supernova dust mass maps is consistent with the optical extinction measurements ($A_{\text{V}}$=0.20-0.34\,mag) from \citet{2017A&A...599A.110G} for seven dark globules. The dust in the Crab is predominantly found in dense filaments south of the pulsar.
\item If we account for the total amount of condensable material predicted to be produced by a 11\,M$_{\odot}$ progenitor star \citep{1995ApJS..101..181W}, we infer a dust condensation efficiency of 8-12$\%$, in line with dust condensation models for core-collapse supernovae \citep{2010ApJ...713..356N} and consistent with dust condensation efficiencies inferred from IR/submm observations for Cas\,A \citep{2019MNRAS.485..440P}. A recently proposed revision of the Crab's distance to 3.37\,kpc \citep{2019ApJ...871...92F} would increase our total dust mass estimate to 0.091-0.138\,M$_{\odot}$, and imply dust condensation efficiencies of 22-33\,$\%$. 
\item The modelled synchrotron power-law spectrum is consistent with an average spectral index $\alpha_{\text{radio}}$=0.297 in the radio domain, a spectral break $\lambda_{\text{break}}$ in the mm-cm wavelength range, and an infrared spectral index $\alpha_{\text{IR}}$=0.429. 
\item We are unable to fit the Crab's total integrated spectrum without accounting for the observed millimetre excess emission. We investigate whether the emission is produced by electric or magnetic dipole emission originating from small spinning or magnetised grains. The spatial correlation of the excess emission with the Crab's synchrotron radiation, whereas the dust is mostly distributed along the dense filaments, does not support scenarios of spinning or magnetised grains. At the Crab's dust temperature, we would furthermore expect to observe electric and/or magnetic dipole emission from spinning grains predominantly at far-infrared wavelengths. Although we cannot rule out that spinning and/or magnetised grains are responsible for the excess emission, we deem it unlikely. 
\item Instead, we argue that the mm-wave excess emission could be affected by spatial and secular variations in the Crab's synchrotron spectrum and, more specifically, the energy distributions of relativistic particles producing the synchrotron emission. Although it is unclear how these spatial and secular variations could give rise to a bump in the spectrum. It is more plausible that the excess emission results from two distinct populations of synchrotron emitting particles driven by two different acceleration mechanisms. Future modelling efforts, with an increased number of resolved mm, cm and longer wavelength radio observations, may be able to shed light on the origin of this excess emission.
\end{itemize}

In line with recent studies of other SNRs, we conclude that the Crab's efficient dust condensation (8-12$\%$) provides further evidence for a scenario whereby supernovae could provide considerable contributions to the interstellar dust budgets in galaxies. In the future, with upcoming facilities such as the James Webb Space Telescope (JWST, \citealt{2006SSRv..123..485G}), SPICA \citep{2018PASA...35...30R} and the Origins Space Telescope, we will be able to expand on the current sample of PWNe with dust mass detections and, at the same time, look for dust features characteristic of the type of grain species formed in PWNe, which will help to further bring down uncertainties on current supernova dust masses and to provide observational input to test the nucleation of various grain species during the first couple of hundred days post-explosion in the current generation of core-collapse supernova dust models (see  \citealt{2018SSRv..214...63S} for a recent review).

\section*{Acknowledgements}
The authors would like to thank the anonymous referee for her/his useful suggestions that have significantly improved the discussion and the presentation of our results. The authors would like to thank Rick Arendt for kindly sharing his mm data (GISMO 2\,mm and MUSTANG 3.3\,mm) of the Crab Nebula; and Isabella Lamperti, Anthony Jones, Bruce Draine and Boris Leistedt for fruitful discussions. 
IDL gratefully acknowledges the support of the Research Foundation Flanders (FWO). M.J.B., A.B. and R.W. acknowledge support from European Research Council (ERC) Advanced Grant {\sc SNDUST} 694520. M.M. acknowledges support from STFC Ernest Rutherford fellowship (ST/L003597/1). HLG and HC acknowledges support from the European Research Council (ERC) in the form of Consolidator Grant {\sc CosmicDust}. F.P. acknowledges
support from the UK Science and Technology Funding Council (STFC).

PACS was developed by a consortium of institutes led by MPE (Germany) and including UVIE (Austria); KU Leuven, CSL, IMEC (Belgium); CEA, LAM (France); MPIA (Germany); INAFIFSI/ OAA/OAP/OAT, LENS, SISSA (Italy); IAC (Spain). This development has been sup- ported by the funding agencies BMVIT (Austria), ESA- PRODEX (Belgium), CEA/CNES (France), DLR (Ger- many), ASI/INAF (Italy), and CICYT/ MCYT (Spain). SPIRE was developed by a consortium of institutes led by Cardiff University (UK) and including Univ. Lethbridge (Canada); NAOC (China); CEA, LAM (France); IFSI, Univ. Padua (Italy); IAC (Spain); Stockholm Observatory (Sweden); Imperial College London, RAL, UCL-MSSL, UKATC, Univ. Sussex (UK); and Caltech, JPL, NHSC, Univ. Colorado (USA). This development has been supported by national funding agencies: CSA (Canada); NAOC (China); CEA, CNES, CNRS (France); ASI (Italy); MCINN (Spain); SNSB (Sweden); STFC and UKSA (UK); and NASA (USA).


\bibliographystyle{mnras}
\bibliography{Crab_dust}
\clearpage







\appendix
\section{Bayesian SED modelling approach}
\label{BayesianSED.sec}
Due to the vast number of parameters, we have employed a Bayesian inference method coupled to a Markov Chain Monte Carlo (MCMC) algorithm to search the entire parameter space in an efficient way, and to reveal any parameter degeneracies. More specifically, we have sampled the N-dimensional parameter space through an affine invariant ensemble sampler \citep{2010CAMCS...5...65G} through the use of the $``$emcee$"$ package for Markov Chain Monte Carlo (MCMC) in \texttt{Python} \citep{2013PASP..125..306F}. The sampling of the N-dimensional parameter space is achieved through random walks of a collection of walkers ($N_{\text{walkers}}$). During each step, the position of the walker in the N-dimensional space is aimed to change to a new position with a higher likelihood. We apply a likelihood function based on the commonly used $\chi^{2}$ statistic with $\chi^{2}=\sum_{i=1}^{N}\left(\frac{f_{\text{i,obs}}-f_{\text{i,model}}}{\sigma_{\text{i,obs}}}\right)$, where $f_{\text{i,obs}}$ and $f_{\text{i,model}}$ are the observed and model fluxes in waveband i, and $\sigma_{\text{i,obs}}$ corresponds to the observational uncertainty. The position of each walker (and thus the values of each parameter) is recorded after an initial warm-up phase of $N_{\text{burn}}$ steps, and used to construct the posterior probability distribution functions for each model parameter. We have used $N_{\text{walkers}}$=100, $N_{\text{burn}}$=50,000 and $N_{\text{steps}}$=200,000 for the total integrated model runs (see Section \ref{GlobalSED.sec}), and $N_{\text{walkers}}$=200, $N_{\text{burn}}$=4,000 and $N_{\text{steps}}$=10,000 for the resolved modelling in individual pixels (see Section \ref{ResolvedSED.sec}). We verified that these number of steps were sufficient for convergence by estimating the integrated autocorrelation time, $\tau_{\text{int}}$, of the chain. The effective sample size $N_{\text{eff}}$, defined as $N_{\text{chain}}$/$\tau_{\text{int}}$ was always (significantly) higher than 10 for all parameters. We furthermore looked at the acceptance fraction of walkers, which (ideally) should vary between 0.2 and 0.5, and adjusted the scale parameter accordingly if the acceptance rate of walkers was too low initially. The priors assumed for each of the parameters are outlined in the respective model descriptions.

\begin{table*}
\caption{Overview of the Bayesian model parameters and the assumed priors for the total integrated and resolved SED fitting models presented in Sections \ref{GlobalModel.sec} and \ref{ResolvedSED.sec}.}
\label{BayesianModelParameters}
\begin{tabular}{lcccc} 
\hline
\hline
\multicolumn{5}{|c|}{\textbf{Total integrated SED model}} \\
\hline 
\hline
\textbf{Synchrotron model parameters:} & $\alpha_{\text{radio}}$ & $\alpha_{\text{IR}}$ & $\lambda_{\text{break}}$ & $F_{\text{1.4\,GHz}}$ \\
\hline
Units: & & & [$\mu$m] & [Jy] \\ 
Prior range: & [0.1,0.4] & [0.3,1.0] & [20,20000] & [300,3000]  \\
\hline 
\textbf{SN dust model parameters:} & $T_{\text{cold}}$ & $\log$ $M_{\text{cold}}$ & $T_{\text{warm}}$ & $\log$ $M_{\text{warm}}$  \\
\hline
Units: & [K] & [$\log$ M$_{\sun}$] & [K] & [$\log$ M$_{\sun}$] \\ 
Prior range: & [12,60] & [-4.0,1.0] & [40,100] & [-7.0,1.0] \\
\hline 
\textbf{Mm excess model parameters:} & Peak $\nu_{\text{mm}}$ & Width $\sigma$ & $F_{\text{mm,peak}}$ &   \\
\hline
Units: & [GHz] &  & [Jy] &  \\ 
Prior range: & [30,400] & [0.1,1.0] & [1,300] & \\
\hline 
\hline
\multicolumn{5}{|c|}{\textbf{Resolved SED model}} \\
\hline 
\hline
\textbf{Synchrotron model parameters:} & $\lambda_{\text{break}}$ & $\alpha_{\text{IR}}$ & $F_{\text{1.4\,GHz}}$ & \\
\hline
Units: & [mm] & & [Jy] & \\ 
Prior range: & [0.01,60] & [0.3,1.0] & [1,1000] & \\
\hline 
\textbf{SN dust model parameters:} & $T_{\text{cold}}$ & $\log$ $M_{\text{cold}}$ & $T_{\text{warm}}$ & $\log$ $M_{\text{warm}}$  \\
\hline
Units: & [K] & [$\log$ M$_{\sun}$] & [K] & [$\log$ M$_{\sun}$] \\ 
Prior range: & [12,60] & [-6.0,-2.0] & [40,100] & [-6.0,0.0] \\
\hline 
\end{tabular}
\end{table*}

\section{Resolved Bayesian model residuals}
\label{BayesianSEDresiduals.sec}
\begin{figure*}
	\includegraphics[width=13.7cm]{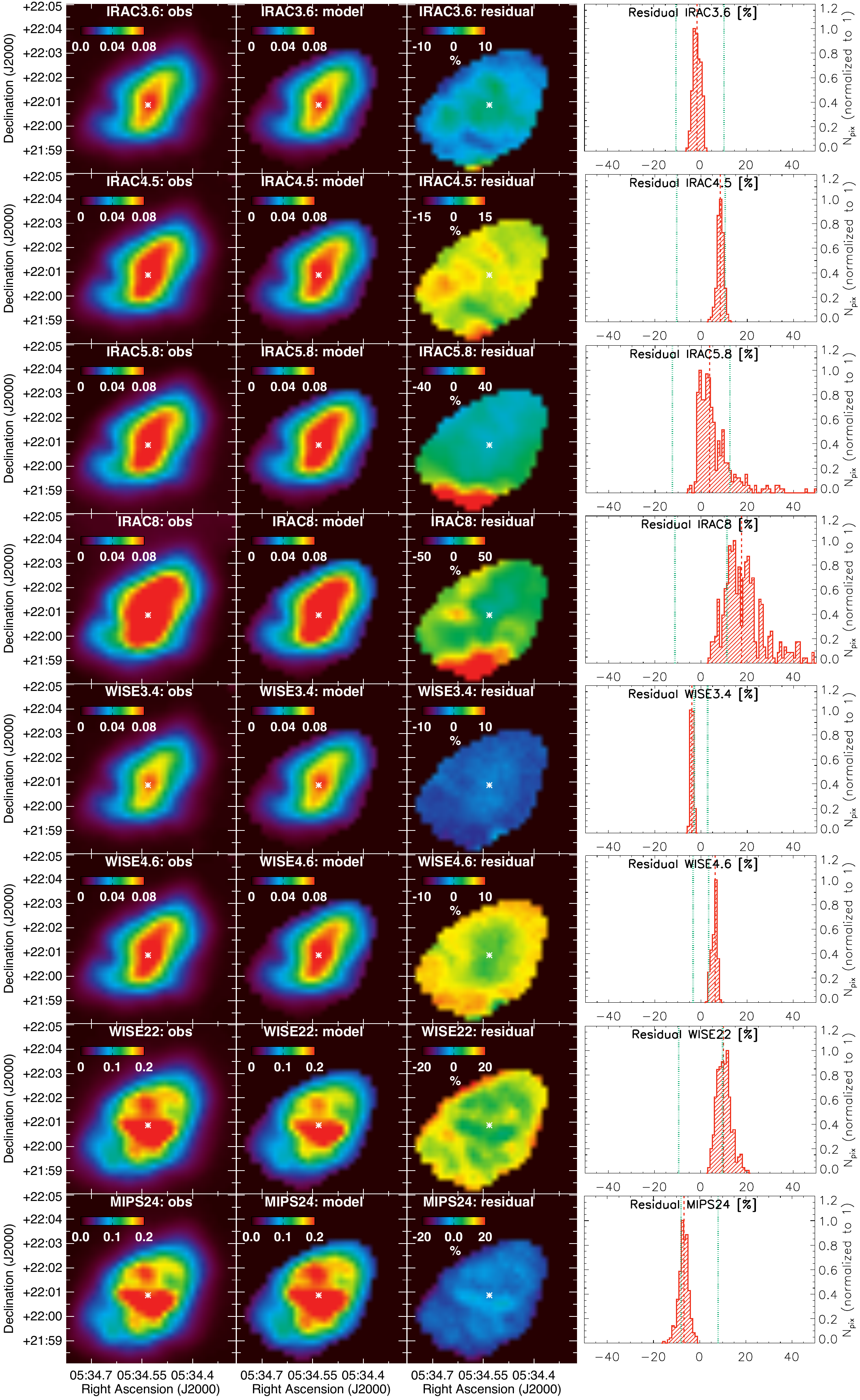}
	    \caption{The observed (left column) and modelled (second column) IRAC\,3.6, 4.5, 5.8, 8.0\,$\mu$m, WISE\,3.4, 4.6, 22\,$\mu$m and MIPS 24\,$\mu$m images. The observed maps were corrected for line contamination and ISM dust emission (both of which are not included in the Bayesian model). The third column shows the residual image, $\frac{F_{\text{obs}}-F_{\text{model}}}{F_{\text{obs}}}$, as percentage deviations to highlight deviations of the model from the observations. The scale of the residual images was adjusted to cover the entire range of residuals. In the right-hand column, the histogram of residuals is shown for every waveband. The median residual (vertical red dashed lines) are compared to the median uncertainty on the flux density in every pixel (green dashed line).}
    \label{Crab_SED_residuals_part1}
\end{figure*}

\begin{figure*}
	\includegraphics[width=13.8cm]{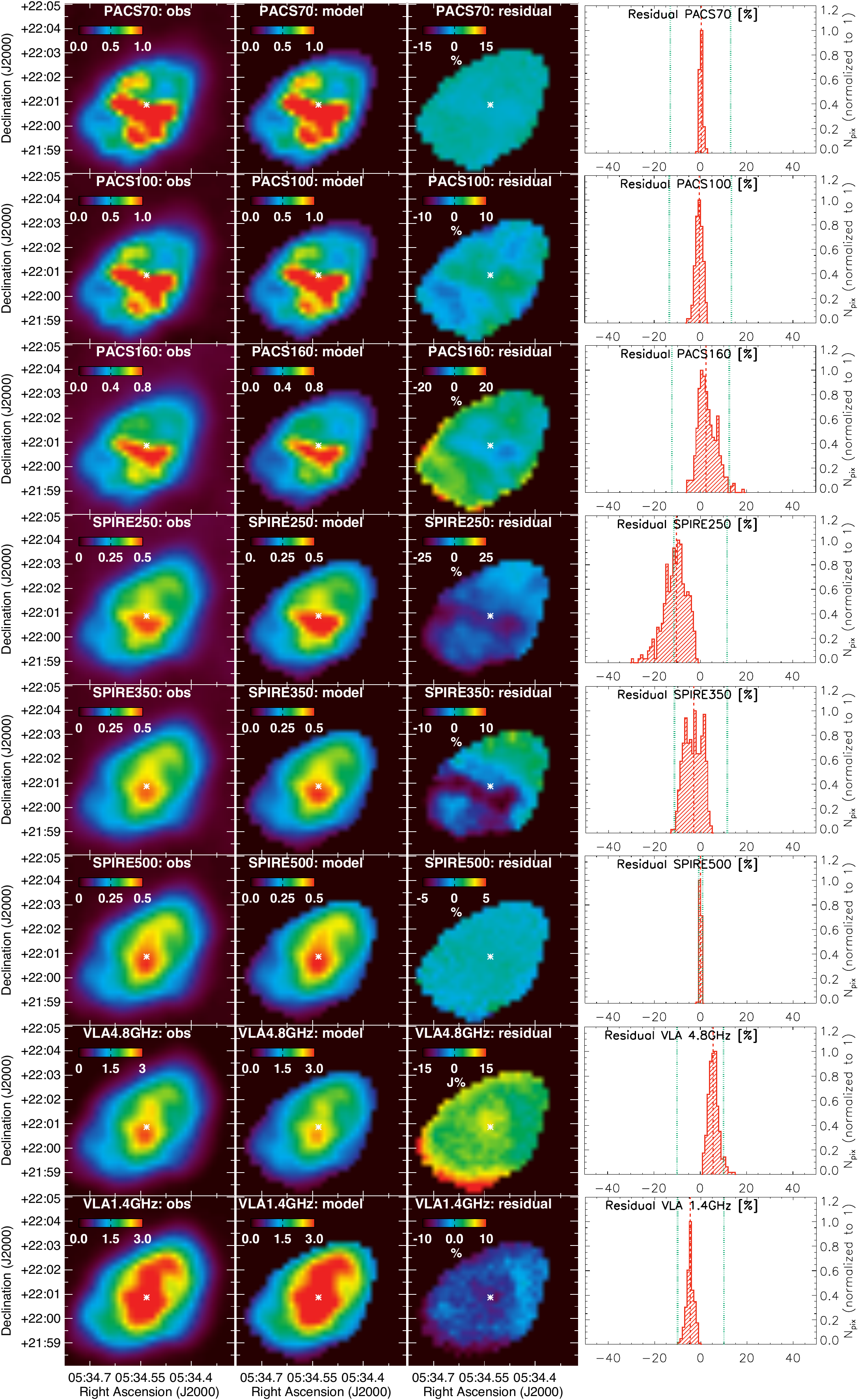}
	    \caption{Same as Figure \ref{Crab_SED_residuals_part1} but with the PACS\,70, 100, 160\,$\mu$m, SPIRE\,250, 350, 500\,$\mu$m, and VLA\,4.8\,GHz and 1.4\,GHz images.}
    \label{Crab_SED_residuals_part2}
\end{figure*}

The goodness of fit of the resolved Bayesian model (see Section \ref{ResolvedSED.sec}) is discussed here based on Figures \ref{Crab_SED_residuals_part1} and \ref{Crab_SED_residuals_part2}, displaying the residuals for the NIR to radio images that were used to constrain the model in each pixel. In all wavebands, the median uncertainties in each band encompass the histograms of residuals (with the exception of a few pixels), which demonstrates that our model is adequate to fit the resolved NIR to radio emission of the Crab. The longer tail of positive IRAC\,5.8 and IRAC\,8.0\,$\mu$m residuals originates from the southern (visible) edge of the Crab, and is likely a consequence of a stronger contribution from warm interstellar dust (as suggested by the increased ISRF at that location, see Fig. \ref{Crab_interstellar_maps}), compared to the average ISM dust contribution which was subtracted from the Crab images. 

The flux density levels in the VLA\,4.8\,GHz image tend to be higher than our model (although within the uncertainty), while an opposite trend is observed for the VLA\,1.4\,GHz image. This suggests that the radio spectral index might be somewhat steeper compared to the average value of 0.297 inferred from the total integrated SED modelling and assumed for the resolved modelling. Alternatively, the model solutions might be pushed to the limits of observational uncertainties to enable a fit to the SPIRE fluxes with a broken power law synchrotron model with variable spectral break. 

\section{Galactic dust emission}
\label{GalacticDust.sec}
To estimate the contribution from IS dust emission, both in the foreground and background of the Crab Nebula, we modelled the \textit{Herschel} emission in the field around the Crab Nebula. For this, we convolved all PACS and SPIRE images to the SPIRE\,500\,$\mu$m resolution, and rebinned them to the standard pixel size of 14$\arcsec$ in the SPIRE\,500\,$\mu$m waveband. We only fitted pixels with 3$\sigma$ detections in the \textit{Herschel} PACS\,160\,$\mu$m and SPIRE\,250, 350, and 500$\mu$m wavebands. In practice, the signal-to-noise ratio in the PACS\,160\,$\mu$m band is the limiting factor. We required the PACS\,160\,$\mu$m fluxes to better capture the peak of the dust SED and thus to constrain the ISM dust temperature. 

To model the dust SED emission, we employed the THEMIS (The Heterogeneous dust Evolution Model for Interstellar Solids) dust model \citep{2013A&A...558A..62J,2014A&A...565L...9K,2017A&A...602A..46J} including amorphous hydrocarbon grains (a-C(:H)) and silicates with iron nano-particle inclusions (a-Sil$_{\text{Fe}}$). The optical properties for these grains have been derived from laboratory studies, and the size distribution and abundances of grain species were calibrated to reproduce the extinction and emission observed in the diffuse interstellar regions in the Milky Way. We use the SED fitting tool \texttt{DustEm} \citep{2011A&A...525A.103C} to generate a library of model dust SEDs for a range of scaling factors (G) of the far-ultraviolet (FUV) interstellar radiation field (ISRF), which correspond to steps in $T_{\text{dust}}$ of 0.1\,K. The shape of the ISRF is set to the radiation field in the solar neighbourhood \citep{1983A&A...128..212M} normalised to $G$=1.0\,$G_{\text{0}}$\footnote{Here G refers to the average FUV ISRF normalised to the units of the \citet{1968BAN....19..421H} field, i.e., $G_{\text{0}}$=1.6$\times$10$^{-3}$ erg s$^{-1}$ cm$^{-2}$ \citep{2005pcim.book.....T}, which relates to the \citet{1978ApJS...36..595D} field, $\xi_{\text{0}}$, as $G_{\text{0}}$=1.7$\xi_{\text{0}}$.}. We derive an estimate of $T_{\text{dust}}$ for each model $G$ based on the mean equilibrium temperatures for large carbon grains and amorphous silicates in the THEMIS dust model weighted by the abundance of grains in each grain size bin. During the fitting procedure, the model SED is convolved with the PACS and SPIRE filter response curves. The best fitting SED model is inferred from a Bayesian inference method (see Appendix \ref{BayesianSED.sec}). Figure \ref{Crab_interstellar_SED} shows an example fit of the ISM dust SED for a randomly chosen pixel (left), with the associated posterior distributions (right) for each model parameter.   

\begin{figure*}
	\includegraphics[width=9.5cm]{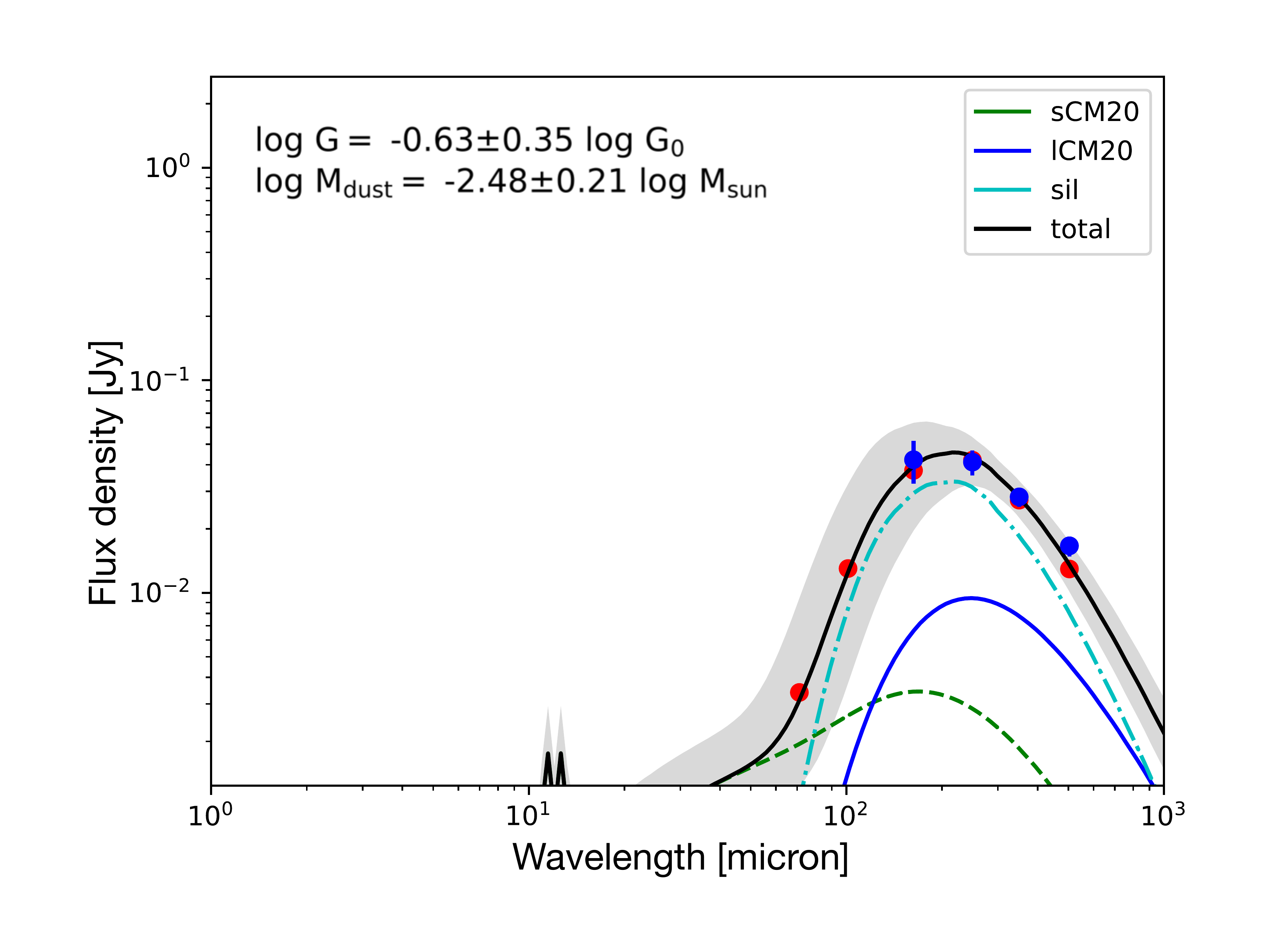}
	\includegraphics[width=7.0cm]{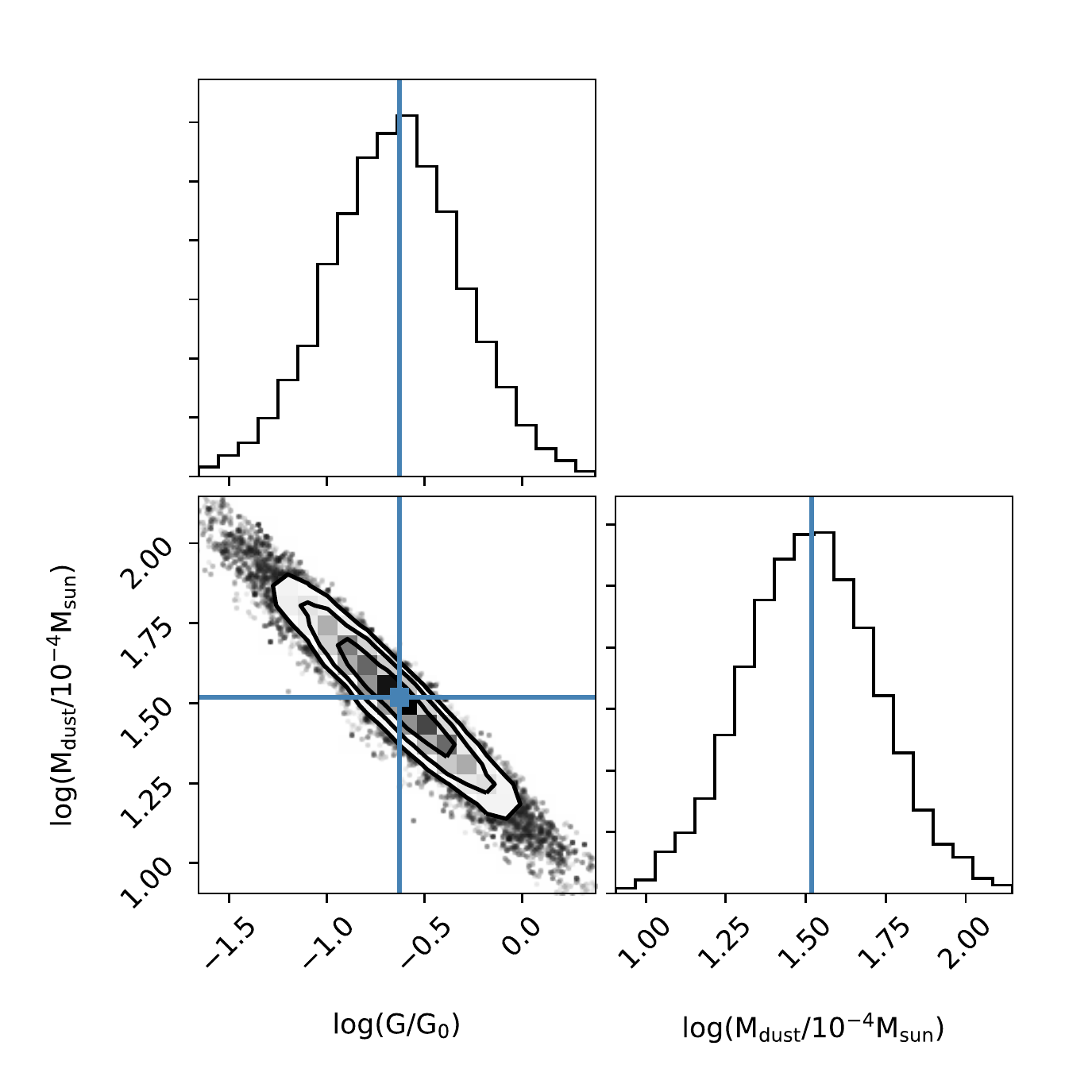}	    \caption{Left: best-fit THEMIS ISM dust SED model as inferred from our Bayesian modelling with observed data points indicated in blue, and modelled data points in red. The legend explains the curves used to represent individual THEMIS dust species: small (sCM20) and large (lCM20) amorphous hydrocarbon grains and amorphous silicates with iron nano-particle inclusions (sil). The uncertainty on the best-fit SED is indicated with a grey shaded zone. Right: 1D and 2D posterior distributions for the starlight intensity (G) and the dust mass ($M_{\text{dust}}$).}
    \label{Crab_interstellar_SED}
\end{figure*}

\begin{figure*}
	\includegraphics[width=17.5cm]{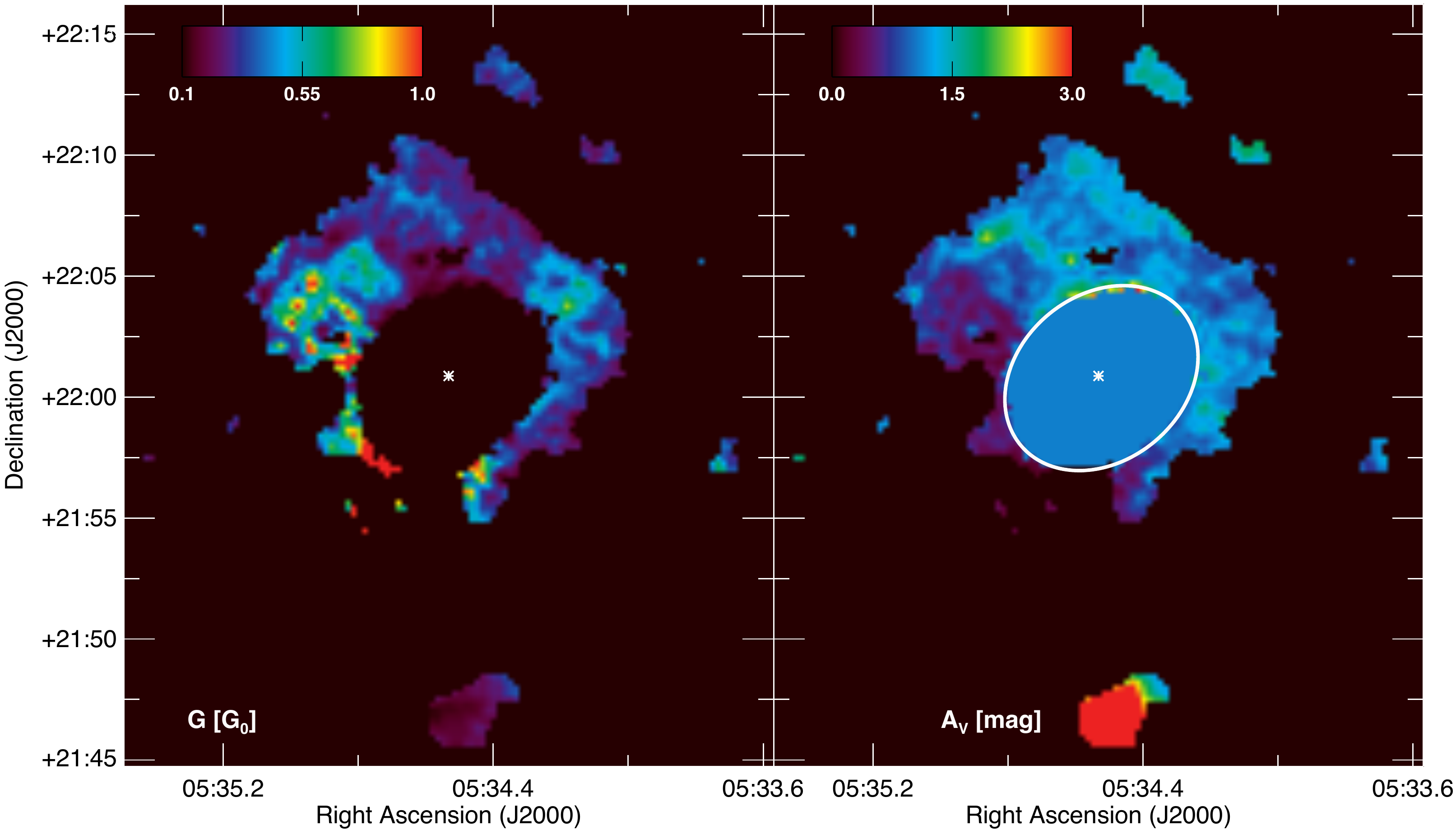} \\
	\includegraphics[width=8.5cm]{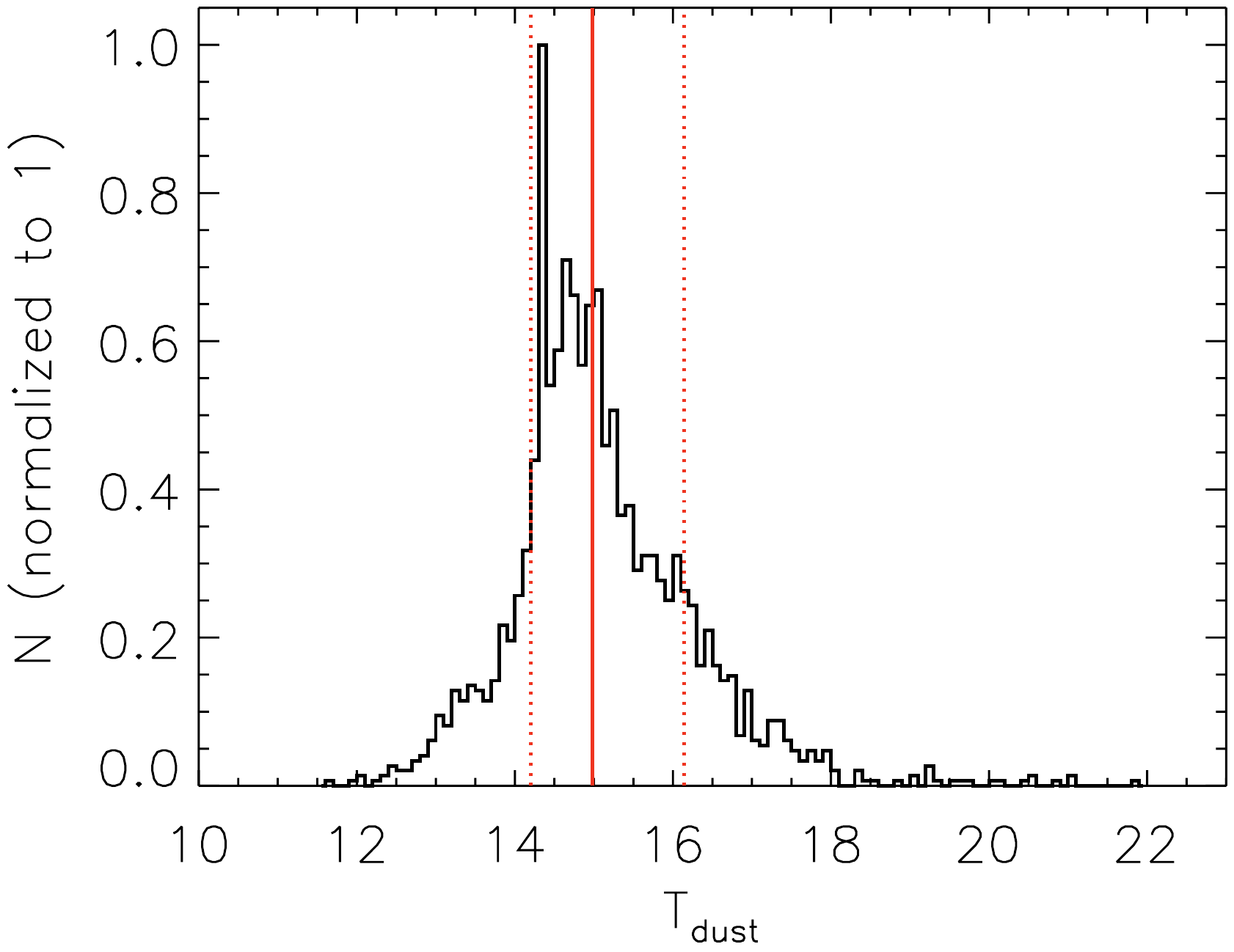}
	\includegraphics[width=8.5cm]{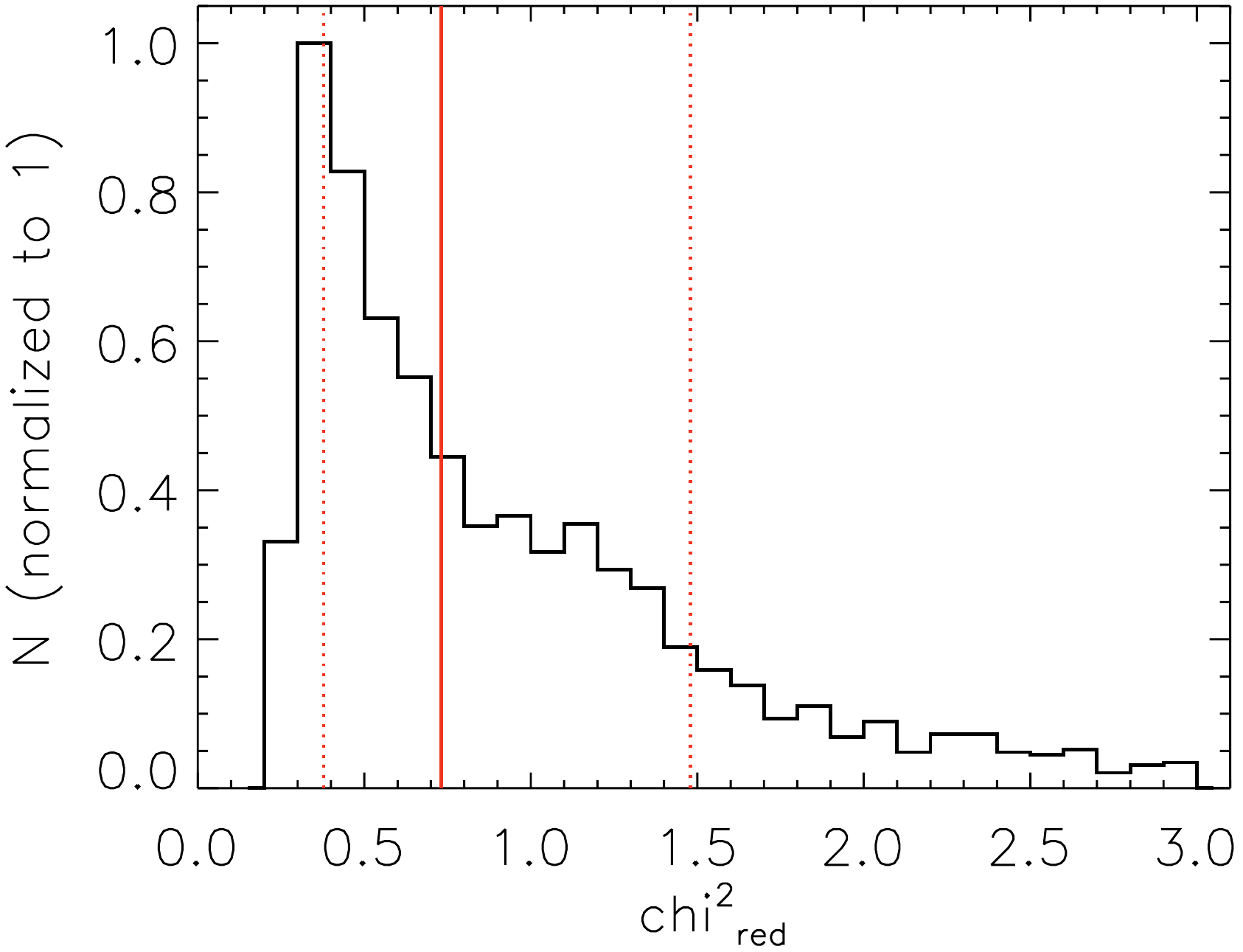}
    \caption{Top row: Map of the radiation field strength (G, left) and $V$ band extinction ($A_{\text{V}}$, right) for regions in the vicinity of the Crab Nebula. The colour scale in the right-hand panel was adjusted to cover the entire range of $A_{\text{V}}$ values, excluding the blob of high $A_{\text{V}}$ (up to an $A_{\text{V}}$=17) southwards of the Crab. Only pixels with detections in the PACS\,160\,$\mu$m, SPIRE\,250, 350 and 500\,$\mu$m were modelled, with the exception of pixels with possible SNR confusion. These latter pixels (within an ellipse with major- and minor axes of 260$\arcsec$ and 205$\arcsec$) were replaced with the average $A_{\text{V}}$=1.08 inferred from the field surrounding the Crab. In both panels, black-coloured regions correspond to pixels which did not justify the latter criterion and hence were not modelled. The white cross indicates the position of the Crab pulsar. Bottom row: Histograms with the distributions of the dust temperature (as inferred from G) and the reduced $\chi^{2}$ values that quantify the goodness of fit. The vertical solid, and dashed red lines indicate the median, and 16th and 84th percentiles of the distribution.}
    \label{Crab_interstellar_maps}
\end{figure*}

Figure \ref{Crab_interstellar_maps} displays the parameter maps constructed from the median likelihood parameter values in each pixel. Radiation field intensities vary between 0.1 and 1\,G$_{\text{0}}$, which corresponds to ISM dust temperatures between 12 and 18\,K (with a few pixels with $T_{\text{dust}}$ up to 22\,K). The median temperature of $15\pm1$\,K is significantly lower compared to the average Galactic dust temperature from \textit{Planck} ($T_{\text{dust}}$=19.8\,K, \citealt{2014A&A...571A..11P}). Because the average column densities are not particularly high, the low dust temperatures are thought to result from a reduced level of dust heating by starlight. The median $V$ band extinction corresponds to $A_{\text{V}}=1.08\pm0.38$\,mag\footnote{Note that excluding the blob of high $A_{\text{V}}$ south of the Crab hardly changes this median value ($A_{\text{V}}=1.05\pm0.36$\,mag).}. These $A_{\text{V}}$ values are consistent with the reddening $E(B-V)=0.39$ (or $A_{\text{V}}$=1.2\,mag) inferred by \citet{2015ApJ...810...25G}. 


The reduced $\chi^{2}$ values, representative of the goodness of fit in each pixel, are on average equal to 0.7, which suggests that the THEMIS dust model is adequate to reproduce the IS dust emission around the Crab Nebula. For pixels with $\chi^{2}_{\text{red}}$ values below unity, it is possible that the errors on the observed fluxes were slightly overestimated. 

To estimate the contribution of ISM dust emission along the line of sight to the Crab Nebula, we determined the median starlight intensity ($G=0.33^{+0.18}_{-0.09}G_{\text{0}}$) and $V$ band extinction ($A_{\text{V}}=1.08\pm0.38$\,mag) from the modelled pixels shown in Figure \ref{Crab_interstellar_maps}. We then convolved this median SED with the passband filter response curves to obtain the average estimated ISM dust flux density within a 14$\arcsec\times$14$\arcsec$ pixel (see Table \ref{AverageISMdustfluxes}). The median ISM dust emission was subtracted from each pixel along the line-of-sight to the Crab Nebula prior to the resolved SED modelling (see Section \ref{ResolvedSED.sec}). We have also added the lower limit to the uncertainties in each pixel to account for the variation in ISM dust emission from one pixel to another. Even though the upper limits are higher, we have opted to consider the lower limits because our ISM dust estimate is likely biased towards high dust temperatures (due to our constraint to only model pixels with (at least) 3$\sigma$ detections in the PACS\,160\,$\mu$m map).

Total integrated ISM dust contributions are calculated by summing the estimated ISM dust emission for all pixels within the aperture used for photometry. Table \ref{Table_Fluxfraction} gives an overview of the estimated total integrated ISM dust emission in each waveband: 1.2, 4.6, 14.3, 21.6, 13.8 and 6.3$\%$ in the PACS\,70, 100 and 160, and SPIRE\,250, 350 and 500$\mu$m wavebands. We have assumed that the interstellar dust clouds are also located at a distance of 2\,kpc for convenience (in the Perseus arm). In the case where most interstellar dust would be located closer to us or instead further away, the estimated $A_{\text{V}}$ values will become smaller/larger but the relative contributions of ISM dust emission to the \textit{Herschel} bands will remain unaltered.

\begin{table}
\caption{Overview of the estimates for the median ISM dust emission ($F_{\text{ISM dust}}$) in the PACS and SPIRE wavebands within a 14$\arcsec\times$14$\arcsec$ pixel, with the lower and upper limits on these estimates inferred from the 16th and 84th percentiles of the $G$ and $\log M_{\text{dust}}$ distributions, respectively.}
\label{AverageISMdustfluxes}
\begin{tabular}{lc} 
\hline
Waveband &  $F_{\text{ISM dust}}$ [mJy] \\
\hline
PACS\,70 & 4.01$^{+6.50}_{-2.50}$ \\
PACS\,100 & 14.93$^{+21.66}_{-9.16}$ \\
PACS\,160 & 37.26$^{+38.95}_{-20.84}$\\
SPIRE\,250 & 37.39$^{+29.91}_{-19.27}$\\
SPIRE\,350 & 22.91$^{+15.68}_{-11.23}$\\
SPIRE\,500 & 10.43$^{+6.32}_{-4.92}$\\
\hline 
\end{tabular}
\end{table}

\section{Alternative explanations for the mm excess emission in the Crab}
\label{App_mmexcess}
In this Appendix, we briefly discuss alternative scenarios (other than the effects of a secondary synchrotron component, see Section \ref{OriginMillimetreExcess.sec}) that could play a role in explaining the observed mm excess emission in the central regions of the Crab Nebula.

\textbf{(2.) Free-free emission:} Using the same method (based on the attenuation corrected H$\beta$ emission) as outlined by \citet{2012ApJ...760...96G}, we infer an upper limit for the contribution of free-free emission at 2\,mm of 0.65\,Jy, which is well below 1$\%$ of the excess emission level at that wavelength. This demonstrates that free-free emission can not be responsible for the observed excess at mm wavelengths.
\smallskip \\
\textbf{(3.) Spinning dust grains:} Anomalous Microwave Emission (AME) is the name given to the mm excess emission observed in a handful of Galactic dust clouds \citep{2013ApJ...770..122T} and extragalactic sources \citep{2010A&A...523A..20B,2010ApJ...709L.108M,2018ApJ...862...20M}, and to explain mm-excess emission in the cosmic microwave background \citep{1998ApJ...494L..19D,2010A&A...509L...1Y}. The origin of this AME has been argued to be rotational emission from ultra-small dust grains \citep{1998ApJ...508..157D}, which contribute mostly in the frequency range between 10 and 60\,GHz (or 500\,$\mu$m and 3\,mm) (see \citealt{2018NewAR..80....1D} for a recent review). Such electric dipole radiation from small grains was also postulated to have been observed in (at least) two Galactic SNRs, 3C\,396 \citep{2007MNRAS.377L..69S} and IC\,443 \citep{2017AJ....153...32O}, in which excess emission was detected near frequencies of 33\,GHz and 20-70\,GHz, respectively. We note that both of these Galactic SNRs are also classified as composite SNRs, with a pulsar wind nebula emitting synchrotron radiation that originates from relativistic non-thermal electrons in a magnetised plasma, and from a shell of expanding supernova ejecta material.

Intuitively, AME from spinning dust grains is expected to correlate with far-infrared thermal emission as both should be arising from the same dust component. It has been demonstrated that this is not always the case, and that AME correlates better with the total far-infrared radiance than with the \textit{Planck} 353\,GHz optical depth \citep{2016ApJ...827...45H}. This result was taken as evidence that polycyclic aromatic hydrocarbons (PAHs) are not necessarily the dominant source of spinning grain emission, and that instead nanosilicates (with a minor contribution from iron nanoparticles) could be responsible for the bulk of AME \citep{2017ApJ...836..179H}. Specifically for the Crab, we do not find a correlation between the dust mass column density (see Fig.\ref{Images_Crab_SED_resolved}, bottom left panel) and the location of the mm excess emission (see Fig.\ref{Images_Crab_mm_resolved}, right column). Since dust is present in the central regions which show the excess, we can not rule out that spinning dust grains are responsible for the mm excess based on positional arguments alone.

\begin{figure}
	\includegraphics[width=8.25cm]{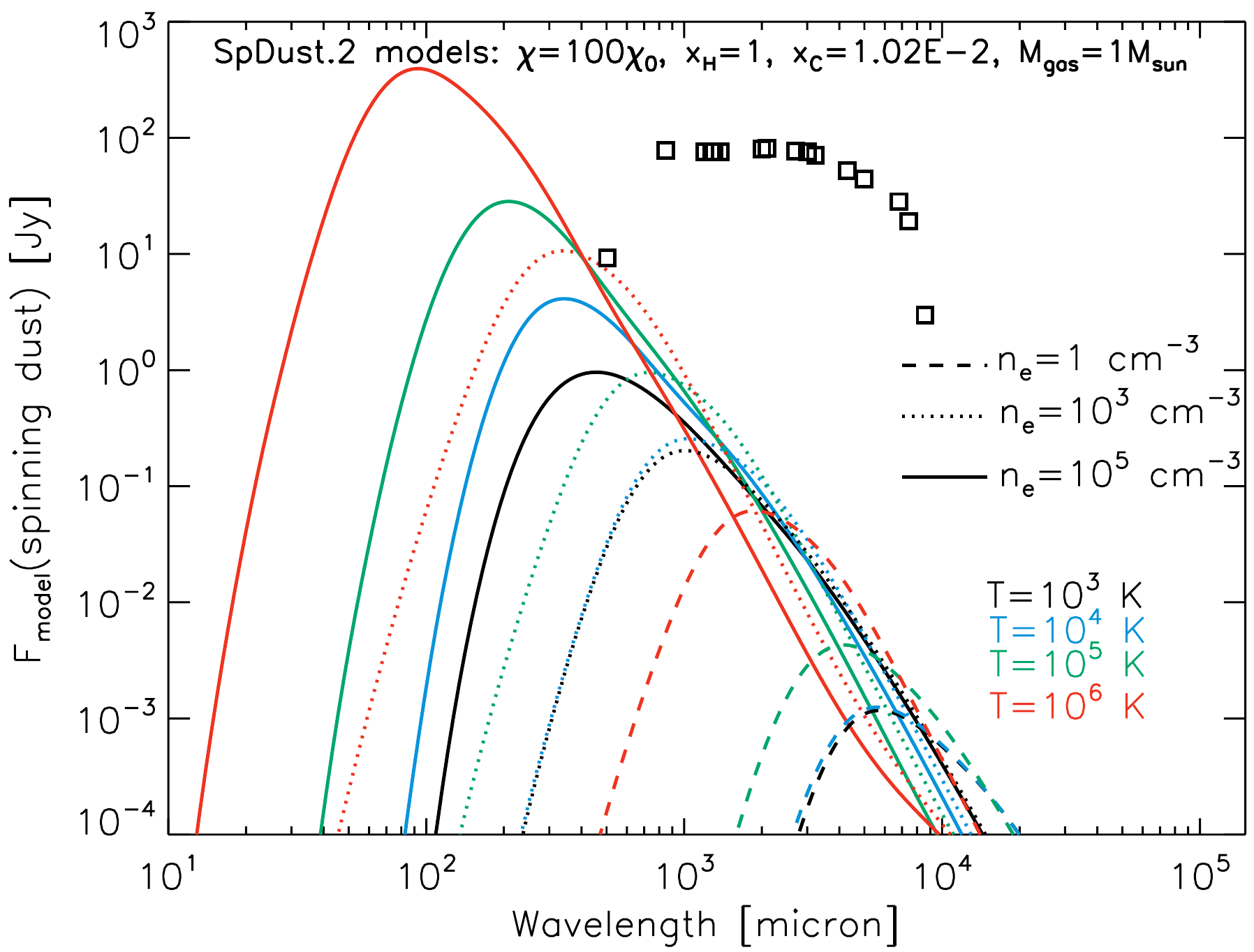}	
    \caption{The excess emission observed at mm wavelengths in the Crab Nebula (black squares) is compared to spinning dust grain models for varying gas conditions. The legend explains the model curves corresponding to various gas temperatures (T) and densities (n).}
    \label{SED_Crab_spdust}
\end{figure}

To verify whether the mm excess emission observed in the Crab can be attributed to spinning dust grains, we have used the spinning dust models from \citet{2009MNRAS.395.1055A} and \citet{2011MNRAS.411.2750S} and their associated \texttt{SpDust2}\footnote{http://cosmo.nyu.edu/yacine/spdust/spdust.html} code to predict the expected contributions for a variety of gas temperatures, gas densities and radiation field intensities. The grain size distribution for carbon grains (up to 12\AA, as larger grains hardly produce AME) was adopted from \citet{2001ApJ...548..296W}. We have assumed a scaling factor for the \citet{1983A&A...128..212M} UV radiation field of $\xi=100\,\xi_{\text{0}}$, in agreement with the values found by \citet{2017MNRAS.472.4444P}. The ionisation fraction is assumed to be $x_{\text{H}}=1$ as most material is ionised in the vicinity of the pulsar wind nebulae. The fractional ionised carbon abundance $x_{\text{C}}=n_{\text{C+}}/n_{\text{H}}$ is assumed to be $1.02\times10^{-2}$ which was adopted from the gas phase carbon abundance inferred from combined photo-ionisation and radiative transfer modelling by \citet{2015ApJ...801..141O}. Models were scaled to a gas mass of $M_{\text{gas}}=1\,M_{\odot}$ which corresponds to a dust mass of 0.03-0.04\,$M_{\odot}$ assuming a gas-to-dust mass ratio of 26-39 as inferred by \citet{2015ApJ...801..141O}\footnote{These gas-to-dust mass ratios were inferred based on different estimates of the Crab's gas and dust masses, and therefore only provide us with a rough estimate of the Crab's gas-to-dust mass ratio that would be applicable here.}. The residual total integrated mm fluxes were compared to the predicted spinning grain emission for gas densities of 1, 10$^{3}$ and 10$^{5}$ cm$^{-3}$ and gas temperatures T=10$^{3}$, 10$^{4}$, 10$^{5}$, and 10$^{6}$\,K (see Fig. \ref{SED_Crab_spdust}). The emission for models with $n$=1\,cm$^{-3}$ (typical of warm ionised gas) peaks at mm wavelengths, but the amplitude of these models is at least three orders of magnitude too low to account for the observed mm excess. The amplitude of models with higher densities ($10^{3}-10^{5}$\,cm$^{-3}$, as inferred for the ejecta filaments, \citealt{2017MNRAS.472.4444P}) is of the order of 10$^2$\,Jy, similar to the residual mm fluxes. But due to the increased density, the AME shifts towards the infrared wavelength domain, and hardly contributes to the mm flux. These models have already assumed the maximum possible dust mass (at the limit of what our SN dust models predict for carbon and/or silicate grains), and have adopted an optimistic value for the ionised carbon abundance. Within the uncertainty limits remaining in the model\footnote{The model employs the \citet{1983A&A...128..212M} UV radiation field, and does not account for more energetic radiation shortwards of the Lyman limit.}, we argue that it is unlikely that spinning dust grains are responsible for the observed mm-wave excess emission from the Crab Nebula based on (a.) the positional offset between the mm excess emission and the dominant grain population in the Crab, and (b.) the incompatibility between the observed and predicted excess emission levels using spinning dust grain models.
\smallskip \\
\textbf{(4.) Magnetic dipole emission from magnetic nanoparticles:} Other than electric dipole emission from nanoparticles, magnetic dipole emission due to thermal fluctuations in the magnetisation of magnetic grain materials can also produce AME with expected contributions at (sub-)mm wavelengths \citep{2012ApJ...757..103D,2013ApJ...765..159D}. These magnetic grains can be pure metallic iron grains, or iron nanoparticles with typical grain sizes $a<$0.01$\mu$m, with the latter being inclusions in other larger grains. The size of dust grains in the Crab Nebula is still open for debate. Most previous studies have suggested that grain size distributions are dominated by large grains \citep{2013ApJ...774....8T,2015ApJ...801..141O}, but there are strong variations in the inferred grain size distributions for different species. \citet{2013ApJ...774....8T} suggest maximum sizes of 0.1-0.6\,$\mu$m for carbon grains, while silicate grains could grow up to 5\,$\mu$m in size, with a non-negligible contribution from smaller grains for both dust species. \citet{2015ApJ...801..141O}, on the other hand, favour a size distribution dominated by large grains with maximum grain sizes up to 0.5-1\,$\mu$m as inferred from fitting dust radiative transfer models to the Crab's infrared and submm dust SED. Similarly, Priestley et al.\,(in prep.) retrieve a power-law grain distribution with a slope between -2.5 and -3.3, flatter than the typical power-law distribution assumed for the interstellar medium (-3.5, \citealt{1977ApJ...217..425M}), and suggest that the supernova dust mass is dominated by grains with sizes $a>$0.1\,$\mu$m.

The presence of a powerful pulsar, and the detection of polarised emission from the Crab, with an average polarisation fraction of 8.8$\%$ with a peak value 30$\%$ \citep{2010A&A...514A..70A}, supports the idea that a strong magnetic field is present in the Crab Nebula, which could be responsible for the magnetisation of small grains. At the same time, the Crab's volume is dominated by a highly ionised plasma which provides a hostile environment for the smallest grains ($a<$0.01$\mu$m) that are most easily magnetised. We therefore speculate that magnetic dipole radiation is unlikely to account for the bulk of the mm-wave excess emission in the Crab Nebula. However, a more detailed assessment of the dust properties and grain sizes in the Crab Nebula might be required before this scenario can be fully ruled out. 

\section{Evolutionary synchrotron spectrum}
\label{Alternative_synchrotron_model.sec}

The near-infrared to radio spectral range coincides with the location of a spectral break in the Crab Nebula, associated with a change in the evolutionary regime that affects the emitting electrons. The evolution of the electrons emitting at longer wavelengths is thought to be dominated by adiabatic losses while that of those emitting at shorter wavelengths is dominated by synchrotron losses, and this causes a change of spectral slope \cite[see, e.g.][]{1973ApL....13..103P}. The intermediate range of energies between these two asymptotic regimes is rather broad, and this leads to a smooth transition of the spectral break, combined with effects like the presence of a spectral bump associated with this transition \cite[see, e.g.][]{2009ApJ...703..662R}.

We have attempted to model this spectral transition, whose shape depends both on the slope of the energy distribution at injection and on some details of the evolutionary phase of the pulsar wind nebula. Hereto, we have assumed that the magnetic field $B(t)$ does not depend on position (but has a time-dependence), and there is a homogeneous injection of new particles $J(E,t) = K(t) E^{-s}$ (where $s=2*\alpha_{\text{radio}}+1$ depends on the shock compression ratio, and is linked to the radio spectral index $\alpha_{\text{radio}}$), distributed throughout the nebula. In order to model the effects of the nebular evolution, we have assumed that the nebular radius and magnetic field can be described by power laws of time (respectively $R(t)\propto t^\delta$, $B(t)\propto t^{-\beta}$). This has allowed us to verify that the dependence of the shape of the spectral transition is only moderately dependent on these parameters, while the main effect is introduced by the slope of the energy distribution at injection. We have used a monochromatic approximation for synchrotron emission, i.e., we have assumed that electrons emit only at a given frequency, identified by the peak of the synchrotron spectrum for a mono-energetic particle distribution. This approximation works well in the case of a smooth particle energy distribution.

With the aim of minimizing the number of free parameters, we have modelled the evolution of the Crab pulsar and its nebula in detail using a one-zone model, and the assumption that the pulsar spin-down power is converted (assuming fixed fractional contributions) into newly injected particles and into the magnetic field. We have then used this model to constrain the evolutionary parameters $\delta$ and $\beta$ (i.e., $\delta$=1.16, $\beta$=1.50), while varying the slope of the injected spectrum. 

\begin{figure*}
	\includegraphics[width=16.5cm]{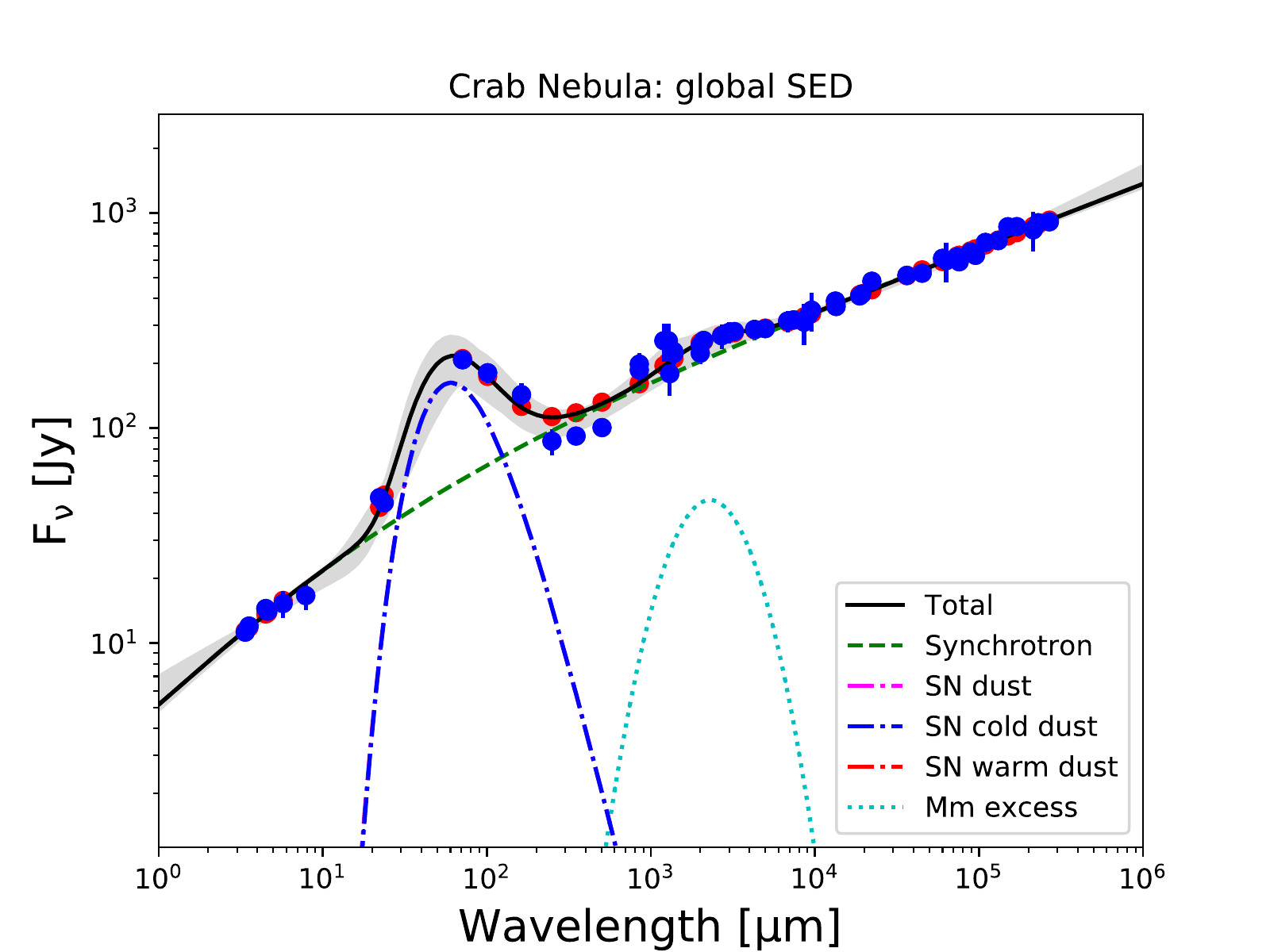}
    \caption{Total integrated spectral energy distribution for the Crab Nebula extending from near-infrared to radio wavebands, assuming a synchrotron spectrum with evolutionary break instead of a broken power-law. The best-fit SED model is indicated with a black solid line, with the grey shaded region corresponding to the 16th and 84th percentiles of the N-dimensional likelihood (i.e., corresponding to the 1$\sigma$ upper and lower bounds to the model). The blue dots correspond to the observed datapoints (with the uncertainties shown as vertical lines), while the red circles indicate the model fluxes in those wavebands. We have also included the emission of individual model components: synchrotron radiation (green dashed curve), mm excess emission (blue dashed curve), cold and warm SN dust SEDs (blue and red dot-dashed curves, respectively), and the combined SN dust emission (purple dot-dashed curve).} 
    \label{Crab_global_evolutive_spectrum}
\end{figure*}

\begin{figure*}
	\includegraphics[width=18.0cm]{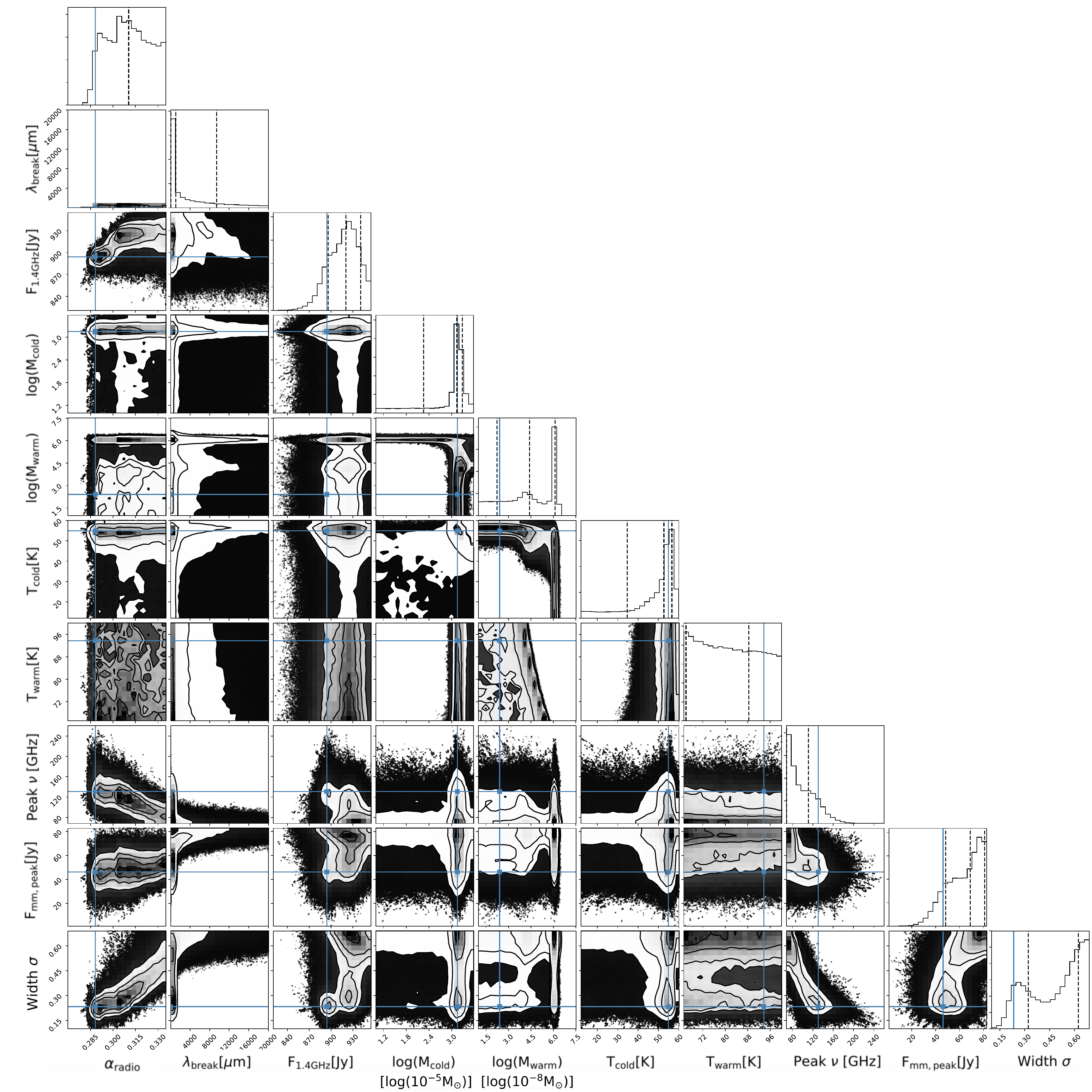}
    \caption{Corner plot resulting from a total integrated SED fitting procedure assuming a synchrotron spectrum with a smooth evolutionary break. The contour plots correspond to 2D posterior distributions indicating the probability of two parameters in a 2D plane, where contours represent the 0.5$\sigma$, 1.0$\sigma$, 1.5$\sigma$ and 2.0$\sigma$ likelihoods. The histograms correspond to 1D marginalised posterior distribution plots showing the likelihood that a certain value will be assigned to a given parameter. The maximum likelihood (blue solid curve) corresponds to the ``best-fit" solution. The black dashed lines correspond to the ``median likelihood", and the 16th and 84th percentiles to reflect the lower and upper bound due to uncertainties on these parameter values.}
    \label{Crab_global_evolutive_spectrum_errors}
\end{figure*}

Based on this model, we have produced a grid of evolutionary synchrotron models with radio spectral indices $\alpha_{\text{radio}}$ between 0.27 and 0.38 (in steps of 0.005) and break wavelengths between 20\,$\mu$m and 2\,cm under the assumption of flat priors. Figures \ref{Crab_global_evolutive_spectrum} and \ref{Crab_global_evolutive_spectrum_errors} present the best-fit model SED that reproduces the Crab's multi-wavelength observations, and the 1D and 2D posterior distributions for the 10 model parameters, respectively. Table \ref{DustMassesOverview} summarises the median likelihood parameter values inferred from this fit for the supernova dust model (where amorphous carbon a-C grains with size a=1\,$\mu$m were assumed for the supernova dust), while Table \ref{SynchrotronOverview} reviews the synchrotron and mm excess model parameters.

The total supernova dust mass (0.014$^{+0.018}_{-0.012}$\,M$_{\odot}$) and the dominant cold dust temperature (53$^{+4}_{-18}$\,K) agree very well with the supernova dust model parameters inferred from a model with a broken power-law synchrotron spectrum. However, the smooth break in the synchrotron spectrum overestimates the SPIRE flux densities. To compensate for this, the Bayesian approach favours models with a steep radio spectral index ($\alpha_{\text{radio}}$=0.359$^{+0.016}_{-0.049}$), and attempts to fit any residual mm, cm and radio emission with a broader mm excess component. The best-fit model (shown in Fig. \ref{Crab_global_evolutive_spectrum}) with $\alpha_{\text{radio}}$=0.288 appears to still provide a reasonable fit to the data, but the 1D posterior distribution (see Fig. \ref{Crab_global_evolutive_spectrum_errors}, top left panel) shows that models with steeper synchrotron slopes have an increased likelihood, which is inconsistent with measurements from the radio spectral index for the Crab (\citealt{1977A&A....61...99B,1997ApJ...490..291B,2012ApJ...760...96G}). The mm excess component is less prominent in the best-fit model with a peak flux density of 46\,Jy at 131\,GHz (compared to 69\,Jy at 163\,GHz for a broken power-law synchrotron model), which might suggest that an evolutionary synchrotron model is able to account for part of the mm excess emission observed in the Crab Nebula. However, our model with $\alpha_{\text{radio}}$=0.288 does not have a strong spectral bump related to this evolutionary break, and can therefore not (fully) explain the origin of the mm-excess emission. Further refinement of our synchrotron model with evolutionary break and its contribution to the mm emission of the Crab is beyond the scope of this work, and will be deferred to future work. We remind the reader that our model assumptions for the synchrotron spectrum do not affect the inferred model supernova dust masses.   

\newpage

\section{Figures}
\label{Figures.sec}
This section contains additional figures discussed in the main text of the paper. 

\begin{figure*}
	\includegraphics[width=18.cm]{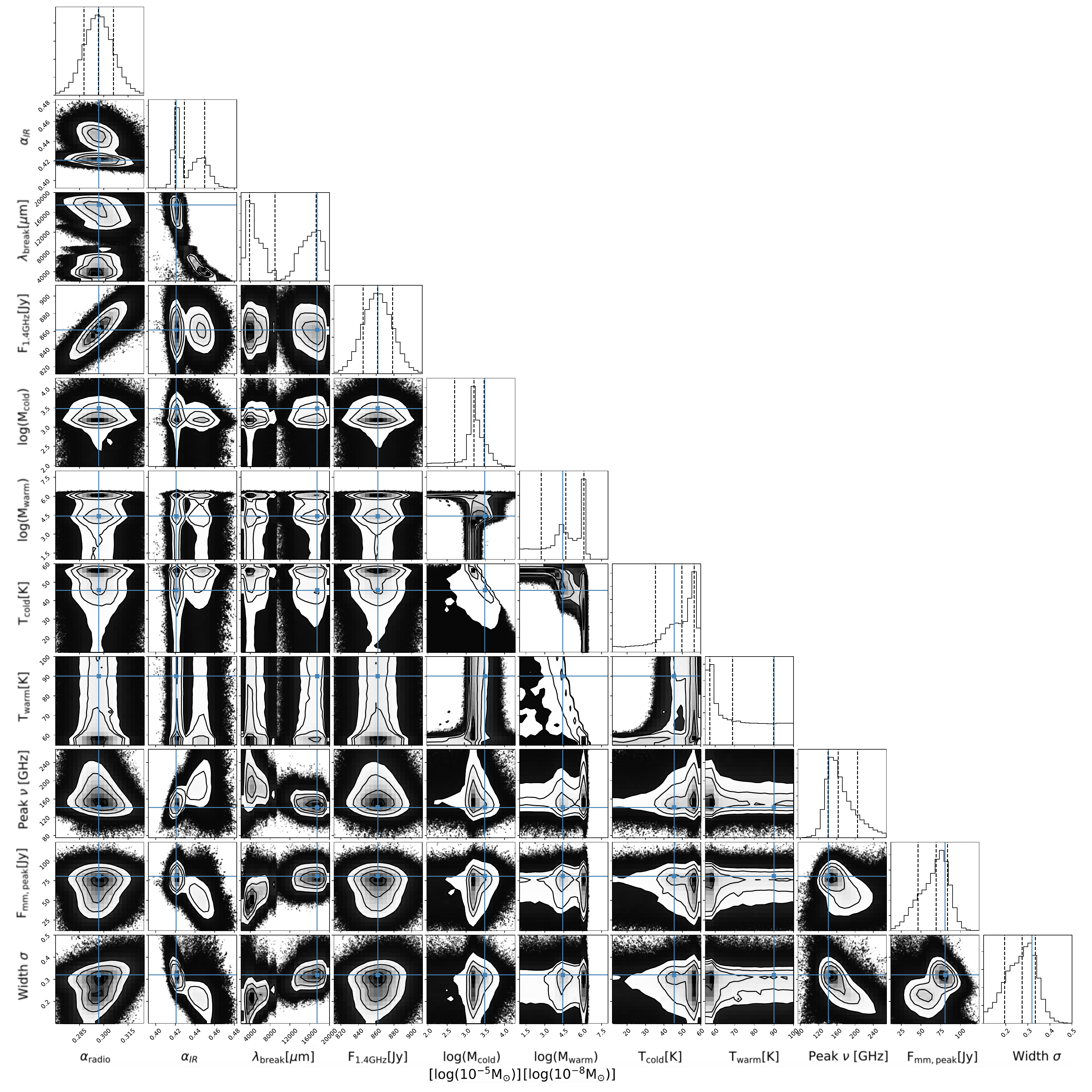}
    \caption{Corner plot resulting from a total integrated SED fitting procedure with a broken power-law synchrotron spectrum. The contour plots correspond to 2D posterior distributions indicating the probability of two parameters in a 2D plane, where contours represent the 0.5$\sigma$, 1.0$\sigma$, 1.5$\sigma$ and 2.0$\sigma$ likelihoods. The histograms correspond to 1D marginalised posterior distribution plots showing the likelihood that a certain value will be assigned to a given parameter. The maximum likelihood (blue solid curve) corresponds to the ``best-fit" solution. The black dashed lines correspond to the ``median likelihood", and the 16th and 84th percentiles to reflect the lower and upper bound due to uncertainties on these parameter values.}
    \label{Crab_global_spectrum_errors}
\end{figure*} 

\begin{figure*}
	\includegraphics[width=18.0cm]{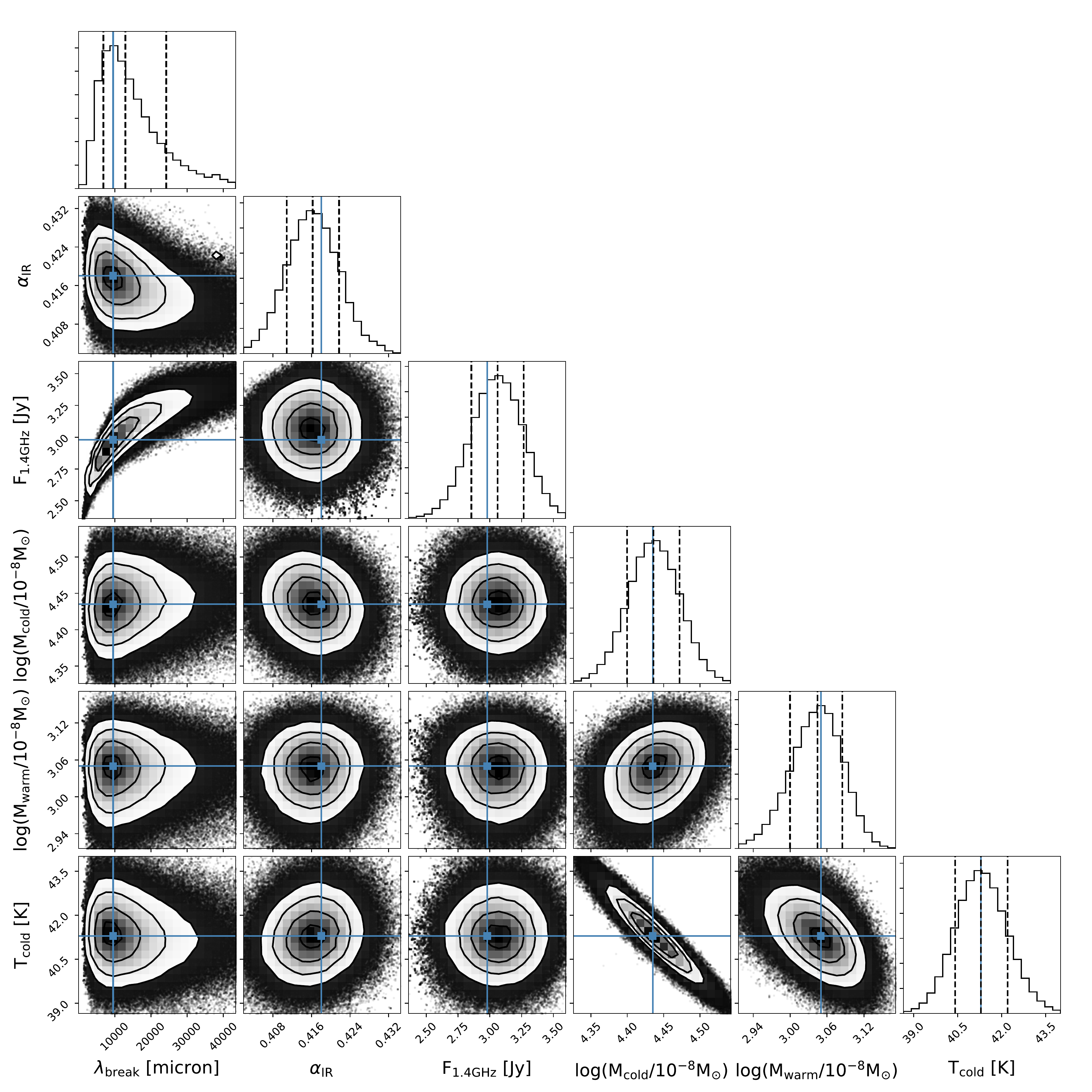}
    \caption{Corner plot resulting from a spatially resolved SED fitting procedure for a centrally located pixel. The contour plots correspond to 2D posterior distributions indicating the probability of two parameters in a 2D plane, where contours represent the 0.5$\sigma$, 1.0$\sigma$, 1.5$\sigma$ and 2.0$\sigma$ likelihoods. The histograms correspond to 1D marginalised posterior distributions demonstrating the likelihood that a certain value will be assigned to a given parameter. The maximum likelihood (blue solid curve) corresponds to the ``best-fit" solution. The black dashed lines correspond to the ``median likelihood", and the 16th and 84th percentiles to reflect the lower and upper bound due to uncertainties on these parameter values.}
    \label{Crab_resolved_spectrum_errors}
\end{figure*} 

\bsp	
\label{lastpage}
\end{document}